%% file: main.tex
\documentclass{lmcs}
\pdfoutput=1

\usepackage{lastpage}
\lmcsdoi{16}{4}{7}
\lmcsheading{}{\pageref{LastPage}}{}{}%
{Feb.~12,~2020}{Oct.~27,~2020}{}


\usepackage{subfigure}
\usepackage{mathtools}
\usepackage{multirow}
\usepackage{booktabs}

\usepackage[utf8]{inputenc}

\theoremstyle{plain}

\input{macroChoreography}
\input{macroProoftree}
\begin{document}

\title[Collaboration vs. Choreography Conformance in BPMN]{Collaboration vs. Choreography Conformance in BPMN}

\author[F.~Corradini]{Flavio Corradini}
\address{Computer Science Division, School of Science and Technology, University of Camerino}
\email{\{flavio.corradini,andrea.morichetta,andrea.polini,barbara.re,francesco.tiezzi\}@unicam.it\vspace{-1cm}}
\author[A.~Morichetta]{Andrea Morichetta}
\author[A.~Polini]{Andrea Polini}
\author[B.~Re]{\texorpdfstring{\\}{}Barbara Re}
\author[F.~Tiezzi]{Francesco Tiezzi}

\begin{abstract}
\noindent The \bpmn{} 2.0 standard is a widely used semi-formal notation to model distributed information systems from different perspectives. The standard makes available a set of diagrams to represent such perspectives. \emph{Choreography} diagrams represent global constraints concerning the interactions among system components without exposing their internal structure.  \emph{Collaboration}  diagrams instead permit to depict the internal behaviour of a component, also referred as process, when integrated with others so to represent a possible implementation of the distributed system.

  This paper proposes a design methodology and a formal framework for checking conformance of choreographies against collaborations. In
  particular, the paper presents a direct formal operational semantics
  for both \bpmn{} choreography and collaboration diagrams. Conformance aspects are proposed through two relations defined on
  top of the defined semantics. The approach benefits from the
  availability of a tool we have developed, named
  \emph{C}$^{\ \!\!4}$, that permits to experiment the theoretical
  framework in practical contexts.  The objective here is to make the
  exploited formal methods transparent to system designers, thus
  fostering a wider adoption by practitioners.
\end{abstract}

\maketitle

\section{Introduction}%
\label{sec:intro}

The \bpmn{} 2.0 standard is a widely used semi-formal notation to model different perspectives of distributed information systems~\cite{omg_business_2011}. \ap{Available diagrams can be used at different stages in the development life-cycle, while focusing on specific aspects of the system under construction, and also according to different development strategies
~\cite{steffen_model-driven_2017,aldazabal_automated_2008}.
In the context of this work, particularly relevant are the diagrams referred as \emph{Choreography}, \emph{Process}, and \emph{Collaboration}.
The first kind of diagram, \emph{Choreography}, permits to represent
 the interactions among cooperating entities (e.g., software components, organisations, services, etc.), without exposing their internal structure. A \emph{Process} diagram, instead, intends to represent the actions and the choices that a single entity puts in place in order to reach specific objectives. Finally, a \emph{Collaboration} diagram permits to represent the compositions of different processes, which include communication actions. In such a way, this latter kind of diagram permits to detail both the communication schema, and the internal actions and choices enabling a correct cooperation.
}

In such a setting, organisations that are willing to cooperate can
refer to a choreography specification detailing how they should interact to reach specific objectives. On the other hand, the
cooperation needs \ap{to involve entities (e.g., software systems) that often are already available, or that will be specifically introduced for the purpose, within each single organisation.
In the first case, the process to be integrated can be considered as a sort of legacy element that will directly derive from the specification of the existing entity, while in the latter case it can be convenient to shape it starting from the global choreography specification (e.g., using projection mechanisms).
The integration of each single process into an overall specification of the system leads to the collaboration diagram previously mentioned. Nevertheless, a collaboration that integrates different processes to reach the objectives specified in a choreography should show a behaviour somehow related to that defined by the global specification, independently from the genesis of the involved processes (legacy or defined for the purpose). Indeed, the choreography acts as a sort of contract among the participating parties, and each organisation expects that the others will follow it.} The {conformance} of a given collaboration with respect to a pre-established choreography becomes then crucial, since it permits to ensure that the integrated components are able to successfully collaborate, \ap{or can be possibly and reasonably adapted}, without invalidating the communication constraints imposed by the global specification, so to reach the objectives defined by the choreography itself.
In the general context of service-oriented systems this problem has
received a lot of attention~\cite{papazoglou2007service,baldoni2005verifying,liu2007static,el2008business,rouached2012web,martens2003compatibility}. Notwithstanding this effort, there is still a lack of frameworks and tools supporting the conformance checking between collaboration and choreography models when the \bpmn\ notation is considered.

Given such a gap and in order to fill it, we provide in this paper
\textbf{a novel framework to check the conformance of \bpmn{}
 choreography  diagrams with respect to \bpmn\ collaborations
  models}. In particular, in our approach we do not resort to a
different intermediate language or formalism;
instead we rely on a direct semantics
describing the behaviour of both models.
\ft{Moreover, we do not impose any syntactic restriction
on the usage of the BPMN modelling notation, i.e.~we allow models
to have an arbitrary topology.}
Clearly, the semantics embeds
the peculiarities of the \bpmn{} standard when used to model
distributed systems (e.g., asynchronous communication among
components). More specifically, to formally describe the behaviour of
a system the operational semantics associates
Labelled Transition Systems (LTSs) to its \bpmn\ models.
A collaboration and a choreography
can then be compared to check the satisfaction of specific behavioural
relations, considering the LTSs resulting from the defined semantic
framework. We rely on a conformance relation (based on
bisimulation~\cite[Sec.~5]{milner_communication_1989}) that is
sensitive to deadlocks and different forms of non-determinism, and on
another relation (based on traces~\cite[Sec.~9.4]{milner_communication_1989}) that instead is more
relaxed on this respect, \ap{and that however can provide useful information in order to possibly drive the
\new{identification of interventions}
permitting to avoid the highlighted issues}. The support of both kinds of relations allows the system
designer to decide the desired trade-off between the strength of the
properties ensured by the system, and the breadth of choice among
available system components.

The developed theoretical framework has
been implemented in the \emph{\name{} (Collaboration vs Choreography
  Conformance Checker for \bpmn)} tool.  Standard input formats for
the \bpmn{} models are accepted by the tool so to enable its
integration with external \bpmn{} modelling environments (e.g.,
Camunda, Signavio and Eclipse \bpmn{}2 Modeller). The tool permits to
hide the underlying formal methods permitting to system designer, not
accustomed with formalisms and formal reasoning, to access and use
well established theories.

Summing up, the major contributions of
this paper are as follows:
\new{(i)} definition, and implementation in Java, of a formal operational
semantics for \bpmn{} choreographies and collaborations,
\ap{in particular making the integration
of existing process specifications a viable option to form collaborations}; %
\new{(ii)} definition, and implementation, of two conformance relations; %
\new{(iii)} implementation of the \name{} tool supporting the proposed methodology and conformance checking framework.
This paper is a revised and extended version of~\cite{DBLP:conf/edoc/Corradini0P0T18}. Specifically, we extend our previous work by proposing a new dedicated design methodology; in consequence, both the formal framework and the related supporting tool have been revised accordingly. \ap{In particular, we now consider and support also more realistic scenarios in which the organisations can integrate  already available artefacts, equipped with corresponding process specifications.}

\medskip
\noindent
\emph{Outline.} Section~\ref{sec:back} provides background notions on the \bpmn\ modelling notation, with a
particular emphasis on choreography and collaboration diagrams.
Moreover, in this section a running example is introduced; this
will be used in the rest of the paper to clarify various
aspects of the proposed framework. Section~\ref{sec:method}
discusses the life-cycle of a choreography specification, and how the
developed theoretical framework, and the related tool, fits in such a
setting.  Section~\ref{sec:formal} introduces formal syntax and
semantics for  both choreographies and collaborations,
and it presents the conformance relations we
have defined. Successively, Section~\ref{sec:tool}  presents the
\name\ tool and illustrates its practical usage, \ft{Section~\ref{sec:discussion}
provides a detailed discussion about subtle points of our conformance checking approach,}
and Section~\ref{sec:related} discusses the scientific works much related to our proposal.
Finally, Section~\ref{conclusions} concludes the paper and discusses directions
for future work.

\section{Background Notions}%
\label{sec:back}
This section first provides some basic notions on elements that can be included in BPMN choreography and collaboration diagrams, then it introduces a scenario that will be used as a running example.

 \begin{figure*}[t]
 	\centering
	\begin{tabular}{ccc}
	\hspace{-0.5cm}
	\includegraphics[height=34mm]{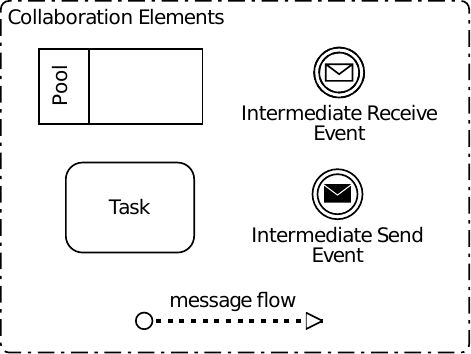}
&
	\includegraphics[height=34mm]{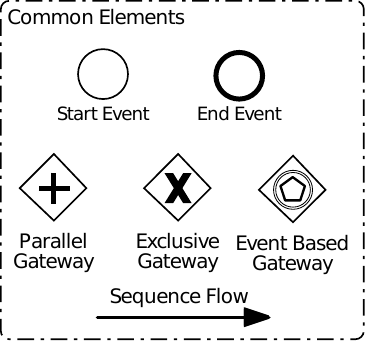}
&
	\includegraphics[height=34mm]{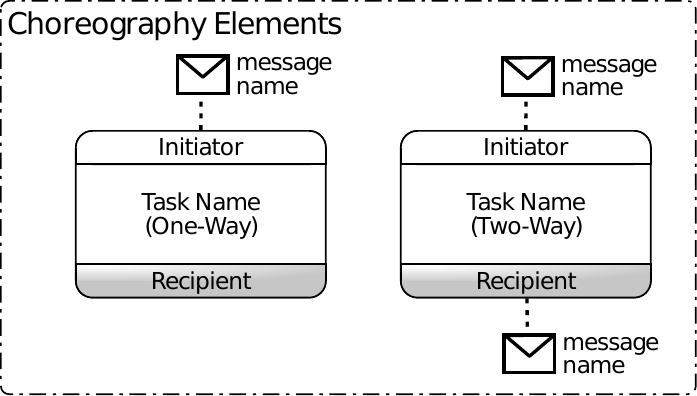} \\
          \footnotesize{a.}          &          \footnotesize{b.} &           \footnotesize{c.}
	\end{tabular}
 	\vspace{-.2cm}
 	\caption{BPMN 2.0 Collaboration and Choreography Elements.}%
 	\label{fig:bpmn_element}
 	\vspace{.2cm}
 \end{figure*}

\medskip
\noindent
\emph{\textbf{The BPMN Standard.}}
This paragraph includes details on the main concepts of BPMN we use in the following, in particular
focusing on the elements reported in Fig.~\ref{fig:bpmn_element}.

Fig.~\ref{fig:bpmn_element}.b depicts the modelling elements that can be included in all the diagrams considered in this paper.
\textbf{Events} are used to represent something that can
 happen. An event can be a \textit{start event}, representing the
 point in which the choreography/collaboration starts, while an \textit{end event} is
 raised when the choreography/collaboration terminates.
 \textbf{Gateways} are used to manage the flow of a choreography/collaboration both
  for parallel activities and choices. Gateways act as either join nodes (merging incoming sequence edges) or split nodes (forking into outgoing sequence edges).  Different types of gateways are available.
A \textit{parallel gateway (AND)} in join mode has to wait to be reached by all its incoming edges to start, and respectively all the outgoing edges are started simultaneously in the split case.
 An \textit{exclusive gateway (XOR)} describes choices; it is activated each time the gateway is reached in join mode and, in split mode, it activates exactly one outgoing edge.
  An \textit{event based gateway} is similar to the XOR-split gateway, but its outgoing branches activation depends on the occurrence of a catching event in the collaboration and on the reception of a message in the choreography; these events/messages are in a race condition, where the first one that is triggered wins and disables the other ones.
\textbf{Sequence Flows} are used to connect collaboration/choreography elements to specify the execution
flow.

In a collaboration or process diagram, also the elements in
Fig.~\ref{fig:bpmn_element}.a can be included.
\textbf{Pools} are used to represent participants involved in the collaboration.
\textbf{Tasks} are used to represent specific works to perform within a collaboration by a participant.
\textbf{Intermediate Events} represent something that happens during the flow of the process, such as sending or receiving of a message.
\textbf{Message Edges} are used to visualize communication flows between different participants, by connecting communication elements within different pools.

Focusing on the choreography diagram, we underline its ability to specify the message exchanges between two or more participants. This is done by means of  \textbf{Choreography Tasks} in
Fig.~\ref{fig:bpmn_element}.c.
They are drawn as rectangles divided in three bands: the central one refers to the name of the task, while the others refer to the involved participants: the white one is the initiator (sender), while the gray one is the recipient.
Messages can be sent either by one participant (One-Way tasks) or by both participants (Two-Way tasks).

\ft{In selecting the considered BPMN elements, we have mainly focused on the control flow and communication views. In doing that, we have followed a pragmatic approach to provide a precise characterization, and tool support, for a subset of BPMN elements that are largely used in practice.
\new{
Indeed, even though the BPMN specification is quite wide, only a limited part of its vocabulary is used regularly in designing BPMN models. This is witnessed by the models included in the BPM Academic Initiative repository\footnote{http://bpmai.org/}.
Moreover, a previous study we did~\cite{MPRWTPattern2018} also confirms that
our selection of BPMN elements is expressive enough to support %60\%
the majority of the interaction patterns proposed in~\cite{Barros2005}
(see,~\cite[Sec.~8.5]{Muzi19} for more details).
}
It is also worth noticing that most of the elements we have intentionally left out can be
expressed in terms of the elements we include.
Indeed, it is common to represent the inclusive gateways as a combination of exclusive
and parallel gateways enumerating all possible combinations of
outgoing edges activation~\cite{carmona2018conformance}.
Tasks with the loop marker can be easily represented by embedding standard tasks in
a looping behaviour expressed using two exclusive gateways (one in the split mode and the
other one in the join mode).
Moreover, we have left out timing elements, as they are not precisely specified in choreography diagrams~\cite[p. 341]{omg_business_2011}.
Other elements, such as those concerning error and compensation handling, are instead left out
in order to keep the formal framework more manageable.
Finally, for what concerns data objects and gateway conditions, we abstract them since we
aim at exhaustively analyzing all executions of a given model, and not only those resulting
from specific input values. The introduction of data, indeed, can only restrict the behaviour
of considered models. Notably, data modelling is optional in BPMN, as
the notation mainly focuses on control flow and, hence, only modelling constructs of this type are
mandatory. In summary, for the BPMN standard data-related elements remain somehow second class modelling constructs~\cite{meyer2011data}.
}

 \begin{figure*}[t]
 	\centering
	\begin{tabular}{cc}
	&
	\begin{tabular}{c}
  	\includegraphics[width=0.79\textwidth]{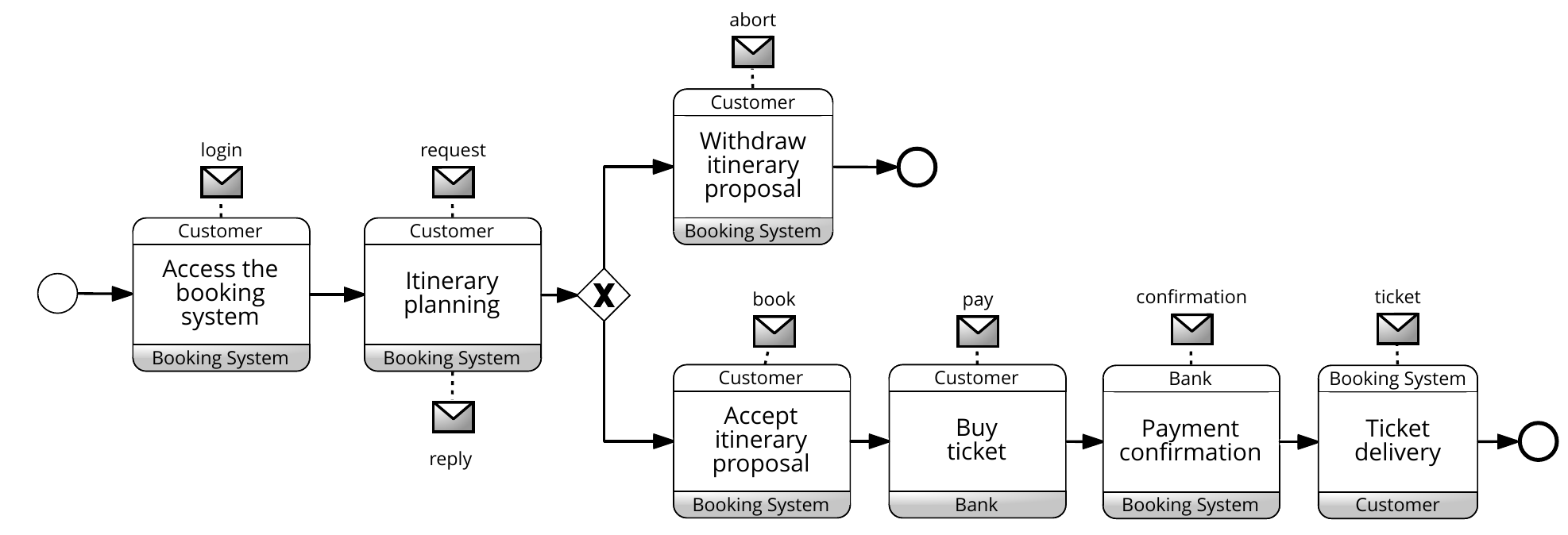}
	\\
	\footnotesize{a. Booking choreography}
	\\[4mm]

	\includegraphics[width=0.8\textwidth]{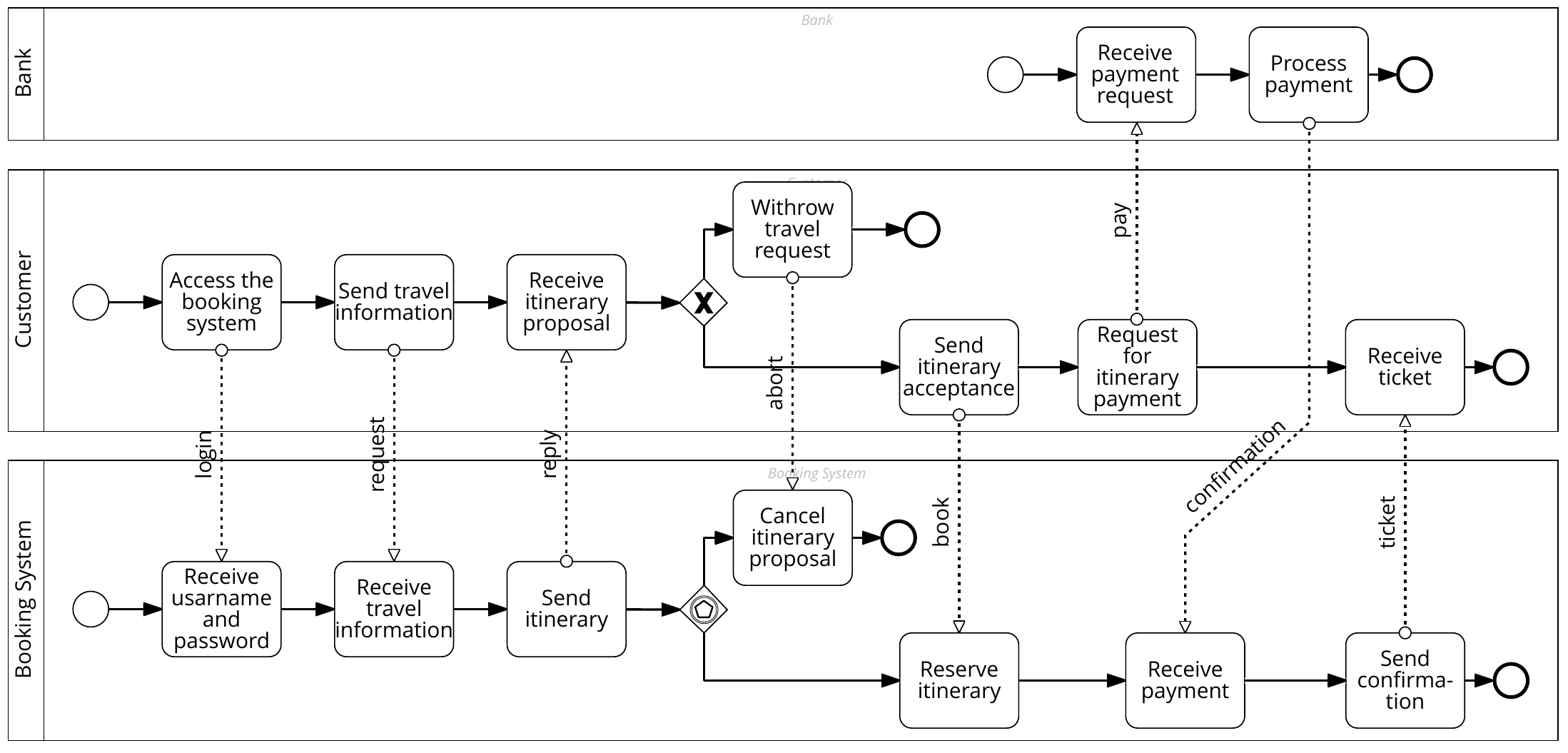}
	\\
	\footnotesize{b. Booking collaboration}
	\end{tabular}
	\end{tabular}
 	\caption{Booking Running Example.}%
 	\label{fig:example}
 	\vspace{-.2cm}
 \end{figure*}

\medskip
\noindent
\emph{\textbf{Running Example.}}
The collaboration and the choreography diagrams regarding a booking system introduced here are successively used in the paper to illustrate the various aspects of the proposed framework.

\smallskip
\noindent
\emph{Choreography Example}. The choreography in Fig.~\ref{fig:example}.a  combines the work-activities
of a booking system, a customer and a bank. They interact in order to book and pay for travel.
After accessing to the booking system, the customer requests an itinerary, and then he/she receives a tentative planning. Then, the choreography can proceed following two different paths according to the customer decision. The upper path is triggered when the customer decides to withdraw the travel proposal; while the lower path is used to accept the proposal. In particular, when the proposal is accepted, the customer interacts with the bank for the payment of the ticket, and then the bank sends the confirmation to the booking system. The latter completes the procedure by sending the ticket to the customer.

\smallskip
\noindent
\emph{Collaboration Example}.
The collaboration in Fig.~\ref{fig:example}.b
shows the behaviour of the same participants of the choreography to reach the same goals.
After the customer logs into the booking system, he/she will request some travel information, and then he/she will receive a proposal from the booking system.
The customer then decides whether to withdraw or accept the proposal; this is represented through a XOR gateway. According to the decision, either the upper path, for the proposal withdraw, or the lower path, for the confirmation, is activated.
The booking system waits for the decision of the customer and behaves accordingly. This is represented through an event-based gateway. In the case of withdrawing, the two participants terminate with end events. In case of confirmation, the customer sends the itinerary acceptance to the booking system and asks for payment to the bank. As soon as the bank processes the payment, and confirms it to the booking system, the customer receives the ticket.

\section{The \texorpdfstring{{\normalfont\emph{C}$^{\ \!\!4}$}}{C4} Methodology}%
\label{sec:method}

Choreographies have emerged in the context of distributed computing, and in particular of Service Oriented Computing, as an approach to describe/specify application-level protocols to be adopted by different services willing to cooperate.
In particular, the specification defines the messages and their mutual
dependencies, that \textit{are}, or \textit{have to} be, exchanged by
the participating parties in order to fulfil the choreography
objectives.

In the literature, different approaches to choreography specifications, \ap{and their usage},
have been adopted~\cite{4785057,mandell2003bottom}. The first one considers choreographies as emerging
artefacts that relate to the integration of services
composed
to collaboratively reach an objective. The
emerging models, that can be derive to represent the exchanged messages and their order, can be successively considered to analyse
properties of the composition and to possibly reason on it.
Symmetrically, besides such a bottom-up approach, a top-down approach
has also been proposed. In this case, a choreography acts as a blueprint
defining which are the messages that \textit{have to} be
exchanged. The specification can be used to drive the development of
services that will take part in a possible choreography
enactment. In particular, in such a case the participants, and their
composition, have to abide by the communication constraints
defined in the specification.
\ap{Finally, a \textit{hybrid} approach can be conceived. In this case, as in a top-down approach, a choreography is defined to act as a blueprint for parties interested in collaborating to reach shared objectives. However, as it is for the case of a bottom-up approach, the collaborating parties will employ and integrate already available resources,
\new{possibly making adjustments}, if feasible, so to correctly reproduce the behaviour expected by the considered choreography specification.
}

 \begin{figure}[h]
  \includegraphics[scale=.5]{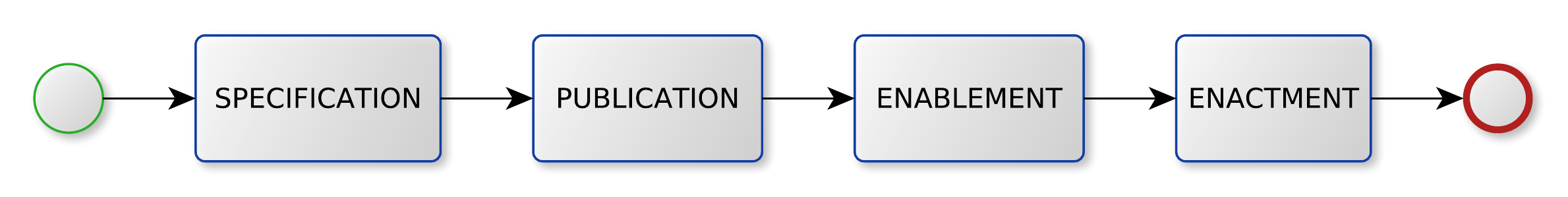}
  \caption{\name{} methodology life process.}%
  \label{fig:lc}
\end{figure}

The \name{} methodology that is presented here \ap{fits best with a hybrid approach}. Fig.~\ref{fig:lc} represents an ideal life process of a choreography
specification in
this setting. In the first state
(\textit{specification}), given an application domain, a `super
partes' organisation defines a choreography specification that will act
as a blueprint for those organisations interested in participating in
possible choreography enactments. The specification reports the
expected message exchanges, and the objectives that partners
participating to the choreography can reach, both singularly and
collectively. For instance, the choreography in Fig.~\ref{fig:example}.a
defines precise prescriptions for a \textit{booking
  system}, a \textit{customer} and a \textit{bank} that would like to
cooperate to perform a business transaction that will include a
reservation for a resource in change of the corresponding price, to be
paid via a bank transaction.

Once the choreography has been completely specified, it is made
available to interested parties (\textit{publication}), and can be
retrieved in order to implement services able to play the possibly
foreseen roles. The publication can be as simple as a pdf file stored on
a precise location, or be managed using specific service repositories
(see, e.g.,~\cite{DBLP:conf/sose/AliAP13,DBLP:conf/bpm/CorradiniFP0T19}).

Once published, choreography can be seen as an opportunity for an
organisation to integrate with others so to do business together (\textit{enablement}). In such a sense, an organisation is generally interested in choreographies in which employing its services, possibly already available, it can play a role. Then it looks for potential
partners in order to pursue the objectives of the choreography. The goal of such a phase is to enable the
choreography so that an instance of it can be successively enacted. Notably, all the roles
foreseen by the specification must be played by a different
organisation. To do this, each organisation employ a service among
the one foreseen by the choreography, and proposes it as a possible
candidate. Different processes can be conceived to select the
participants so to fill all the roles. In this paper, we do not make
any assumption on such a selection; anyway, when all the roles are filled the choreography is enabled. Nevertheless, before
enacting the choreography it is necessary to check that the composition of the
selected services will conform to the one specified by the
choreography. In a BPMN setting, the choreography specification will
have been defined using a BPMN choreography diagram, while the
different services will have been described using process diagrams, reflecting the actual implementation, that however when they will be composed will form a BPMN collaboration diagram.

Finally, the transition to the next state of the \name{} methodology (\textit{enactment})
will result in the execution of a choreography instance, and then in
the exchange of the prescribed messages. This should be made possible only
if the resulting collaboration diagram can reach the
objectives specified by the choreography. Instead, if this is not the
case, the enactment should be prevented, and the participants informed of
possible issues in the composition. In fact, as it will be clarified
in the next sections, this is where the \name{} tool comes into play permitting
to check if the participants involved in an enabled choreography
produce \ap{a behaviour ``fully respecting'' the prescriptions of the
choreography model, or instead if issues have been identified. In particular, the \name{} tool supports different kinds of checks, and the corresponding results can help to
\new{identify interventions on the collaboration}
to possibly solve the highlighted issues.
}
\new{A common intervention could be the definition of adapters for the non-conforming parties.
An adapter is indeed an artefact that carries out a series of activities that are generally used to fix interoperability issues in a component-based system. In our case, the adapter will aim at solving issues in a collaboration that lead to the violation of a conformance relation with respect to a given choreography. Different works have been proposed in the literature concerning adaptation in BPMN (see, e.g.,~\cite{ASGPT18,RH15, HW04, RS16}).}
\new{Other possible interventions could also be related to the refinement of the high-level specification of collaborations with the introduction of lower-level information (e.g., guards driving the decisions taken by XOR gateways).
In fact, it may happen that the issues in the specification are caused by the inherent non-determinism present in choreography model for which guards are left unspecified. In such a case, when concrete data and expressions are added these issues could simply disappear.}

\begin{figure}[t!]
  \begin{tabular}{cc}

    \multicolumn{2}{c}{\includegraphics[scale=.2]{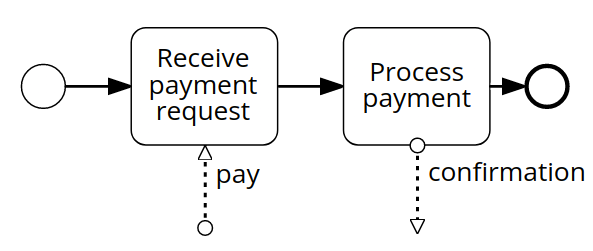}}\\
    \multicolumn{2}{c}{a. Bank 1}\\		\hline 	\\ % chktex 44

    \multicolumn{2}{c}{\includegraphics[scale=.25]{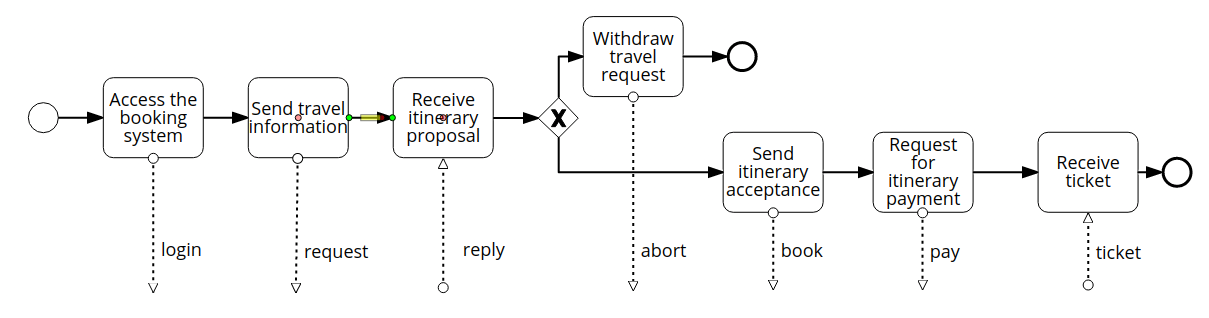}}\\
    \multicolumn{2}{c}{b. Customer 1}\\
    \multicolumn{2}{c}{\includegraphics[scale=.25]{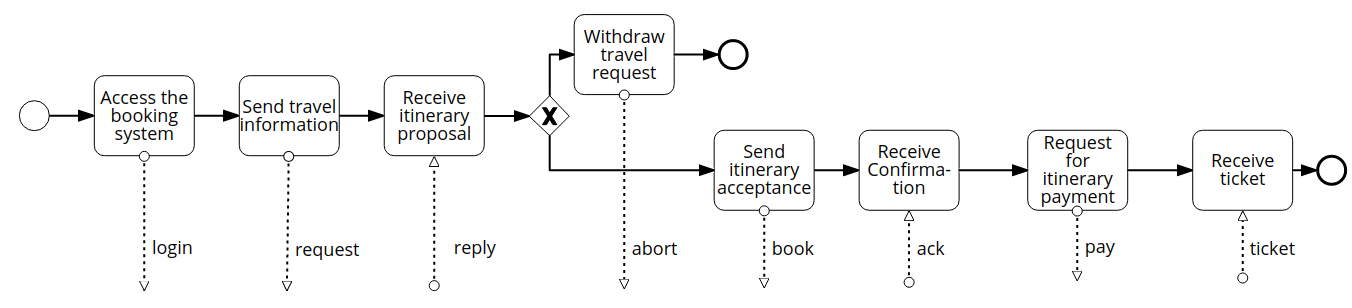}}\\
    \multicolumn{2}{c}{c. Customer 2}\\
    \hline % chktex 44
    \\
    \multicolumn{2}{c}{\includegraphics[scale=.25]{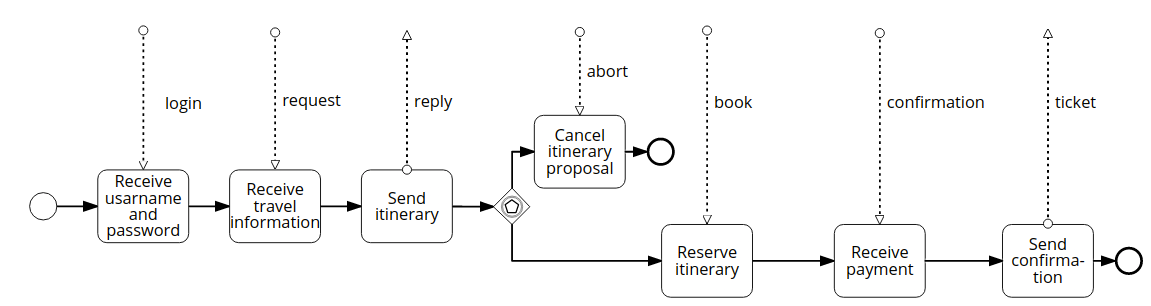}}\\
    \multicolumn{2}{c}{d. Booking System 1}\\
    \multicolumn{2}{c}{\includegraphics[scale=.25]{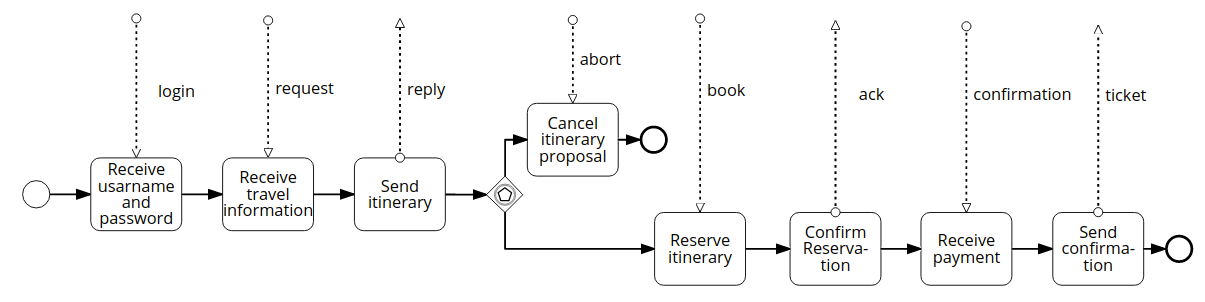}}\\
    \multicolumn{2}{c}{e. Booking System 2}\\
    \multicolumn{2}{c}{\includegraphics[scale=.25]{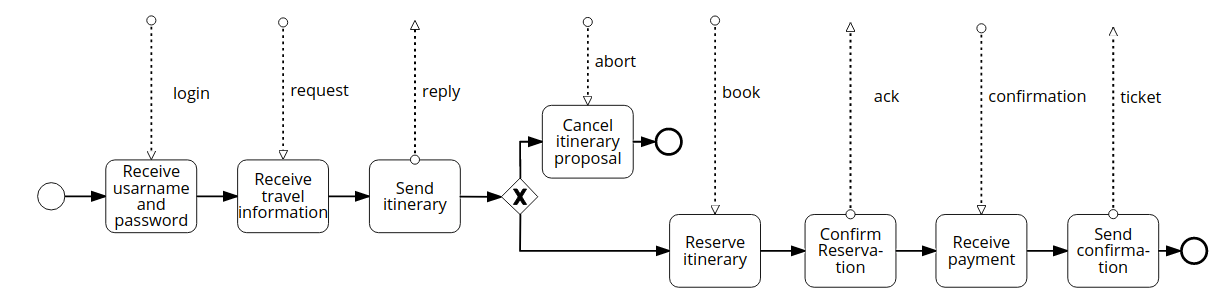}}\\
    \multicolumn{2}{c}{f. Booking System 3}\\

  \end{tabular}
  \caption{Possible processes for the roles in the
    choreography of Fig.~\ref{fig:example}.}%
  \label{fig:p_repo}
\end{figure}

Fig.~\ref{fig:p_repo} helps to clarify the steps of the \name{} methodology in practice. In the figure, six
different processes are represented, possibly implemented by
services delivered by different organisations. In particular, in
reference to the choreography reported in Fig.~\ref{fig:example}, different options could be admitted for each role to complete the specification.
The  process in Fig.~\ref{fig:p_repo}.a has been defined to play
the role of the bank, while the two processes
in Fig.~\ref{fig:p_repo}.b and Fig.~\ref{fig:p_repo}.c have been both
selected to play the role of the customer. Finally, the three
processes in Fig.~\ref{fig:p_repo}.d, Fig.~\ref{fig:p_repo}.e, and
Fig.~\ref{fig:p_repo}.f have been selected to play the role of a
booking system. In such a situation, six possible different
compositions of processes are possible. Nevertheless, only few of them
 actually permit to achieve the objectives of the corresponding
choreography, as it is detailed in Section~\ref{sec:conformance}
(see Table~\ref{tab:combination}). In particular, possible composition problems can relate to
structural issues (e.g., non correspondence on the sets of exchanged
messages by the involved processes) or behavioural issues (e.g., non conformance with respect to
specific behavioural relations of the resulting composition).
In such a setting, the \name{} tool then permits to prevent the enactment of
enabled choreography instances that will successively result in the
emergence of problems during the running stage. \new{In these cases the engineer could evaluate possible interventions, such as the introduction of adapters or the refinement of models so to solve the issues highlighted by \name{}.}

\section{The formal framework}%
\label{sec:formal}

This section presents the formalisations at the basis of our framework,
concerning the semantics of \bpmn{} choreography and collaboration
diagrams, and their conformance.

\subsection*{Choreography and Collaboration Semantics}%
\label{sec:formal_semantics}
We first summarise the distinctive aspects of the semantics of the choreographies and collaborations
in relation to the \bpmn{} modelling principles, and then we illustrate their formal definitions.

\subsubsection*{Linguistic Aspects and Design Choices.}%
\label{sec:design}
Concerning choreography diagrams, we made some specific design choices. In relation to the \emph{Two-Way choreography task}, the OMG standard states that it is ``an atomic activity in a choreography process'' execution~\cite[p. 323]{omg_business_2011}.  However, this does not mean that the task blocks the whole execution of the choreography. In fact, participants are usually distributed, and we assume that other choreography tasks involved in different parallel paths of the choreography can be executed. Thus, here we intend atomicity to mean that both messages exchanged in a Two-Way task have to be received before triggering the execution along the sequence flow outgoing from the task.
Therefore, even if we allow Two-Way tasks in the choreography models, we safely manage them as pairs of One-Way tasks preserving the same meaning.

A further distinctive aspect of our formal semantics concerns the \emph{communication model} that, to be compliant with the \bpmn{} standard, is different for choreographies and collaborations.
\ft{
A BPMN choreography diagram is, indeed, a description at a high level of abstraction of message exchanges among participants of a system. The basic element of this kind of model is the choreography task, which specifies a single message exchange from a sender participant to a receiver one.
According to the BPMN standard~\cite[p. 315]{omg_business_2011}, a choreography task completes when the receiver participant reads the message. Hence, a choreography task is a blocking activity, which resumes the execution only when an exchanged message is received. For this reason, the communication model of choreographies is deemed to be synchronous.}
The communication model of collaborations, instead, is asynchronous. This means that a message sent by one participant is enqueued by the receiving one, which can then consume and process it subsequently, while the sender is free to proceed with its execution.  This reflects the distributed nature of collaborations.
The use of two different communication models also impacts on the definition of the conformance relations as illustrated below.

\ft{Finally, we do not require BPMN models to satisfies any structural constraint, such as
the well-structuredness~\cite{WS} property required by many approaches in the Business Process Management domain.
In fact, we consider models with an arbitrary topology, so that the models can have unbalanced workflows. Therefore,
the class of models we consider is the most comprehensive one that can be designed in BPMN from the structural
point of view.}

\subsubsection*{Semantics of \bpmn{} Choreographies.}%
\label{sec:globalcho}
To enable a formal treatment of a \bpmn{}  choreography we defined a Backus Normal Form (BNF) syntax of its model structure (Fig.~\ref{fig:syntaxGlobalCho}).
\begin{figure}[h]
\setlength{\belowcaptionskip}{-5pt}
\centering
\footnotesize
$
\begin{array}{|@{\ }r@{\ }c@{\ }l@{\ }|} % chktex 44
\hline % chktex 44
&&\\[-0.2cm]
\chos  & ::=  &
\choevStartG{\edgeG{\edgename_o}}
  \mid\  \choevEndG{\edgeG{\edgename_i,\edgename_c}}
 \mid\  \andSplitG{\edgeG{\edgename_i}}{\edgeList_o}
  \mid\ \andJoinG{\edgeList_i}{\edgeG{\edgename_o}}
  \mid\ \xorSplitG{\edgeG{\edgename_i}}{\edgeList_o}
 \mid\   \xorJoinG{\edgeList_i}{\edgeG{\edgename_o}}
\\[.1cm]
& \mid &
 \chotaskOWG{\edgeG{\edgename_i}}{\edgeG{\edgename_o}}{\orgname_1}{\orgname_2}{\messagename}{\taskname}
\mid\
 \eventbasedG{\edgeG{\edgename_i}}{\taskList_1, \taskList_2}
\mid\  \chos_1 \!\! \chopar\! \chos_2
\\[0.01cm]
&&\\[-0.2cm]
\taskList  & ::=  &
\chotasklistOWG{\edgeG{\edgename_o}}{\orgname_1}{\orgname_2}{\messagename}{\taskname}
\  \mid\ \  \taskList_1, \taskList_2
\\[0.1cm]
\hline % chktex 44
\end{array}
$
\vspace{-.1cm}
\caption{Syntax of \bpmn{} Choreography Structures.}
\label{fig:syntaxGlobalCho}
\end{figure}
In the proposed grammar, the non-terminal symbol $\mathit{\chos}$ represents \emph{Choreography Structures}, while the terminal symbols, denoted by the ${\sf sans\ serif}$ font, are the considered elements of a \bpmn{} model, i.e.~events, tasks and gateways.
We are not proposing a new modelling formalism, but we are only using a textual notation for the \bpmn{} elements. With respect to the graphical notation, the textual one is more manageable
for supporting the formal definition of the semantics and its implementation.
Notably, even if our syntax would allow to write terms that cannot be expressed in BPMN, we consider here only those terms of the syntax that can be derived from BPMN models.

Let $\edgeSetE$ be the set of edge names, in the following $\edgename \in \edgeSetE$ denotes a sequence edge, while $\edgeList \in 2^\edgeSetE$ a set of edges; we require $ |\edgeList|>1$ when $\edgeList$ is used in joining and splitting gateways. For the convenience of the reader we refer with $\edgename_i$ the edge incoming into an element and with  $\edgename_o$ the edge outgoing from an element.
 $\orgname$ and $\messagename$ denote names uniquely identifying a participant and a message, respectively.
The correspondence between the syntax used here and the graphical notation of \bpmn{} illustrated in
Section~\ref{sec:back} is as follows.
\begin{itemize}[noitemsep,topsep=0pt]
\item $\choevStartG{\edgeG{\edgename_o}}$ represents a start event with outgoing edge~$\edgename_o$.
\item $\choevEndG{\edgeG{\edgename_i,\edgename_c}}$ represents an end event with incoming edge $\edgename_i$ and a (spurious) edge $\edgename_c$ representing the complete status of the end event.
\item $\choandSplitG{\edgeG{\edgename_i}}{\edgeList_o}$
\ (resp. $\choxorSplitG{\edgeG{\edgename_i}}{\edgeList_o}$)\
represents an AND (resp. XOR) split gateway with incoming edge $\edgename_i$
and outgoing edges $\edgeList_o$.
\item $\choandJoinG{\edgeList_i}{\edgeG{\edgename_o}}$
\ (resp. $\choxorJoinG{\edgeList_i}{\edgeG{\edgename_o}}$)\
represents an AND (resp. XOR) join gateway with incoming edges $\edgeList_i$
and outgoing edge $\edgename_o$.
\item $\chotaskOWG{\edgeG{\edgename_i}}{\edgeG{\edgename_o}}{\orgname_1}{\orgname_2}{\messagename}{\taskname}$ represents a one-way task
with incoming edge $\edgename_i$ and outgoing edge $\edgename_o$ sending a message $\messagename$ from $\orgname_1$ to $\orgname_2$.
As explained above, the two-way tasks are rendered in our formal framework as pairs of one-way tasks, hence they are not explicitly included in the syntax.

\item $\choeventbasedG{\edgeG{\edgename_i}}{\taskList_1}{\taskList_2}$ represents an event-based gateway with incoming edge $\edgename_i$, and a list of (at least two) tasks $\taskList_1,\taskList_2$ to be processed. It is worth noticing that the definition of the task list $\taskList$  is composed by elements of the same structure of the one-way task except for the
incoming edge, which is subsumed in the structure of the event-based gateway.
When convenient, we shall regard a task list simply as a set.
\item $\chos_1 \!\!\chopar\! \chos_2$ represents a composition of elements in order to
render a choreography structure in terms of a collection of elements.
\end{itemize}
To achieve a compositional definition, each sequence edge of the \bpmn{} model is split in two parts:
the part outgoing from the source element and the part incoming into the target element.
The two parts are correlated by means of unique sequence  edge names in the \bpmn{} model.

\begin{exa}
\label{running_ex:syntax}
Let us consider the BPMN choreography model in Fig~\ref{fig:example}.a. The textual representation of its structure is as follow (for reader's convenience, we use $\edgename_i$, with $i$ a natural number, to denote sequence edges, and $\orgname_c$ for \emph{Customer}, $\orgname_{bs}$  for \emph{Booking System} and $\orgname_{bk}$ for \emph{Bank}):
\[
\begin{array}{l}
\choevStartG{\edgeG{\edgename_1}} \ \chopar\  \chotaskOWG{\edgeG{\edgename_1}}{\edgeG{\edgename_2}}{\orgname_{c}}{\orgname_{bs}}{\mathsf{login}}{} \ \chopar\
\chotaskOWG{\edgeG{\edgename_2}}{\edgeG{\edgename_3}}{\orgname_{c}}{\orgname_{bs}}{\mathsf{request}}{} \ \chopar\\
\chotaskOWG{\edgeG{\edgename_3}}{\edgeG{\edgename_4}}{\orgname_{bs}}{\orgname_{c}}{\mathsf{reply}}{} \ \chopar\
\xorSplitG{\edgeG{\edgename_4}}{\{\edgename_5, \edgename_6\}} \ \chopar \
\chotaskOWG{\edgeG{\edgename_5}}{\edgeG{\edgename_7}}{\orgname_{c}}{\orgname_{bs}}{\mathsf{abort}}{} \ \chopar\
\choevEndG{\edgeG{\edgename_7,\edgename_8}} \ \chopar \\
\chotaskOWG{\edgeG{\edgename_6}}{\edgeG{\edgename_9}}{\orgname_{c}}{\orgname_{bs}}{\mathsf{book}}{} \ \chopar\
\chotaskOWG{\edgeG{\edgename_9}}{\edgeG{\edgename_{10}}}{\orgname_{c}}{\orgname_{bk}}{\mathsf{pay}}{} \ \chopar\
\chotaskOWG{\edgeG{\edgename_{10}}}{\edgeG{\edgename_{11}}}{\orgname_{bk}}{\orgname_{bs}}{\mathsf{confirmation}}{} \ \chopar\ \\
\chotaskOWG{\edgeG{\edgename_{11}}}{\edgeG{\edgename_{12}}}{\orgname_{bs}}{\orgname_{c}}{\mathsf{ticket}}{} \ \chopar\
\choevEndG{\edgeG{\edgename_{12},\edgename_{13}}}
\end{array}
\]
\end{exa}

The operational semantics we propose is given in terms of configurations of the form
$\langle \chos, \stateedge
\rangle$, where
$\chos$ is a choreography structure, and
$\stateedge$ is the execution state storing for each edge the current number of tokens marking it.
Specifically, a state $\stateedge : \edgeSetE \rightarrow \mathbb{N}$ is a function mapping edges to numbers of tokens.
The state obtained by updating in the state $\stateedge$ the number of tokens of the edge $\edgename$ to $\token$, written as
$\stateupd{\stateedge}{\edgename}{\token}$, is defined as follows:
$(\stateupd{\stateedge}{\edgename}{\token})(\edgename')$ returns $\token$ if $\edgename'=\edgename$,
otherwise it returns $\stateedge(\edgename')$.
The \emph{initial state}, where all edges are unmarked is denoted by $\initstateedge$ formally, $\initstateedge(\edgename)=0 \ \ \forall \edgename \in \edgeSetE$.
The transition relation over configurations, written  $\transitionCho{l}{}$ and defined by the rules in Fig.~\ref{fig:semanticsChoreographyGlobal}, formalizes the execution of a choreography in terms of marking evolution and message exchanges. Labels $l$ represent computational steps and are defined as:
$\tau$, denoting internal computations; and
\mbox{$\orgname_1 \rightarrow \orgname_2:\messagename$}, denoting an exchange of message $\messagename$ from participant $\orgname_1$ to $\orgname_2$.
Notably, despite the presence of labels, this has to be thought of as a reduction semantics,
because labels are not used for synchronization (as instead it usually happens in labeled
semantics),  but only for keeping track of the exchanged messages in order to enable the
conformance checking discussed later on.
Since choreography execution only affects the current state, for the sake of presentation, we omit the choreography structure from the target configurations of transitions.  Thus, a transition
$\langle \chos, \stateedge \rangle   \transitionCho{l}{} \langle  \chos, \stateedge' \rangle$ is written as $\langle \chos, \stateedge \rangle   \transitionCho{l}{}   \stateedge' $.

Before commenting on the rules, we introduce the auxiliary functions they exploit. Specifically, function
$\mathit{\incTokenName}: \statesSet \times \edgeSetE \rightarrow \statesSet$
(resp. $\mathit{\decTokenName}: \statesSet \times \edgeSetE \rightarrow \statesSet$), where $\statesSet$ is the set of states, allows updating a state by incrementing (resp.\ decrementing) by one the number of tokens marking an edge in the state. Formally, they are defined as follows:
$\incToken{\stateedge}{\edgename} =\stateupd{\stateedge}{\edgename}{\statefunc{\edgename}+1}$ and
$\decToken{\stateedge}{\edgename} =\stateupd{\stateedge}{\edgename}{\statefunc{\edgename}-1}$.
These functions extend in a natural ways to sets of edges as follows:
$\incToken{\stateedge}{\emptyset}= \stateedge $
and  $\incToken{\stateedge}{\{\edgename\}\cup \edgeList}= \incToken{\incToken{\stateedge}{\edgename}}{\edgeList}$;
the cases for $\decTokenName$ are similar.

%%%%%%%%%
\begin{figure}[!t]
\footnotesize
\[
\begin{array}{|@{\hspace*{0.2cm} \ }l@{\hspace*{0.3cm} }l@{\hspace*{-0.2cm}}l|} % chktex 44
  \hline % chktex 44
&&\\[-.2cm]
                     \begin{array}{l}\!\!\!\!
     	\langle \choevStartG{\edgeG{\edgename_o}}, \initstateedge \rangle
  	\gtransitionCho
	 \incToken{\initstateedge}{\edgename_o}
	 \end{array}
	&  &
  \quad  (\mathit{Ch\textrm{-}Start})
   \\[.6cm]
\begin{array}{l}\!\!\!\!
\langle \choevEndG{\edgeG{\edgename_i,\edgename_c}}, \stateedge \rangle
  	\gtransitionCho
	    \incToken{\decToken{\stateedge}{\edgename_i}}{\edgename_c}
\end{array}
	&  \statefunc{\edgename_i}>0 &
		\quad (\mathit{Ch\textrm{-}End})
   \\[.6cm]
    \begin{array}{l}\!\!\!\!
        \langle \choandSplitG{\edgeG{\edgename_i}}{\edgeList_o}, \stateedge \rangle
  	\gtransitionCho
	  \incToken{\decToken{\stateedge}{\edgename_i}}{\edgeList_o}
	\end{array}
	&  \statefunc{\edgename_i}>0 &
	\quad (\mathit{Ch\textrm{-}AndSplit})
 \\[.6cm]
    \begin{array}{l}\!\!\!\!
   \langle \choandJoinG{\edgeList_i}{\edgeG{\edgename_o}},  \stateedge \rangle
      \gtransitionCho
      \incToken{ \decToken{\stateedge}{\edgeList_i}}{\edgename_o}
     	\end{array}
	&\!\!\!\!
       \forall \edgename \in \edgeList_i \qsep   \statefunc{\edgename}>0
     & \quad  (\mathit{Ch\textrm{-}AndJoin})
 \\[.6cm]
                   \begin{array}{l}\!\!\!\!
       \langle \choxorSplitG{\edgeG{\edgename_i}}{\{\edgeG{\edgename}\} \cup\edgeList_o},  \stateedge \rangle
      \gtransitionCho
    \incToken{\decToken{\stateedge}{\edgename_i}}{\edgename}
         	\end{array}
               &
                                \statefunc{\edgename_i}>0
     & \quad (\mathit{Ch\textrm{-}XorSplit})
     \\[.6cm]
                   \begin{array}{l}\!\!\!\!
       \langle \choxorJoinG{\{\edgeG{\edgename}\} \cup\edgeList_i}{\edgeG{\edgename_o}}, \stateedge \rangle
      \gtransitionCho
  \incToken{\decToken{\stateedge}{\edgename}}{\edgename_o}
       	\end{array}
               &
       \statefunc{\edgename}>0
     & \quad (\mathit{Ch\textrm{-}XorJoin})
     \\[.6cm]
                   \begin{array}{l}\!\!\!\!
   \langle  \chotaskOWG{\edgeG{\edgename_i}}{\edgeG{\edgename_o}}{\orgname_1}{\orgname_2}{\messagename}{\taskname},  \stateedge \rangle
\transitionCho{\orgname_1 \rightarrow \orgname_2:\messagename}{}
  \incToken{ \decToken{\stateedge}{\edgename_i}}{\edgename_o}
              \end{array}
     &
                 \begin{array}{l}
      \statefunc{\edgename_i}>0\\
      %\\
            \end{array}
     & \quad  (\mathit{Ch\textrm{-}Task})
      \\[.6cm]
                   \begin{array}{l}\!\!\!\!
     	\langle \eventbasedG{\edgeG{\edgename_i}}{\chotasklistOWG{\edgeG{\edgename_o}}{\orgname_1}{\orgname_2}{\messagename}{\taskname}  \cup  \taskList}, \stateedge \rangle
\transitionCho{\orgname_1 \rightarrow \orgname_2:\messagename}{}
	\incToken{ \decToken{\stateedge}{\edgename_i}}{\edgename_o}
	                             \end{array}
       &
      \statefunc{\edgename_i}>0
      %\\
      %\\
     & \quad  (\mathit{Ch\textrm{-}EventG})
      \\[.6cm]
    \multicolumn{3}{|c|}{
           \begin{array}{l}
      \prooftree
   \langle \chos_1, \stateedge \rangle %\gtransitionCho
   \transitionCho{l}{}  \stateedge'
    \justifies
     \langle \chos_1 \!\! \mid\! \chos_2 ,\stateedge \rangle
      \transitionCho{l}{} %\gtransitionCho
       \stateedge'
      \using \ \  (\mathit{Ch\textrm{-}Int_1})
       \endprooftree
         \\
    \end{array}
    \qquad
     \begin{array}{l}
      \prooftree
   \langle \chos_2, \stateedge \rangle %\gtransitionCho
   \transitionCho{l}{}   \stateedge'
    \justifies
     \langle \chos_1 \!\! \mid\! \chos_2 ,\stateedge \rangle
      \transitionCho{l}{} %\gtransitionCho
       \stateedge'
      \using  \ \ (\mathit{Ch\textrm{-}Int_2})
       \endprooftree
         \\
    \end{array}
}
     \\ [.6cm]
   \hline % chktex 44
\end{array}
\]
\vspace*{-.5cm}
\caption{Choreography Semantics.}%
\label{fig:semanticsChoreographyGlobal}
\vspace{-3mm}
\end{figure}

We now briefly comment on the operational rules in Fig.~\ref{fig:semanticsChoreographyGlobal}.
Rule $\mathit{Ch\textrm{-}Start}$ starts the execution of a choreography when it is in its initial state (i.e., all edges are unmarked). The effect of the rule is to increment the number of tokens in the edge outgoing from the start event.
 Rule  $\mathit{Ch\textrm{-}End}$ instead is enabled when there is at least a token in the incoming edge of the end event, which is then moved to the spurious edge to keep track that a token ended up in the event.
Rule $\mathit{Ch\textrm{-}AndSplit}$ is applied when there is at least one token in the incoming edge of an AND split gateway; as result of its application the rule decrements the number of tokens in the incoming edge and increments that in each outgoing edge.
Rule $\mathit{Ch\textrm{-}AndJoin}$ decrements the tokens in each incoming edge and increments the number of tokens of the outgoing edge, when each incoming edge has at least one token.
Rule $\mathit{Ch\textrm{-}XorSplit}$ is applied when a token is available in the incoming edge of an XOR split gateway, the rule decrements this token and increments the tokens in one of the outgoing edges.
Rule $\mathit{Ch\textrm{-}XorJoin}$ is activated every time there is a token in one of the incoming edges, which is then moved to the outgoing edge.
Rule $\mathit{Ch\textrm{-}Task}$ is activated when there is a token in the incoming edge of a choreography task, so that the application of the rule produces a message exchange label and moves the token from the incoming edge to the outgoing one.
 Rule $\mathit{Ch\textrm{-}EventG}$ is activated each time there is a token in the incoming edge, which is moved to the outgoing edge of one task in the enclosed list, and produces a message exchange label.
Different message exchanges can take place; the selection of the executed task from the
 list specified in the gateway is non-deterministic, in order to properly model the race condition regulating
 the behaviour of the event-based gateway.
Finally, rules  $\mathit{Ch\textrm{-}Int_1}$ and  $\mathit{Ch\textrm{-}Int_2}$ deal with interleaving.

\begin{exa}\label{running_ex:semantics}
Let $\chos_{i}$ be the choreography structure defined in Example~\ref{running_ex:syntax}.
The initial configuration of the collaboration is
$\langle \chos_{i}, \initstateedge \rangle$, where
$\stateedge_0(\edgename_{i})=0 \ \ \forall\,i\in\{1,\ldots,13\}$.
The state $\stateedge_1$ obtained by applying the rule  $\mathit{Ch\textrm{-}Start}$, which marks the edge
$\edgename_{1}$, is obtained as follows:
$\stateedge_1=\incToken{\initstateedge}{\edgename_1}$.
\end{exa}

\subsubsection*{From Processes to Collaborations.}
According to the \name{} methodology described in Section~\ref{sec:method},
a choreography is enacted by collaborations resulting from the combination
of sets of processes.
We formalise here the notions of process, collaboration, and process
composition.

The BNF syntax of the process model structure is given in Fig.~\ref{fig:syntaxGlobalProcesses}. The non-terminal symbol $\proc$ represents \emph{Process Structures}, while terminal symbols denote, as usual, the considered \bpmn{} elements.
In a process model there are three types of tasks, i.e.
non-communicating (${\sf task}$),
receiving (${\sf taskRcv}$)
and sending (${\sf taskSnd}$),
and also two intermediate events, i.e.\ receiving (${\sf interRcv}$) and sending (${\sf interSnd}$).
Each receiving/sending elements specifies an exchanged message, while
an event-based gateway specifies a list of (at least two) messages, each one enriched with the outgoing edge enabled by the message reception. When convenient, we shall regard a message
list simply as a set.

\begin{exa}
Let us consider the BPMN process model in Fig~\ref{fig:p_repo}.b. The textual representation of its structure is as follow:
\[
\begin{array}{l}
\evStartG{\edgeG{\edgename_{1}'}} \ \collabspar \
 \tasksendG{\edgeG{\edgename_{1}'}}{\edgeG{\edgename_{2}'}}{{\sf login}}\ \collabspar \
  \tasksendG{\edgeG{\edgename_{2}'}}{\edgeG{\edgename_{3}'}}{{\sf request}}\ \collabspar \
  \taskreceiveG{\edgeG{\edgename_{3}'}}{\edgeG{\edgename_{4}'}}{{\sf reply}} \ \collabspar \ \\
  \xorSplitG{\edgeG{\edgename_{4}'}}{\{\edgename_5', \edgename_{6}'\}}\ \collabspar \
    \tasksendG{\edgeG{\edgename_{5}'}}{\edgeG{\edgename_{7}'}}{{\sf abort}}\ \collabspar \
     \evEndG{\edgeG{\edgename_{7}',\edgename_{8}'}}\ \collabspar \
      \tasksendG{\edgeG{\edgename_{6}'}}{\edgeG{\edgename_{9}'}}{{\sf book}}\ \collabspar \ \\
     \tasksendG{\edgeG{\edgename_{9}'}}{\edgeG{\edgename_{10}'}}{{\sf pay}}\ \collabspar \
       \taskreceiveG{\edgeG{\edgename_{11}'}}{\edgeG{\edgename_{12}'}}{{\sf ticket}} \ \collabspar \
            \evEndG{\edgeG{\edgename_{12}',\edgename_{13}'}}
\end{array}
\]
\end{exa}

\begin{figure}[t]
\centering
\setlength{\belowcaptionskip}{-7pt}
%\vspace{-.0002cm}
\footnotesize
$
\begin{array}{|@{\ }rcl@{\ }|} % chktex 44
\hline % chktex 44
&&\\[-0.2cm]
\proc  & ::=  &
\evStartG{\edgeG{\edgename_o}}
 \mid  \evEndG{\edgeG{\edgename_i,\edgename_c}}
 \mid  \andJoinG{\edgeList_i}{\edgeG{\edgename_o}}
 \mid  \xorSplitG{\edgeG{\edgename_i}}{\edgeList_o}
 \mid  \andSplitG{\edgeG{\edgename_i}}{\edgeList_o}
\mid   \xorJoinG{\edgeList_i}{\edgeG{\edgename_o}}
   \\[.1cm]
 & \mid &
\taskG{\edgeG{\edgename_i}}{\edgeG{\edgename_o}}
 \mid\  \taskreceiveG{\edgeG{\edgename_i}}{\edgeG{\edgename_o}}{{\messagename}}
\mid\
 \tasksendG{\edgeG{\edgename_i}}{\edgeG{\edgename_o}}{{\messagename}}
  \mid\
 \evInterRcvG{\edgeG{\edgename_i}}{\edgeG{\edgename_o}}{{\messagename}}
 \mid\
 \evInterSndG{\edgeG{\edgename_i}}{\edgeG{\edgename_o}}{{\messagename}}
\\[.1cm]
& \mid &   \eventbasedG{\edgeG{\edgename_i}}{\msgList_1,\msgList_2}
\ \mid\ \proc_1\!\!  \collabspar\! \proc_2
\\[0.1cm]
\msgList & ::=  &   \msgEdgeEventP{\orgname_1}{\orgname_2}{\messagename}{\edgename_o} \ \ \mid\ \ \msgList_1,\msgList_2
\\[.1cm]
\hline % chktex 44
\end{array}
$
\vspace{-.3cm}
\caption{Syntax of \bpmn{} Processes Structures.}
\label{fig:syntaxGlobalProcesses}
\end{figure}

The BNF syntax of the collaboration model structure is given in Fig.~\ref{fig:syntaxGlobal}, where the non-terminal symbol $\collabs$ represents \emph{Collaboration Structures}.
Each process involved in a collaboration is identified by a participant name, denoted by $\orgname$.
The exchange of messages in a collaboration is modeled by means of \emph{message edges}. Here, they are represented by triples of the form $\msgEdge{\orgname_1}{\orgname_2}{\messagename}$ indicating, in order, the sending participant, the receiving participant and the message.
Accordingly, an event-based gateway specifies a list of (at least two) message edges.

\begin{exa} \label{ex4}
Let us consider the BPMN collaboration model in Fig~\ref{fig:example}.b. The textual representation of its structure is as follow:
\[
C_{bk}
\ \collabspar \
C_{bs}
\ \collabspar \
C_{c}
\]
where:
\[
\begin{array}{l@{\ }l}
\\
C_{bk}   ::=  &
\evStartG{\edgeG{\edgename_1'''}} \ \collabspar \
\taskreceiveG{\edgeG{\edgename_1'''}}{\edgeG{\edgename_2'''}}{\msgEdge{\orgname_{c}}{\orgname_{bk}}{\sf pay}}\ \collabspar \\
& \tasksendG{\edgeG{\edgename_2'''}}{\edgeG{\edgename_3'''}}{\msgEdge{\orgname_{bk}}{\orgname_{bs}}{\sf confirmation}}\ \collabspar \
 \evEndG{\edgeG{\edgename_3''',\edgename_4'''}}   \\
\end{array}
\]
%&\\
\[
\begin{array}{l@{\ }l}
C_{bs}  ::=  &
\evStartG{\edgeG{\edgename_1''}} \ \collabspar \
\taskreceiveG{\edgeG{\edgename_1''}}{\edgeG{\edgename_2''}}{\msgEdge{\orgname_{c}}{\orgname_{bs}}{\sf login}}\ \collabspar \
\taskreceiveG{\edgeG{\edgename_2''}}{\edgeG{\edgename_3''}}{\msgEdge{\orgname_{c}}{\orgname_{bs}}{\sf request}}\ \collabspar \ \\
&
\tasksendG{\edgeG{\edgename_3''}}{\edgeG{\edgename_4''}}{\msgEdge{\orgname_{bs}}{\orgname_{c}}{\sf reply}} \ \collabspar \\
&  \eventbasedG{\edgeG{\edgename_4''}}{\msgEdgeEvent{\orgname_{c}}{\orgname_{bs}}{\sf abort}{\edgename_5''},\msgEdgeEvent{\orgname_{c}}{\orgname_{bs}}{\sf book}{\edgename_6''}}\ \collabspar \
  \evEndG{\edgeG{\edgename_5'',\edgename_7''}}\ \collabspar \\
& \taskreceiveG{\edgeG{\edgename_{6}''}}{\edgeG{\edgename_8''}}{\msgEdge{\orgname_{bk}}{\orgname_{bs}}{\sf confirmation}} \ \collabspar \
\tasksendG{\edgeG{\edgename_8''}}{\edgeG{\edgename_9''}}{\msgEdge{\orgname_{bs}}{\orgname_{c}}{\sf ticket}} \ \collabspar \\
& \evEndG{\edgeG{\edgename_9'',\edgename_{10}''}}\\
&\\
C_{c} ::=  &
\evStartG{\edgeG{\edgename_1'}} \ \collabspar \
\tasksendG{\edgeG{\edgename_1'}}{\edgeG{\edgename_2'}}{\msgEdge{\orgname_{\orgname_{bk}}}{\orgname_{bs}}{\mathsf{login}}}\ \collabspar \
\tasksendG{\edgeG{\edgename_2'}}{\edgeG{\edgename_3'}}{\msgEdge{\orgname_{\orgname_{bk}}}{\orgname_{bs}}{\mathsf{request}}}\ \collabspar \ \\
&
\taskreceiveG{\edgeG{\edgename_3'}}{\edgeG{\edgename_4'}}{\msgEdge{\orgname_{bs}}{\orgname_{\orgname_{bk}}}{\mathsf{reply}}} \ \collabspar \
\xorSplitG{\edgeG{\edgename_4'}}{\{\edgename_5', \edgename_6'\}}\ \collabspar \ \\
&
\tasksendG{\edgeG{\edgename_5'}}{\edgeG{\edgename_7'}}{\msgEdge{\orgname_{\orgname_{bk}}}{\orgname_{bs}}{\mathsf{abort}}}\ \collabspar \
\evEndG{\edgeG{\edgename_7',\edgename_8'}}\ \collabspar \
\tasksendG{\edgeG{\edgename_6'}}{\edgeG{\edgename_9'}}{\msgEdge{\orgname_{\orgname_{bk}}}{\orgname_{bs}}{\mathsf{book}}}\ \collabspar \ \\
&
\tasksendG{\edgeG{\edgename_9'}}{\edgeG{\edgename_{10}'}}{\msgEdge{\orgname_{\orgname_{bk}}}{\orgname_{bk}}{\mathsf{pay}}}\ \collabspar \
\taskreceiveG{\edgeG{\edgename_{11}'}}{\edgeG{\edgename_{12}'}}{\msgEdge{\orgname_{bk}}{\orgname_{bs}}{\mathsf{ticket}}} \ \collabspar \
\evEndG{\edgeG{\edgename_{12}',\edgename_{13}'}}
\end{array}
\]
\end{exa}

\begin{figure}[t]
\centering
\setlength{\belowcaptionskip}{-7pt}
\footnotesize
$
\begin{array}{|@{\ }rcl@{\ }|} % chktex 44
\hline % chktex 44
&&\\[-0.2cm]
\collabs  & ::=  &
\evStartG{\edgeG{\edgename_o}}
 \mid  \evEndG{\edgeG{\edgename_i,\edgename_c}}
 \mid  \andJoinG{\edgeList_i}{\edgeG{\edgename_o}}
 \mid  \xorSplitG{\edgeG{\edgename_i}}{\edgeList_o}
 \mid  \andSplitG{\edgeG{\edgename_i}}{\edgeList_o}
 \mid   \xorJoinG{\edgeList_i}{\edgeG{\edgename_o}}
   \\[.1cm]
 & \mid &
\taskG{\edgeG{\edgename_i}}{\edgeG{\edgename_o}}
\mid\ \taskreceiveG{\edgeG{\edgename_i}}{\edgeG{\edgename_o}}{\msgEdge{\orgname_1}{\orgname_2}{\messagename}}
\mid\
 \tasksendG{\edgeG{\edgename_i}}{\edgeG{\edgename_o}}{\msgEdge{\orgname_1}{\orgname_2}{\messagename}}
\\[.1cm]
& \mid &
 \evInterRcvG{\edgeG{\edgename_i}}{\edgeG{\edgename_o}}{\msgEdge{\orgname_1}{\orgname_2}{\messagename}}
\mid\
 \evInterSndG{\edgeG{\edgename_i}}{\edgeG{\edgename_o}}{\msgEdge{\orgname_1}{\orgname_2}{\messagename}}
 \mid\
 \eventbasedG{\edgeG{\edgename_i}}{\msgListE_1,\msgListE_2}
\ \mid\ \collabs_1\!\!  \collabspar\! \collabs_2
\\[0.1cm]
\msgListE & ::=  &   \msgEdgeEvent{\orgname_1}{\orgname_2}{\messagename}{\edgename_o} \ \ \mid\ \ \msgListE_1,\msgListE_2
\\[.1cm]
\hline % chktex 44
\end{array}
$
\vspace{-.3cm}
\caption{Syntax of \bpmn{} Collaboration Structures.}
\label{fig:syntaxGlobal}
\end{figure}

When composing processes in order to form collaborations, we must properly connect
via a message edge each task (or intermediate event) of a process sending a given
message with a corresponding receiving element belonging to another process, and vice
versa. Hence, the resulting collaboration should not contain disconnected communicating
elements. To formalise this property, we need to introduce the auxiliary functions $\outF{\collabs}$
and $\inF{\collabs}$, which return, respectively, the (multi)sets of message edges outgoing from
and incoming into a communicating element in the collaboration $C$:
\[
\begin{array}{l}
\outF{C} = \left\{
\begin{array}{@{}l@{\ }l}
\{\msgEdge{\orgname_1}{\orgname_2}{\messagename}\} &
\begin{array}{l}
    \mathrm{if}\ C\!=\!\tasksendG{\edgeG{\edgename_i}}{\edgeG{\edgename_o}}{\msgEdge{\orgname_1}{\orgname_2}{\messagename}} \\
\ \mathrm{or}\
C\!=\!\evInterSndG{\edgeG{\edgename_i}}{\edgeG{\edgename_o}}{\msgEdge{\orgname_1}{\orgname_2}{\messagename}}\end{array}
\\[.4cm]
\outF{\collabs_1} \!\uplus\! \outF{\collabs_2} & \begin{array}{l}\mathrm{if}\ C\!=\!\collabs_1\!\!  \collabspar\! \collabs_2\end{array}
\\[.2cm]
\emptyset & \begin{array}{l}\mathrm{otherwise}\end{array}
\end{array}
\right.
\\[1.3cm]
%%%%
\inF{C} = \left\{
\begin{array}{@{}l@{\ }l}
\{\msgEdge{\orgname_1}{\orgname_2}{\messagename}\} &\begin{array}{l}
\mathrm{if}\ C\!=\!\taskreceiveG{\edgeG{\edgename_i}}{\edgeG{\edgename_o}}{\msgEdge{\orgname_1}{\orgname_2}{\messagename}} \\
\ \mathrm{or}\
C\!=\!\evInterRcvG{\edgeG{\edgename_i}}{\edgeG{\edgename_o}}{\msgEdge{\orgname_1}{\orgname_2}{\messagename}}\end{array}
\\[.4cm]
\inF{\msgListE} &
\begin{array}{l}\mathrm{if}\ C\!=\! \eventbasedG{\edgeG{\edgename_i}}{\msgListE}\end{array}
\\[.2cm]
\inF{\collabs_1} \uplus \inF{\collabs_2} & \begin{array}{l}\mathrm{if}\ C\!=\!\collabs_1\!\!  \collabspar\! \collabs_2\end{array}
\\[.2cm]
\emptyset &
\begin{array}{l}\mathrm{otherwise}\end{array}
\end{array}
\right.
\\[1.5cm]
\inF{\msgEdgeEvent{\orgname_1}{\orgname_2}{\messagename}{\edgename_o}}=
\{\msgEdge{\orgname_1}{\orgname_2}{\messagename}\}
\qquad\qquad
\inF{\msgListE_1,\msgListE_2} = \inF{\msgListE_1} \uplus \inF{\msgListE_2}
\end{array}
\]
where $\uplus$ denotes the multiset union operator.

We can now formally define the well-composedness property for collaborations.

\begin{defi}[Well-composed collaboration]\label{def1:composition}
Let $\collabs$ be a collaboration, $C$ is well-composed if
$\outF{C}=\inF{C}$
and
$\forall \msgEdge{\orgname_1}{\orgname_2}{\messagename} \in \outF{C}\cup\inF{C} \ . \
\orgname_1 \neq \orgname_2$.
\end{defi}

\begin{exa}
Let consider the processes \emph{a}, \emph{b} and \emph{d} in Fig.~\ref{fig:p_repo}.
By associating them the participant names $\orgname_{bk}$, $\orgname_{c}$
and $\orgname_{bs}$, respectively, and by composing them we obtain the collaboration model $C_1$ in Fig~\ref{fig:example}.b. This is well-composed, because we have:
\[
\begin{array}{l@{}l@{}l}
\outF{C_1} = \inF{C_1} &=& \{\msgEdge{\orgname_{bk}}{\orgname_{bs}}{\mathsf{confirmation}}, \msgEdge{\orgname_{bs}}{\orgname_{c}}{\mathsf{reply}}, \msgEdge{\orgname_{bs}}{\orgname_{c}}{\mathsf{ticket}}, \msgEdge{\orgname_{c}}{\orgname_{bs}}{\mathsf{login}}, \\
&&\ \ \msgEdge{\orgname_{c}}{\orgname_{bs}}{\mathsf{request}},  \msgEdge{\orgname_{c}}{\orgname_{bs}}{\mathsf{abort}}, \msgEdge{\orgname_{c}}{\orgname_{bs}}{\mathsf{book}}, \msgEdge{\orgname_{c}}{\orgname_{bk}}{\mathsf{pay}}\}
% \\[4mm]
 \end{array}\]
Now, let us consider to replace the process \emph{d} by the process \emph{e} in Fig~\ref{fig:p_repo},
which adds an extra behaviour sending an acknowledge message that we suppose targeted to the customer
participant. In this case, however, the resulting composition is the collaboration $C_2$ that is not well-composed.
Indeed, the set $\outF{C_2}$ contains the message edge
$\msgEdge{\orgname_{bs}}{\orgname_c}{ack}$ that is not in $\inF{C_2}$.
Graphically, this corresponds to a malformed BPMN model with a message edge
outgoing from a task of a pool but not connected with a task of another pool.
\end{exa}

Before presenting our formalisation of collaboration creation via
processes composition, we need to introduce few notations and
auxiliary functions.
Firstly, notation $\tuple{\cdot}$ stands for tuples, with
$\tuplel{\cdot}$ denoting the tuple length and $\cdot\proj{i}$
denoting the i-th element of the tuple. For example,
$\tuple{\orgname}$ represents a tuple of participant names
$\langle \orgname_1,\ldots,\orgname_n\rangle$, with $n\geq 0$,
and hence $\tuplel{\tuple{\orgname}}=n$ and $\tuple{\orgname}\proj{i}=\orgname_i$
for $i\in\{1,\ldots,n\}$.
Secondly, since the operator $\collabspar$ is associative, we can generalise
it to an n-ary operator and use the notation $\prod$ to represent its iterated
version. For example, $\prod_{i=1}^n \collabs_i = \collabs_1 \collabspar
\ldots \collabspar \collabs_n$.
Finally, we will resort to the auxiliary functions $\sndFunction$ and
$\rcvFunction$ that, given a message, return the sender and the receiver
participant, respectively.
$\sndFunction$ and $\rcvFunction$ are written as collections of pairs of the form
$\mpair{\messagename}{\orgname}$.
We use $\emptyset$ to denote the empty message
function (i.e., $\emptyset(\messagename)$ is undefined for any $\messagename$),
and $\sndFunction_1 \sqcup \sndFunction_2$
(resp. $\rcvFunction_1 \sqcup \rcvFunction_2$)
to denote the union of $\sndFunction_1$ and $\sndFunction_2$
(resp. $\rcvFunction_1$ and $\rcvFunction_2$)
when they have disjoint domain. We will also generalise $\sqcup$
to the n-ary operator $\bigsqcup$.

Let us now to formally define how a set of processes can be associated
to process names and composed together in order to form a collaboration,
which is defined according to the syntax given in Fig.~\ref{fig:syntaxGlobal}.

\begin{defi}[Processes composition function]\label{def:compositionFunc}
Let $\tuple{\proc}$ and $\tuple{\orgname}$ be a tuple of processes
and a tuple of participant names, respectively; their composition is defined by the
function $\compFunction$ as follows:
\[
\compFunction(\tuple{\proc},\tuple{\orgname})=
\prod_{i=1}^n \namingFunction{
\tuple{\proc}\proj{i},
\sndFunctionC{\tuple{\proc}}{\tuple{\orgname}},
\rcvFunctionC{\tuple{\proc}}{\tuple{\orgname}}
}
\]
with $\tuplel{\tuple{\proc}}=\tuplel{\tuple{\orgname}}=n$.
Functions $\mathcal{S}$ and $\mathcal{R}$, computing the message
functions for the given processes and the corresponding names, are
defined as follows:
\[
\begin{array}{l}
\sndFunctionC{\tuple{\proc}}{\tuple{\orgname}}=\bigsqcup_{i\in\{1,\ldots,n\}}
\sndFunctionC{\tuple{\proc}\proj{i}}{\tuple{\orgname}\proj{i}}
\quad \mathrm{with}\ \tuplel{\tuple{\proc}}=\tuplel{\tuple{\orgname}}=n
\\[.3cm]
\sndFunctionC{\proc_1\mid\proc_2}{\orgname}=
\sndFunctionC{\proc_1}{\orgname} \sqcup \sndFunctionC{\proc_2}{\orgname}
\\[.3cm]
\sndFunctionC{\tasksendG{\edgeG{\edgename_i}}{\edgeG{\edgename_o}}{\messagename}}{\orgname}=
\sndFunctionC{\evInterSndG{\edgeG{\edgename_i}}{\edgeG{\edgename_o}}{{\messagename}}}{\orgname}=
\{\mpair{\messagename}{\orgname}\}
\\[.3cm]
\sndFunctionC{\proc}{\orgname}=\emptyset
\quad \mathrm{for\ any}\ \proc\ \mathrm{different\ from}\ {\sf taskSnd}\ \mathrm{and}\ {\sf interSnd}
\end{array}
\]
\[
\begin{array}{l}
\rcvFunctionC{\tuple{\proc}}{\tuple{\orgname}}=\bigsqcup_{i\in\{1,\ldots,n\}}
\rcvFunctionC{\tuple{\proc}\proj{i}}{\tuple{\orgname}\proj{i}}
\quad \mathrm{with}\ \tuplel{\tuple{\proc}}=\tuplel{\tuple{\orgname}}=n
\\[.3cm]
\rcvFunctionC{\proc_1\mid\proc_2}{\orgname}=
\rcvFunctionC{\proc_1}{\orgname} \sqcup \rcvFunctionC{\proc_2}{\orgname}
\\[.3cm]
\rcvFunctionC{\taskreceiveG{\edgeG{\edgename_i}}{\edgeG{\edgename_o}}{\messagename}}{\orgname}=
\rcvFunctionC{\evInterRcvG{\edgeG{\edgename_i}}{\edgeG{\edgename_o}}{{\messagename}}}{\orgname}=
\{\mpair{\messagename}{\orgname}\}
\\[.3cm]
\rcvFunctionC{\eventbasedG{\edgeG{\edgename_i}}{\msgList} }{\orgname}=
\rcvFunctionC{\msgList}{\orgname}
\\[.3cm]
\rcvFunctionC{\msgEdgeEventP{\orgname_1}{\orgname_2}{\messagename}{\edgename_o}}{\orgname}=
\{\mpair{\messagename}{\orgname}\}
\qquad
\rcvFunctionC{(\msgList_1,\msgList_2)}{\orgname}=\rcvFunctionC{\msgList_1}{\orgname} \sqcup \rcvFunctionC{\msgList_2}{\orgname}
\\[.3cm]
\rcvFunctionC{\proc}{\orgname}=\emptyset
\quad \mathrm{for\ any}\ \proc\ \mathrm{different\ from}\ {\sf taskRcv},\ {\sf interRcv}\ \mathrm{and}\ {\sf eventBased}
\end{array}
\]
Finally, the naming function $\mathcal{N}$ is defined by the
following relevant cases (in the remaining cases, the function acts as an homomorphism):
\[
\begin{array}{l}
\namingFunction{\proc_1\mid\proc_2,\sndFunction,\rcvFunction}=
\namingFunction{\proc_1,\sndFunction,\rcvFunction} \,\mid\,
\namingFunction{\proc_2,\sndFunction,\rcvFunction}
\\[.3cm]
\namingFunction{\tasksendG{\edgeG{\edgename_i}}{\edgeG{\edgename_o}}{\messagename},\sndFunction,\rcvFunction}=
\tasksendG{\edgeG{\edgename_i}}{\edgeG{\edgename_o}}{
(\sndFunction(\messagename),\rcvFunction(\messagename),\messagename)}
\quad \mathrm{if}\ \sndFunction(\messagename)\neq\rcvFunction(\messagename)
\\[.3cm]
\namingFunction{\evInterSndG{\edgeG{\edgename_i}}{\edgeG{\edgename_o}}{{\messagename}},\sndFunction,\rcvFunction}=
\evInterSndG{\edgeG{\edgename_i}}{\edgeG{\edgename_o}}{(\sndFunction(\messagename),\rcvFunction(\messagename),\messagename)}
\quad \mathrm{if}\ \sndFunction(\messagename)\neq\rcvFunction(\messagename)
\\[.3cm]
\namingFunction{\taskreceiveG{\edgeG{\edgename_i}}{\edgeG{\edgename_o}}{\messagename},\sndFunction,\rcvFunction}=
\taskreceiveG{\edgeG{\edgename_i}}{\edgeG{\edgename_o}}{
(\sndFunction(\messagename),\rcvFunction(\messagename),\messagename)}
\quad \mathrm{if}\ \sndFunction(\messagename)\neq\rcvFunction(\messagename)
\\[.3cm]
\namingFunction{\evInterRcvG{\edgeG{\edgename_i}}{\edgeG{\edgename_o}}{{\messagename}},\sndFunction,\rcvFunction}=
\evInterRcvG{\edgeG{\edgename_i}}{\edgeG{\edgename_o}}{(\sndFunction(\messagename),\rcvFunction(\messagename),\messagename)}
\quad \mathrm{if}\ \sndFunction(\messagename)\neq\rcvFunction(\messagename)
\\[.3cm]
\namingFunction{\eventbasedG{\edgeG{\edgename_i}}{\msgList},\sndFunction,\rcvFunction}=
\eventbasedG{\edgeG{\edgename_i}}{
\namingFunction{\msgList,\sndFunction,\rcvFunction}}
\\[.3cm]
\namingFunction{\msgEdgeEventP{\orgname_1}{\orgname_2}{\messagename}{\edgename_o},\sndFunction,\rcvFunction}=
\msgEdgeEvent{\sndFunction(\messagename)}{\rcvFunction(\messagename)}{\messagename}{\edgename_o}
\quad \mathrm{if}\ \sndFunction(\messagename)\neq\rcvFunction(\messagename)
\end{array}
\]
\end{defi}
Intuitively, function $\mathcal{C}$ extracts from the processes to be composed the
information concerning sending and receiving participants for all exchanged messages,
and uses this information for enriching the message edges of each process.
Specifically, function $\mathcal{S}$ identifies as a sender of a message a participant
that performs a sending task or a sending intermediate event with that message as an
argument, while function $\mathcal{R}$ identifies as a receiver of a message a participant
performing a receiving task, a receiving intermediate event or a an event-based gateway
on the message. If the function $\mathcal{C}$
returns a collaboration, this is well-composed; otherwise the function
is undefined, meaning that the processes given in input cannot be correctly composed
to form a collaboration. This is formalised by the following proposition.

\begin{restatable}{prop}{restateprop}
Let $\tuple{\proc}$ and $\tuple{\orgname}$ be a tuple of processes
and a tuple of participant names, respectively; if
$\compFunction(\tuple{\proc},\tuple{\orgname})=\collabs$
then $\collabs$ is well-composed.
\end{restatable}
\begin{proof}
The proof proceeds by contradiction (see the Appendix).
\end{proof}

\begin{exa}
Let $P_a$, $P_b$ and $P_d$ be the textual representation of processes \emph{a}, \emph{b} and \emph{d} in Fig.~\ref{fig:p_repo}. Their composition is formally defined as
$\compFunction(\langle P_a, P_b, P_d \rangle,
\langle \orgname_{bk},\orgname_{c},\orgname_{bs} \rangle)=
(C_{bk}
\ \collabspar \
C_{bs}
\ \collabspar \
C_{c})$,
with
$C_{bk}$, $C_{bs}$ and $C_{c}$ defined in Example~\ref{ex4}.
\end{exa}

\subsubsection*{Semantics of \bpmn{} Collaborations.}%
\label{sec:globalcoll}
The operational semantics we propose for collaborations is given in terms of configurations of the form
$\langle \collabs, \stateedge, \statemsg \rangle$, where:
$\collabs$ is a collaboration structure;
$\stateedge$ is the first part of the execution state,
storing for each sequence edge the current number of tokens marking it;
and $\statemsg$ is the second part of the execution state,
storing for each message edge the current number of message tokens marking it.
Specifically,
$\statemsg : \msgSetMSG \rightarrow \mathbb{N}$ is a function mapping message edges to numbers of message tokens;
so that $\statemsg{\msgEdge{\orgname_1}{\orgname_2}{\messagename}}=\token$
means that there are $\token$ messages of type $\messagename$ sent by $\orgname_1$
and stored in the $\orgname_2$'s queue.
Update and initial state for $\statemsg$ are defined in a way similar to
$\stateedge$'s definitions.

The transition relation $\transitionColl{l}{}$ over collaboration configurations formalizes the execution of a collaboration in terms of edge and message markings evolution. It is defined by the rules in Fig.~\ref{fig:semanticsColl}; for the sake of presentation, we omit the rules concerning start/end events
and gateways, as they are the same of those for choreographies (reported in Fig.~\ref{fig:semanticsChoreographyGlobal}).  As usual, we  omit the collaboration structure  from the target configuration of transitions.

\begin{figure}[!t]
\footnotesize
\[
\hspace*{-1.3cm}
\begin{array}{|@{\hspace*{0.2cm} \ }l@{\hspace*{0.3cm} }l@{\hspace*{-0.2cm}}l|} % chktex 44
  \hline % chktex 44
&&\\[-.2cm]
  \begin{array}{l}
     	\langle \eventbasedG{\edgeG{\edgename_i}}{\msgEdgeEvent{\orgname_1}{\orgname_2}{\messagename}{\edgename_o} \cup \msgList}, \stateedge, \statemsg \rangle
  	\transition{\orgname_1 \rightarrow \orgname_2:\messagename}{}
	\langle
	\incToken{ \decToken{\stateedge}{\edgename_i}}{\edgename_o}, \decToken{\statemsg}{\msgEdge{\orgname_1}{\orgname_2}{\messagename}} \rangle
   \end{array}
	&
		        \begin{array}{l}
          \statefunc{\edgename_i}>0,   \\
           \statemsg{\msgEdge{\orgname_1}{\orgname_2}{\messagename}}\!\!>\!\!0
              \end{array}
	 &
  \quad    (\mathit{C\textrm{-}EventG})
   \\[.6cm]
          \begin{array}{l}
     	\langle \taskG{\edgeG{\edgename_i}}{\edgeG{\edgename_o}}, \stateedge, \statemsg \rangle
  	\transition{\tau}{}
	\langle
	\incToken{ \decToken{\stateedge}{\edgename_i}}{\edgename_o}, \statemsg \rangle
	   \end{array}
	   &
	   		        \begin{array}{l}
	  \statefunc{\edgename_i}>0
	    \end{array}
     &
    \quad      (\mathit{C\textrm{-}Task})
       \\ [.6cm]
             \begin{array}{l}
      	\langle \taskreceiveG{\edgeG{\edgename_i}}{\edgeG{\edgename_o}}{\msgEdge{\orgname_1}{\orgname_2}{\messagename}}, \stateedge, \statemsg \rangle
  	\transition{\orgname_1 \rightarrow \orgname_2:\messagename}{}
		\langle
	\incToken{ \decToken{\stateedge}{\edgename_i}}{\edgename_o}, \decToken{\statemsg}{\msgEdge{\orgname_1}{\orgname_2}{\messagename}} \rangle
  \end{array}
     &
         \begin{array}{l}
   \statefunc{\edgename_i}>0, \\ \statemsg{\msgEdge{\orgname_1}{\orgname_2}{\messagename}}\!\!>\!\!0
    \end{array}
     &
    \quad           (\mathit{C\textrm{-}TaskRcv})
       \\ [.6cm]
                    \begin{array}{l}
    \langle \tasksendG{\edgeG{\edgename_i}}{\edgeG{\edgename_o}}{\msgEdge{\orgname_1}{\orgname_2}{\messagename}}, \stateedge, \statemsg \rangle
\transition{\tau}{}
 \langle \incToken{ \decToken{\stateedge}{\edgename_i}}{\edgename_o},   \incToken{\statemsg}{\msgEdge{\orgname_1}{\orgname_2}{\messagename}}   \rangle
   \end{array}
     &
     	        \begin{array}{l}
  \statefunc{\edgename_i}>0
    \end{array}
     &
    \quad       (\mathit{C\textrm{-}TaskSnd})
       \\ [.6cm]
                       	    \begin{array}{l}
	    \langle \evInterRcvG{\edgeG{\edgename_i}}{\edgeG{\edgename_o}}{\msgEdge{\orgname_1}{\orgname_2}{\messagename}}, \stateedge, \statemsg \rangle
  	\transition{\orgname_1 \rightarrow \orgname_2:\messagename}{}
	\langle
	\incToken{ \decToken{\stateedge}{\edgename_i}}{\edgename_o}, \decToken{\statemsg}{\msgEdge{\orgname_1}{\orgname_2}{\messagename}} \rangle
	  \end{array}
     &
                \begin{array}{l}
           \statefunc{\edgename_i}>0, \\
         \statemsg{\msgEdge{\orgname_1}{\orgname_2}{\messagename}}\!\!>\!\!0
            \end{array}
     &
    \quad      (\mathit{C\textrm{-}InterRcv})
       \\ [.6cm]
                              \begin{array}{l}
   \langle \evInterSndG{\edgeG{\edgename_i}}{\edgeG{\edgename_o}}{\msgEdge{\orgname_1}{\orgname_2}{\messagename}}, \stateedge, \statemsg \rangle
  \transition{\tau}{}
 \langle \incToken{ \decToken{\stateedge}{\edgename_i}}{\edgename_o},   \incToken{\statemsg}{\msgEdge{\orgname_1}{\orgname_2}{\messagename}}   \rangle
             \end{array}
     &
                 \begin{array}{l}
 \statefunc{\edgename_i}>0
   \end{array}
     &
    \quad     (\mathit{C\textrm{-}InterSnd})
       \\ [.6cm]
   \hline % chktex 44
\end{array}
\]
\vspace*{-.5cm}
\caption{Collaboration Semantics (excerpt of rules).}%
\label{fig:semanticsColl}
\vspace{-3mm}
\end{figure}

We now briefly comment on the operational rules.
Rule $\mathit{C\textrm{-}EventG}$ is activated when there is a token in the incoming edge of an event-based
gateway and there is a message $\messagename$ to be consumed, so that the application of the rule moves the token from the incoming edge to the outgoing edge corresponding to the received message, whose number of message tokens in the meantime is decreased (i.e., a message from the corresponding queue is consumed).
Rule $\mathit{C\textrm{-}Task}$  deals with simple tasks, acting as a pass through.
Rule $\mathit{C\textrm{-}TaskRcv}$ is activated not only when there is a token in the incoming edge, like the one related to simple tasks, but also when there is a message to be consumed.
Similarly, rule $\mathit{C\textrm{-}TaskSnd}$, instead of consuming, adds a message in the corresponding queue.
\ft{It is worth noticing that rule $\mathit{C\textrm{-}TaskSnd}$ produces transitions labelled by $\tau$,
meaning that sending actions are not observable. This is because in our conformance checking approach
we compare the execution of a choreography task only with the reception of the corresponding message
in the collaboration. We formalise this below in this section, and we provide a more thorough discussion
about this distinctive aspect of our proposal in Section~\ref{sec:discussion}.
}
Rule $\mathit{C\textrm{-}InterRcv}$ (resp.  $\mathit{C\textrm{-}InterSnd}$) follows the same behavior of rule $\mathit{C\textrm{-}TaskRcv}$ (resp. $\mathit{C\textrm{-}TaskSnd}$).

\subsection*{Conformance Checking}%
\label{sec:conformance}
This section discusses about the relations we propose for checking the conformance
between choreographies and collaborations. We then present how they work in practice.

\subsubsection*{Bisimulation-Based and Trace-Based Conformance.}
Here we present the Bisimulation-Based Conformance (BBC) and the Trace-Based  Conformance (TBC) relations we have defined.
The two relations are inspired by well-established behavioural equivalences~\cite{milner_communication_1989}, largely used in the literature and revised to deal with \bpmn{} characteristics.

Before providing the formal definition of BBC, we introduce the necessary notation. $\choSet$ and $\coSet$  represents the sets of all choreography and collaboration configurations, respectively. Moreover, weak transitions are defined as follows: $\Transition{}{}$ denotes the reflexive and transitive closure of $\transition{\tau}{}$, i.e.\ zero or more $\tau$-transitions; $\Transition{l}{}$ denotes $\Transition{}{} \transition{l}{}  \Transition{}{} $. We exploit functions $labels(C)$ and $labels(\chos)$ returning the sets of all communication labels that can be potentially generated by the collaboration $C$ and the choreography $\chos$, respectively. These functions are inductively defined on the syntax of collaboration and choreography structures in a straightforward way. For example, in case of choreographies we have the definition case $labels(\chotaskOWG{\edgeG{\edgename_i}}{\edgeG{\edgename_o}}{\orgname_1}{\orgname_2}{\messagename}{\taskname})=\{\orgname_1 \rightarrow \orgname_2:\messagename\}$, meaning that if a choreography contains a task element, then its label set contains the label corresponding the message exchange described by the task.

At the collaboration level the definition of conformance requires the use of the hiding operator $\collabs/L$, defined by the rules in Fig.~\ref{fig:operators}. This operator, as usual, transforms into $\tau$ all the actions in the set $L$, in order to consider them as internal actions in the conformance relation.
\begin{figure}
\footnotesize
  \setlength{\belowcaptionskip}{-10pt}
	\[
	\begin{array}{|@{\ \ }r@{\ }|} % chktex 44
	\hline % chktex 44
	\\[-0.3cm]
	\prooftree
	\langle \collabs, \stateedge, \statemsg \rangle \transition{\textit{l}}{} \langle \stateedge', \statemsg' \rangle
	\justifies
	\raisebox{-.1cm}{$
		\langle \collabs/L, \stateedge, \statemsg \rangle \transition{\textit{l}}{} \langle \stateedge', \statemsg' \rangle % chktex 1
		$}
	\endprooftree\ \;
	\textit{l} \notin L
	\qquad
	\prooftree
	\langle \collabs, \stateedge, \statemsg \rangle \transition{\textit{l}}{} \langle \stateedge', \statemsg' \rangle
	\justifies
	\raisebox{-.1cm}{$
		\langle \collabs/L, \stateedge, \statemsg \rangle \transition{\tau}{} \langle \stateedge', \statemsg' \rangle % chktex 1
		$}
	\endprooftree\ \;
	\textit{l} \in L \;\;\;

	\\[.6cm]
	\hline % chktex 44

	\end{array}
	\]
			\vspace{-0.5cm}
		\caption{Hiding Operator.}%
	\label{fig:operators}
	\vspace{-3mm}
\end{figure}

\vspace{2mm}
\begin{defi}[BBC Relation]
A relation $\mathcal{R}\subseteq(\choSet \times \coSet)$  is a weak Bisimulation Conformance if,
for any $\langle \chos, \stateedge_{ch} \rangle \in  \choSet$ and
$\langle \collabs, \stateedge_c, \statemsg \rangle \in \coSet$ such that
$ \langle \chos, \stateedge_{ch} \rangle\ \mathcal{R}\ \langle \collabs, \stateedge_c, \statemsg \rangle$, it holds:
\begin{itemize}[noitemsep,topsep=0pt]
\item for all $\orgname_1, \orgname_2, m$ and $\stateedge_{ch}'$,
if $\langle \chos, \stateedge_{ch} \rangle$ $\transition{\orgname_1 \rightarrow \orgname_2:\messagename}{} \stateedge_{ch}' $\\
then $\langle \collabs, \stateedge_c, \statemsg \rangle$ $\Transition{\orgname_1 \rightarrow \orgname_2:\messagename}{} \langle \stateedge_c', \statemsg' \rangle$
for some $\stateedge_c', \statemsg'$
s.t. $\langle \chos, \stateedge_{ch}' \rangle\ \mathcal{R}\ \langle \collabs, \stateedge_c', \statemsg' \rangle$;

\item for all $\orgname_1, \orgname_2, m, \stateedge_c'$ and $\statemsg'$,
if $\langle \collabs, \stateedge_c, \statemsg \rangle$ $\transition{ \orgname_1 \rightarrow \orgname_2:\messagename}{} \langle \stateedge_c', \statemsg' \rangle$ \\
then $\langle \chos, \stateedge_{ch} \rangle$\ $\Transition{\orgname_1 \rightarrow \orgname_2:\messagename}{}  \stateedge_{ch}'$ for some $\stateedge_{ch}'$
s.t. $\langle \chos, \stateedge_{ch}' \rangle\ \mathcal{R}\ \langle \collabs, \stateedge_c', \statemsg' \rangle$;

\item for all $\stateedge_{ch}'$,
if $\langle \chos, \stateedge_{ch} \rangle$ $\transition{\ \tau \ }{} \stateedge_{ch}' $\\
 then $\langle \collabs, \stateedge_c, \statemsg \rangle$ $\Transition{\quad  \ }{} \langle \stateedge_c', \statemsg' \rangle$
for some $\stateedge_c', \statemsg'$
s.t. $\langle \chos, \stateedge_{ch}' \rangle\ \mathcal{R}\ \langle \collabs, \stateedge_c', \statemsg' \rangle$;

\item for all $\stateedge_c'$ and $\statemsg'$,
if $\langle \collabs, \stateedge_c, \statemsg \rangle$ $\transition{\ \tau \ }{} \langle \stateedge_c', \statemsg' \rangle $\\
then $\langle \chos, \stateedge_{ch} \rangle$ $\Transition{\quad  \ }{} \stateedge_{ch}'$ for some $\stateedge_{ch}'$
s.t. $\langle \chos, \stateedge_{ch}' \rangle\ \mathcal{R}\ \langle \collabs, \stateedge_c', \statemsg' \rangle$.
\end{itemize}
A choreography $\langle \chos, \stateedge_{ch} \rangle$
and a collaboration $\langle \collabs, \stateedge_c, \statemsg \rangle$
\emph{conform}
if there exists a weak Bisimulation Conformance relation $\mathcal{R}$
such that\\
\mbox{$\langle \chos, \stateedge_{ch} \rangle
\ \mathcal{R}\
\langle \collabs/(labels(C)\!\setminus\!labels(\chos)), \stateedge_c, \statemsg \rangle$.}%
	\label{def:conformance}
\end{defi}
The proposed BBC relation considers to conform collaborations that are able to simulate step by step choreographies, and vice versa.
In particular, if the choreography performs a message exchange, in the collaboration we expect
to observe the reception of the message, possibly preceded or followed by any number of internal actions, and then the two continuations have to be in relation.
Analogously, if we observe a message reception in the collaboration, the choreography has to reply with the corresponding weak transition.
Moreover, if one of the two models performs an internal action, the counterpart can react with a weak transition $\Transition{}{}$.
The definition of conformance is quite close to a standard bisimulation relation, except for the use of the hiding operator at the collaboration level.
Specifically, the hiding is used to ignore all additional behaviors in the collaboration that are not explicitly expressed, and hence regulated, in the choreography. In this way, even if a collaboration performs some additional communications, if it is able to (bi)simulate with the given choreography, they do conform. % chktex 36
\ft{The different communication models defined in the semantics of choreographies and collaborations significantly affects the conformance checking.
We thoroughly discuss the motivations underlying
our choice of comparing the execution of tasks in the choreography with message receptions
in the collaboration in Section~\ref{sec:discussion}, when all involved technicalities have been introduced
and practical examples of the application of our conformance checking notion have been shown. }
\new{Notably, as discussed in Section~\ref{sec:method}, our approach is hybrid and its goal
is to encourage the reuse of existing process models as much as possible. For this reason we
have based our conformance checking on weak bismulation rather than a stronger notion
of weak equivalence, as e.g.\ branching bisimulation. Indeed, we do not aim
at ensuring a strong conformance between models, but we just focus on the correct exchange of messages abstracting from internal actions. Hence, for our purposes, a stronger relation would act in a too discriminatory way.}

BBC guarantees that the collaboration takes decisions, concerning the execution flow, exactly as what is specified in the choreography. Sometimes this condition may be too restrictive and the system designer would prefer to adopt a weaker relation
\ft{(examples of these situations are described in the following subsection)}.
To this aim, in our work we also introduce the more relaxed TBC relation.
Intuitively, in this case two models conform if and only if they can perform exactly the
same weak sequences of actions.
In the definition below, we deem a label to be \emph{visible} if it is of
the form $\orgname_1 \rightarrow \orgname_2:\messagename$. Notationally, the transition
$\langle \chos, \stateedge \rangle$\ $\Transition{s\ }{}  \stateedge'$,
where $s$ is a sequence of visible labels $l_1 l_2 \ldots l_n$,
denotes the sequence
$\langle \chos, \stateedge \rangle$\
$\Transition{l_1\ }{}$
$\langle \chos, \stateedge_1 \rangle$\
$\Transition{l_2\ }{}$
$\langle \chos, \stateedge_2 \rangle$\
\ldots
$\Transition{l_n\ }{}$
$\langle \chos, \stateedge' \rangle$\
of weak transitions.
Transition $\langle \collabs, \stateedge, \statemsg \rangle$ $\Transition{s\ }{} \langle \stateedge', \statemsg' \rangle$ is similarly defined.

\begin{defi}[TBC Relation.]\label{def:trace-conformance}
A choreography $\langle \chos, \stateedge_{ch} \rangle$
and a collaboration $\langle \collabs, \stateedge_c, \statemsg \rangle$
\emph{trace conform}
if, given $\collabs'=\collabs/(labels(C)\!\setminus\!labels(\chos))$, for any sequence $s$ of visible labels it holds:
\vspace{-1mm}
\begin{itemize}[noitemsep]
\item
$\langle \chos, \stateedge_{ch} \rangle$\ $\Transition{s\ }{}  \stateedge_{ch}'$
implies
$\langle \collabs', \stateedge_c, \statemsg \rangle$ $\Transition{s\ }{} \langle \stateedge_c', \statemsg' \rangle$
for some $\stateedge_c'$ and $\statemsg'$;
\item
$\langle \collabs', \stateedge_c, \statemsg \rangle$ $\Transition{s\ }{} \langle \stateedge_c', \statemsg' \rangle$
implies
$\langle \chos, \stateedge_{ch} \rangle$\ $\Transition{s\ }{}  \stateedge_{ch}'$
for some $\stateedge_{ch}'$.
\end{itemize}
\end{defi}
\noindent
The TBC relation guarantees that the collaboration is able to produce the same sequences of messages of the choreography, and vice versa, without controlling presence of deadlock states and distinguishing different decision points and non-determinism forms.
Concerning this latter point, BBC can recognize dominated non-determinism, where a participant (non-deterministically) takes a decision using a XOR gateway and the other behaves accordingly, from non-dominated non-determinism, based on a race condition among the messages managed by an event-based gateway. As it usually happens for these classes of behavioural relations,
models that conform according to BBC also conform according to TBC\@.

\subsubsection*{Conformance at work.}%
\label{atWork}
\begin{table}
	\begin{tabular}{ccccccc}
		Cases & Bank & Customer & Booking System & Well-composed & TBC & BBC \\ \toprule
		1    & a                         & b                             & d                                   & yes                         & false                   &  false        \\
		2    & a                         & b                             & e                                   & no                          & ---                    &   ---      \\
		3    & a                         & b                             & f                                   & no                          & ---                     &       --- \\
		4    & a                         & c                             & d                                   & no                          & ---                      &     ---  \\
		5    & a                         & c                             & e                                   & yes                         & true                      &     true \\
		6    & a                         & c                             & f                                   & yes                         & true                      &    false \\
	\end{tabular}
	\caption{Possible Processes Combination}%
	\label{tab:combination}
\end{table}

To demonstrate in practice the characteristics of the conformance relations, focusing on the management of non-determinism and asynchronous messages, we test them considering the various process model presented in Fig.~\ref{fig:p_repo}, where three participants are involved.

Combining the processes reported in Fig.~\ref{fig:p_repo} we have
six different possibilities to enable the choreography, according to
different selections of the processes for the customer and the booking
system roles.  As shown in Table~\ref{tab:combination}, these
combinations will lead to different results with respect to the actual
capability to enact the choreography. In particular, both syntactically and
semantically related issues can emerge.

The combinations of processes in cases
2, 3 and 4 violate the notion of well-composedness. In cases 2 and
3 the Booking system $e$ and $f$ contain an extra \textit{ack} message not
correctly managed by Customer 1. Case 4 instead is in the opposite
situation, where Customer 2 is expecting an ack message not correctly
managed by Booking System 1.  Cases 1, 5 and 6 are different,
as the processes once combined satisfy the notion of well-composedness given in the Definition~\ref{def1:composition}, which guarantees the correct matching of all messages between the processes.
However, when compared to the choreography, just one of these cases satisfies the BBC, while the other two satisfy only the TBC conformance notion.

The conformance checking results reported in the table show in detail the differences between BBC and TBC\@.
The designer can select the more appropriate relation that fits more his needs, taking into account that BBC provides more guarantees on the correct behaviour between the two models, while TBC ensures only that both models produce the same sequences of messages.

\am{The proposed methodological approach, described in Section~\ref{sec:method}, drove our choices for the behavioural equivalences
to use for conformance checking.
Indeed, in a hybrid context, the reuse and the integration of existing processes, or their replacement without altering the global behaviour of the system, are the key factors for a collaborative environment that needs to satisfy an established specification.
In such a situation, the use of the bisimulation equivalence can guarantee the faithful correspondence of the emerging behaviour of the composed collaboration with respect to the choreography specification.
However, such equivalence sometimes could be too restrictive, since the collaboration is not always able to replicate the choreography specification.
At this point, the TBC equivalence plays a fundamental role in discriminating on the possibility to use the collaboration, possibly with some adjustments,  or not. The TBC does not guarantee the same stringent behavioural properties of the BBC (e.g., absence of deadlocks), but still it can distinguish two models according to their admitted sequences of actions.
So, the modeler is conscious that the violation of the BBC establishes in general the incompatibility between the achieved collaboration and the high-level behaviour prescribed by the choreography.
Anyway, having the TBC satisfied leaves some possibilities that the collaboration, after a refactoring process, can satisfy the given choreography.}

\ft{An evident example of this situation is depicted in Table~\ref{tab:combination} at case 6, where the resulting composition of processes \emph{a}-\emph{c}-\emph{f} is TBC but not BBC\@.
The three processes can expose the same sequences of actions defined by the specification, but a deadlock can emerge. In fact, the customer and the booking system processes can make different choices in their exclusive gateways and, hence, the booking system will be blocked waiting for a message forever, consequently invalidating the BBC\@.
Unfortunately, the violation of the BBC indicates that, as it is, the collaboration is not conformant with the choreography specification. However, the satisfaction of the TBC
\new{suggests} that the process behaviours in the collaboration model could still be fixed with a refactoring process.
As mentioned in Section~\ref{sec:method}, this is the case in which it can be possible to solve the issue adapting the behaviour of process \emph{f}, so that it will not take an internal choice different from that taken by \textit{c}. In such a case we could introduce an adapter that shields the process \emph{f} and, at the right time, it uses
an event-based gateway to wait for the message from \emph{c} indicating whether the customer intends to accept or cancel the itinerary proposal. Then, the adapter will interact with process \emph{f} using local and/or lower level mechanisms
(e.g., by setting the value of a data object exploited in the conditions added to the exclusive gateway, or,
at mere implementation level, by invoking a method to provide the information about the customer decision),
so to drive the booking system process towards the right path in the exclusive gateway. Notably, in this way the
structure of \emph{f} remains unchanged.
}

\ft{There are situations in which a collaboration that only satisfies TBC can easily be transformed in order to satisfy BBC by just refining the conditions on gateways, and without making any additional assumption on the knowledge needed by the involved  processes so to take the corresponding paths.} \ft{Fig.~\ref{fig:annotatedBSCH} depicts a model regulating the shopping of alcohol between a Customer and a Bar.
The process starts with the selection by the customer of a type of drink; if it is alcoholic the bar needs further information about the age of the customer before serving the drink. Vice versa, no check is requested in case of nonalcoholic drinks.
Checking this choreography with the corresponding collaboration depicted in Fig.~\ref{fig:annotatedBSC},
we have that only the TBC equivalence holds.
This result
suggests that the collaboration is not respecting the behaviour specified in the choreography,
since the exclusive gateways in the customer and bar processes without any condition could lead to choices that violate the behaviour specification, allowing e.g.\ the customer to skip the age check. With a deeper examination, and a successive refinement consisting in the inclusion of expressions in the gateways, it is possible to improve the model in order to fix the issue and consequently satisfy the BBC equivalence.
\begin{figure}
    \centering
    \includegraphics[scale=.35]{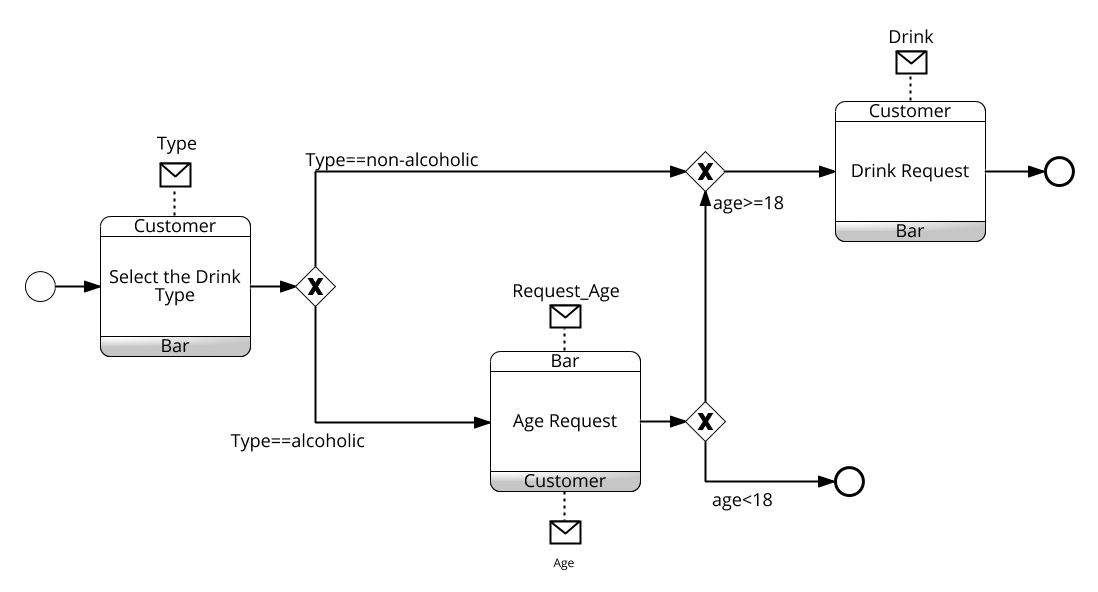}
    \vspace{-.5cm}
        \caption{Alcohol Shopping Choreography.}%
    \label{fig:annotatedBSCH}
\end{figure}
\begin{figure}
    \centering
    \includegraphics[scale=.35]{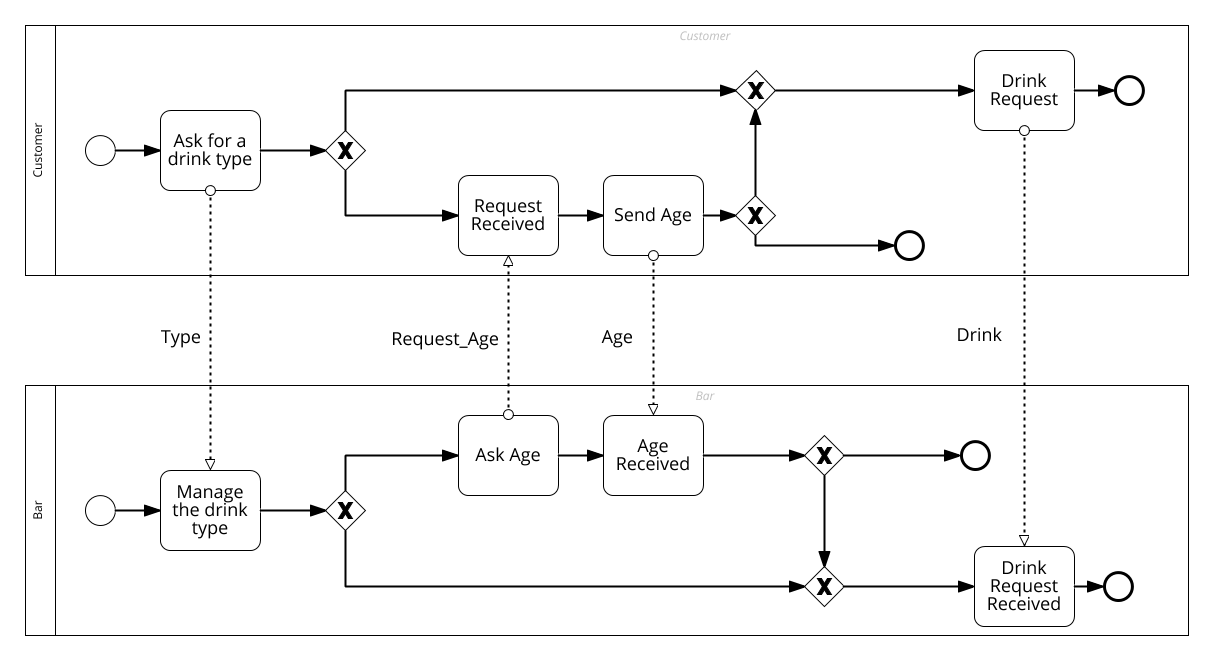}
    \vspace{-.5cm}
        \caption{Alcohol Shopping Collaboration.}%
    \label{fig:annotatedBSC}
\end{figure}
\begin{figure}
    \centering
    \includegraphics[scale=.35]{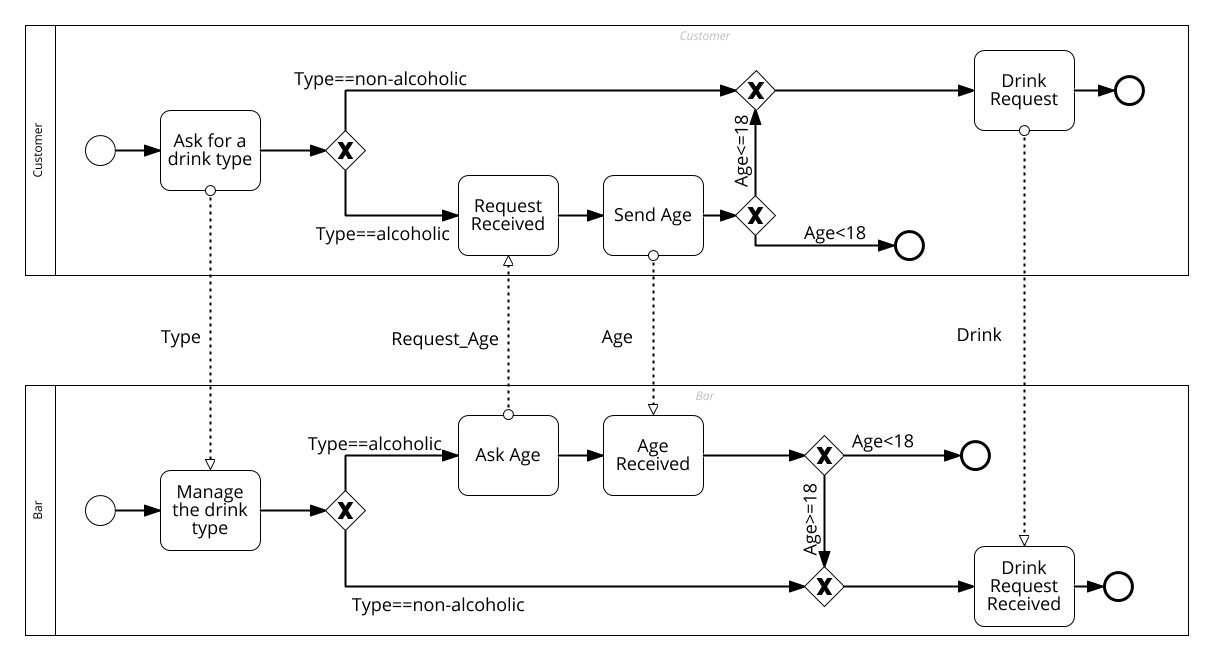}
        \caption{Alcohol Shopping Collaboration Refinement.}%
    \label{fig:refinedC}
\end{figure}
The result of the refinement is reported in Fig.~\ref{fig:refinedC}, which shows the same processes of Fig.~\ref{fig:annotatedBSC} but enriched with expressions based on shared information messages, i.e. \emph{Type} and \emph{Age}.
The introduction of such expressions reduces the
possible behaviours of the collaboration model, thus avoiding those situations
leading to executions that deviate from the ones prescribed by the choreography.}
\ft{It is worth mentioning that the case of processes \emph{a}-\emph{c}-\emph{f} cannot be reasonably solved using the same solution, at least from a pragmatic point of view. Indeed, in such a case an additional communication would have been needed to share, before the two exclusive gateways, the decision of the customer. But, clearly, this would have been absolutely artificial, resulting in a collaboration where the customer has to send twice the same information.}

\section{From Theory to Practice: \texorpdfstring{\emph{C}$^{\ \!\!4}$}{C4}  tool}%
 \label{sec:tool}

The \name\ formal framework presented so far it has been implemented as a web-based
toolchain\footnote{The tool is available on-line at the following link  \emph{\url{http://pros.unicam.it/c4/}}}
that permits to cover the entire choreography life process reported in Section~\ref{sec:method}. The tool supports system designers in modelling diagrams, and in the verification of conformance between a set of modelled processes composed to form a collaboration and a prescribed choreography. A distinctive aspect of the tool is its ability to hide the underlining formal technicalities, so to be usable also to those \bpmn{} designers and software
engineers that, \new{even though they should clearly have high-level modeling and analysis skills}, are not so much familiar with formalisms and verification techniques details.
 \begin{figure}[h]
 \setlength{\belowcaptionskip}{-15pt}
\hspace{-0.3cm}
 	\includegraphics[scale=.28]{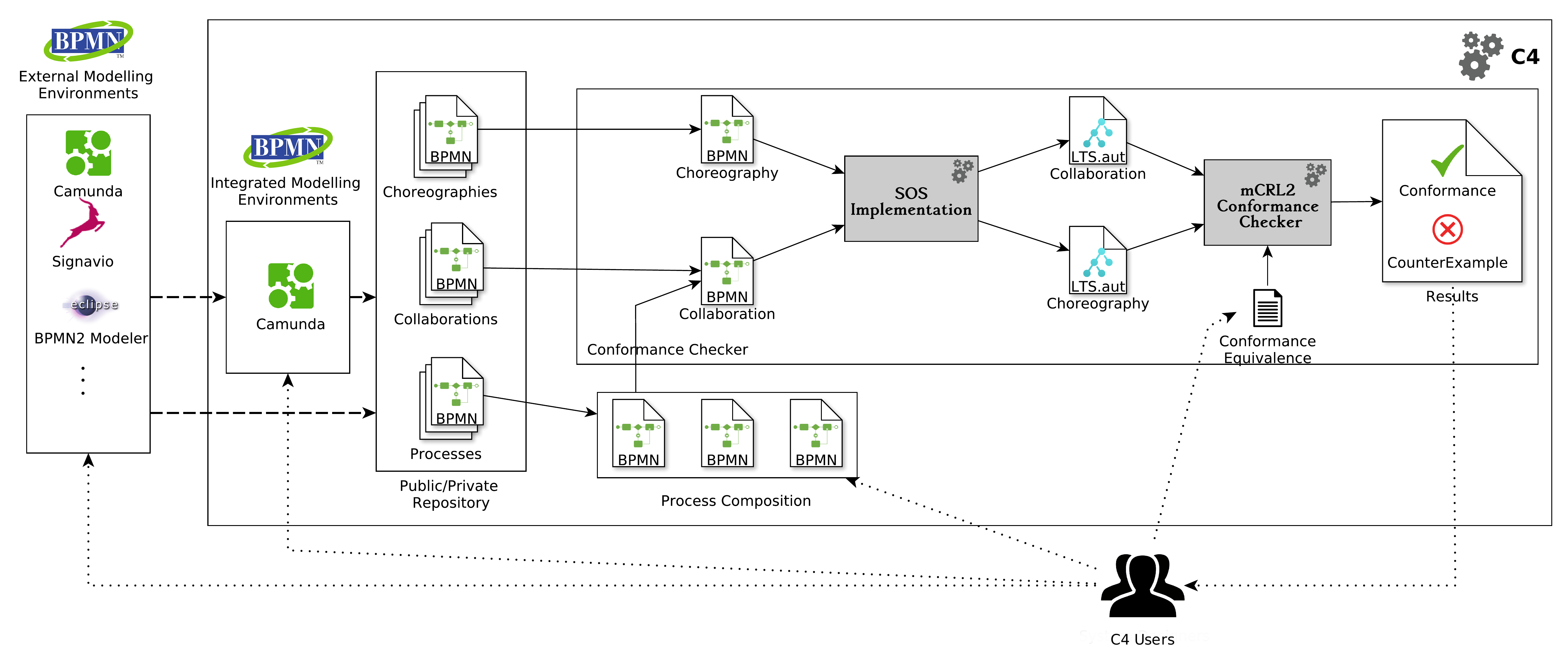}
	\vspace{-0.5cm}
 	\caption{\name\ Supporting Tool.}%
 	\label{fig:framework}
 \end{figure}
\new{It is worth noticing that \name\ has multiple users that, according to their role, are involved
in different moments
of the process. These are modellers, possibly belonging to a super partes organisation, that are in charge of producing choreography diagrams, which in a given application context act as blueprints. Again, the \name{} users are modellers when they represent the (process) behaviours that organisations can bring within compositions.
Moreover, the users are system integrators when they compose, in automatic way, the process models provided by the participating organisations in order to create collaboration diagrams,
which then are automatically compared against choreography diagrams. }

\begin{figure}[t]
  	\centering
  	\subfigure[\protect Models Selection.\label{fig:modeler}]{
	\includegraphics[scale=.2]{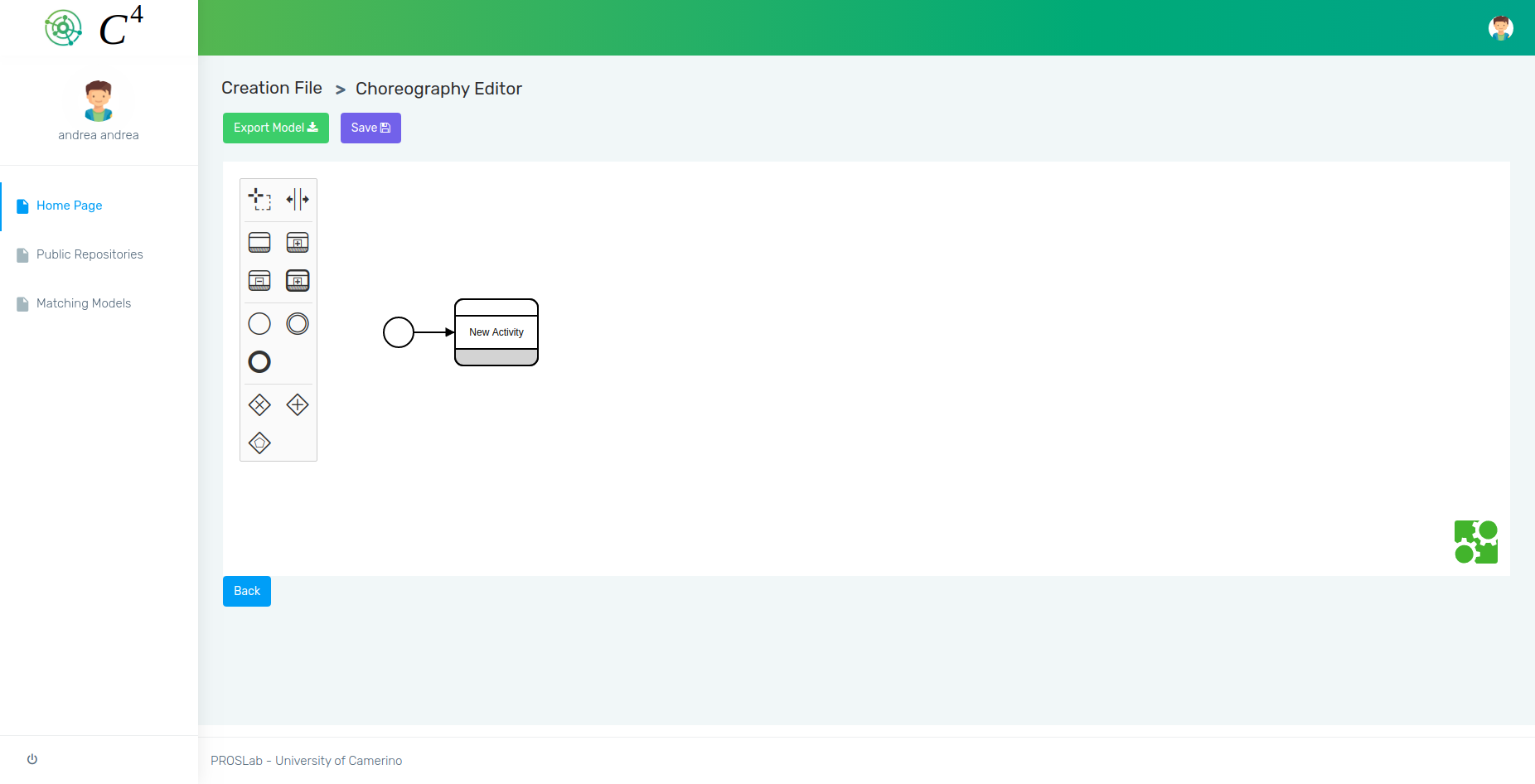}
}
\vspace{0.1cm}
 	\subfigure[\protect Conformance Checker.\label{fig:checker}]{
 	\includegraphics[scale=.3]{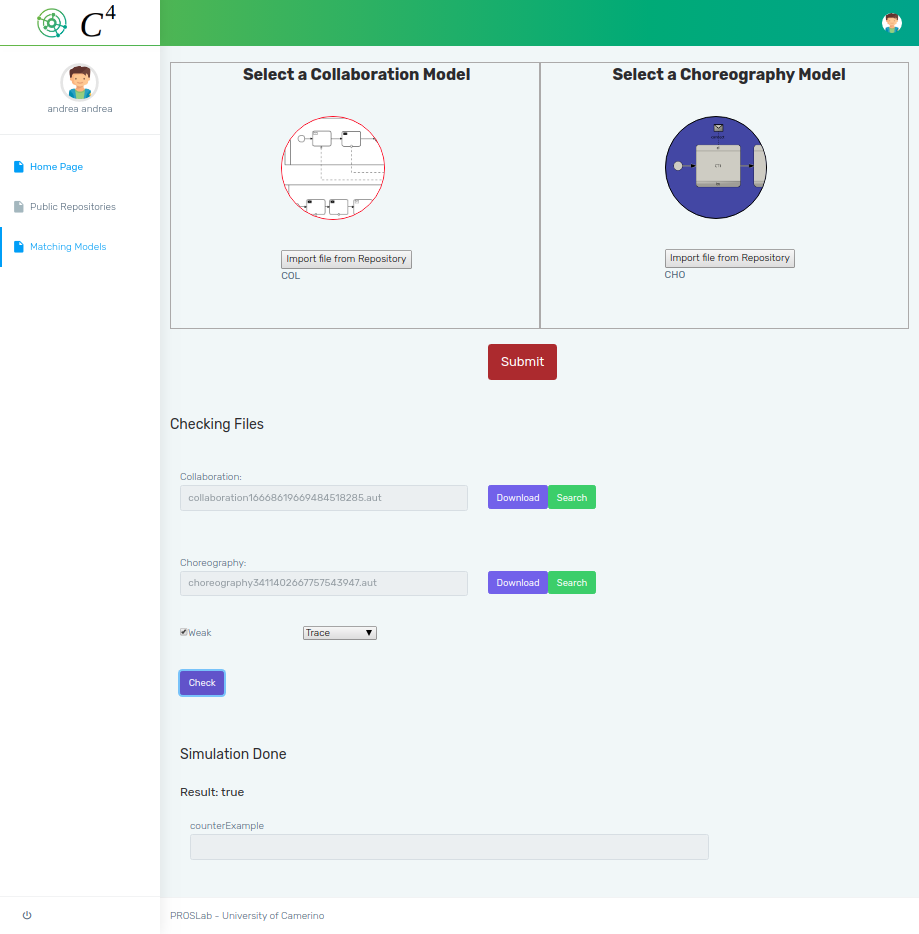}
 	}
         \vspace{-0.4cm}
 	\caption{\name\ Graphical User Interface.}%
 	\label{fig:user_interface}
         \vspace{-0.4cm}
 \end{figure}

Fig.~\ref{fig:framework} depicts the internal components of the \name\ tool, and its interfaces with the user. Specifically, \name\ is able to manage choreography, collaboration and  process models in the \emph{.bpmn} format.  These input models can be generated by a system designer using the Camunda modeller, which has been fully integrated in the \name\ platform (Fig.~\ref{fig:modeler}), or other external \bpmn\ modelling environments. Indeed, the \name\ platform is compatible with the most common modelling environments, such as Eclipse \bpmn{} Modelling, Camunda and Signavio.
Once the user has selected the models to consider, it is possible to further manipulate them. In particular the \name\ framework permits to compose processes according to the approach formalised in Definition~\ref{def:compositionFunc}. The result of such an activity will be a new collaboration diagram containing the selected processes that are connected through messages matched with respect to their names and flow directions.

Generated collaborations can be then checked with respect to selected choreography diagrams. The access to such a functionality is provided via a dedicated GUI (Fig.~\ref{fig:checker}). Here the user can select both the choreography and the collaboration models that have to be compared. Designed models are saved by the editor on a remote folder based repository, and can then be loaded by the user selecting them through the dedicated GUI\@.
At this point, clicking on the submit button, the \name\ tool parses the models and generates the corresponding LTS graphs for both the choreography and the collaboration models.
The parsing of the input files is based on the Camunda APIs, while the LTSs are generated by means of a Java implementation of the direct semantics defined in Section~\ref{sec:formal}.

Once the LTSs are generated, \name\ saves the results in two \emph{.aut} files~\cite{Fernandez1996}. It is now possible to run the conformance checker with respect to a conformance relation that has to be selected by the user through a drop-down list. The conformance checking is implemented using mCRL2~\cite{groote2014modeling}, which has been fully integrated in the \name\ tool. Notably, the standard bisimulation and trace equivalences supported by mCRL2 can be directly used at this stage, as all the specific characteristics of our conformance relations (e.g., the use of hiding) have been already taken into account during the LTS generation.
The verification results consists in a boolean message that reports the value \texttt{true} in case the collaboration conforms the choreography with respect to the selected conformance relation, or \texttt{false} in case the conformance does not hold. In the latter case, the corresponding counterexample is returned.
Notably, the \name\ conformance checker (Fig.~\ref{fig:checker}) allows to have a preview of the LTS graph or download it in the \emph{.aut} format. This enables the possibility to run the verification using other model checkers.
The tool is also available as a stand-alone solution, only with respect to the model checking functionality.

\smallskip
\noindent
\emph{\textbf{\name\ tool at work on the booking example.}}
In order to show the usage of the proposed approach, here we focus on the checking of the well-composed collaborations  depicted in Table~\ref{tab:combination}.
The objective is to check if the generated collaborations are valid implementations of the choreography in Fig.~\ref{fig:example}.a, considering both the BBC and TBC relations.

Analysing the first case in Table~\ref{tab:combination} we realise that by combining the processes \emph{a}-\emph{b}-\emph{d} reported in Fig~\ref{fig:p_repo} we obtain the collaboration in Fig.~\ref{fig:example}.b.
Now checking the collaboration in Fig.~\ref{fig:example}.b with respect to the choreography in Fig.~\ref{fig:example}.a we obtain in return the violations for both conformance relations.
More specifically, considering TBC the following counterexample is reported:
\begin{center}
  { $ \emph{c$\transition{}{}$bs:\,login,\ \
    c$\transition{}{}$bs:\,request,\ \
    bs$\transition{}{}$c:\,reply,\ \
    c$\transition{}{}$bk:\,pay} $}
\end{center}
where $c$, $bs$ and $bk$ stand for the customer, booking system and bank participant names, respectively.
This trace is allowed by the collaboration model and not by the choreography model.
It shows that the reasonable behaviour `book and then pay' is not respected in the collaboration,
which indeed permits to pay a reservation before booking it.
This undesired behaviour is due to the non-blocking nature of the collaboration sending task, which  permits to the customer to send the payment immediately after the booking request, without waiting for any acknowledgement from the booking system.
 This would not be a problem in case of a collaboration with only two participants, or more generally when the receiver of the two messages is the same. In this case the order in which the messages will be processed is managed by the behaviour of the receiver. Instead, in our running scenario the \emph{book} and the \emph{pay} messages are received by two different participants. \ap{Notably, using an adapter that shields process \emph{b} adding the capability of waiting for the \emph{ack} message we could solve the issue. However, this is clearly an artificial situation and in the general case not satisfying TBC is an indication that the development of an adapter is probably a rather complex and expensive activity.}
 \begin{figure}[t]
 \setlength{\belowcaptionskip}{-20pt}
 	\centering
 \includegraphics[width=0.8\textwidth]{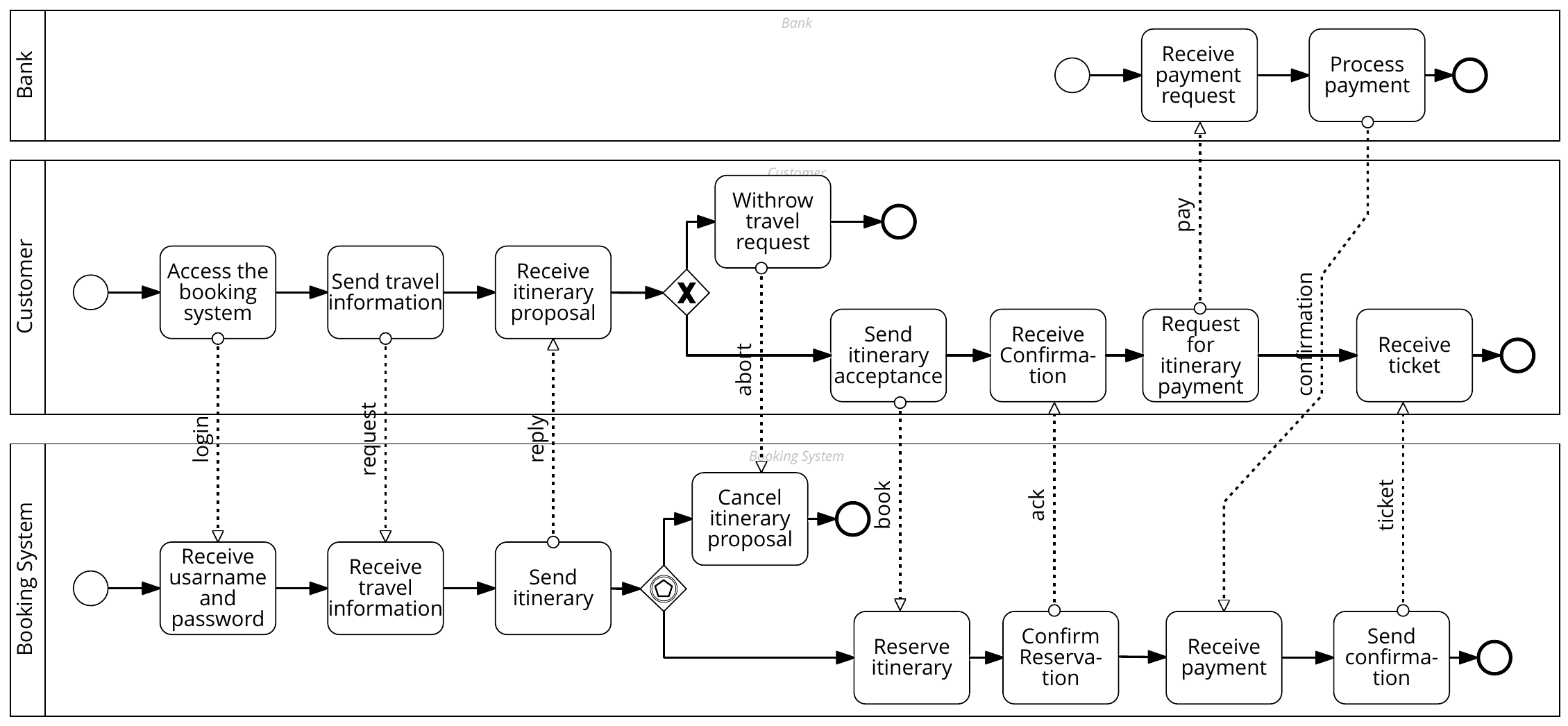}
 	\vspace{-0.2cm}
 	\caption{Booking Collaboration conformant to the Choreography.}%
 	\label{fig:collabotationack}
 	\vspace{-0.2cm}
 \end{figure}

The problem, instead, does not manifest in the composition in case 5 of Table~\ref{tab:combination}. The resulting collaboration is shown in Fig.~\ref{fig:collabotationack}. In this example
  an $ack$ message between the $book$ and $pay$ messages has been included. This guarantees that the booking phase completes before giving to the customer the possibility to proceed with the payment.
In fact, by checking the conformance between this collaboration and the choreography in Fig~\ref{fig:example}.a,
the \name\ tool states that the collaboration is a correct implementation of the choreography, as the two models conform according to both TBC and BBC\@.
Notably, the added message is not foreseen by the choreography specification, nonetheless it permits to further constrain the collaboration so to obtain a behaviour satisfying both conformance relations. In such a case, before the conformance checking, the hiding operator (used in the definition of our conformance relations) will replace the $ack$ message by a $\tau$ action in the composition of the various participants in the collaboration.

The last well-composed collaboration is represented by case 6 of Table~\ref{tab:combination}. This case differs from the previous case 5 just for the gateway in the booking system process. The resulting collaboration is not conformant to the choreography according to the BBC relation. The produced counterexample permits to detect a deadlock related to the presence of two external choices. Notice that the collaboration satisfies instead the TBC relation, since the produced traces are the ones expected in the choreography specification, \ap{and as shown in Section~\ref{sec:conformance} the usage of an adapter can be in this case a viable solution to satisfy the BBC relation.}

\section{Discussion}%
\label{sec:discussion}

\begin{figure}[t]
 \setlength{\belowcaptionskip}{-20pt}
 	\centering
 \includegraphics[scale=0.3]{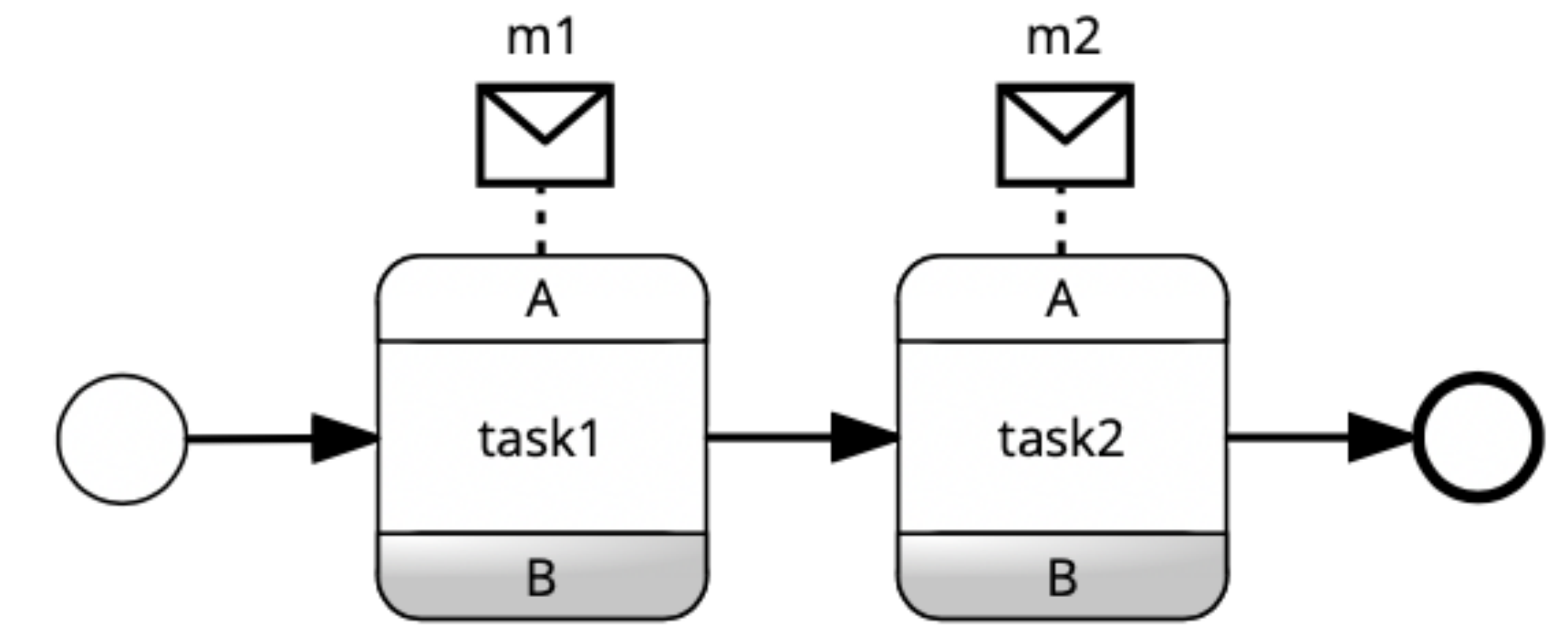}
 	\vspace{-0.2cm}
 	\caption{Simple example of choreography.}%
 	\label{fig:BadSendObservableChoreography}
 	\vspace{0.2cm}
 \end{figure}

\begin{figure}[t]
 \setlength{\belowcaptionskip}{-20pt}
 	\centering
 	\begin{tabular}{c@{\qquad}c}
 	 \includegraphics[scale=0.3]{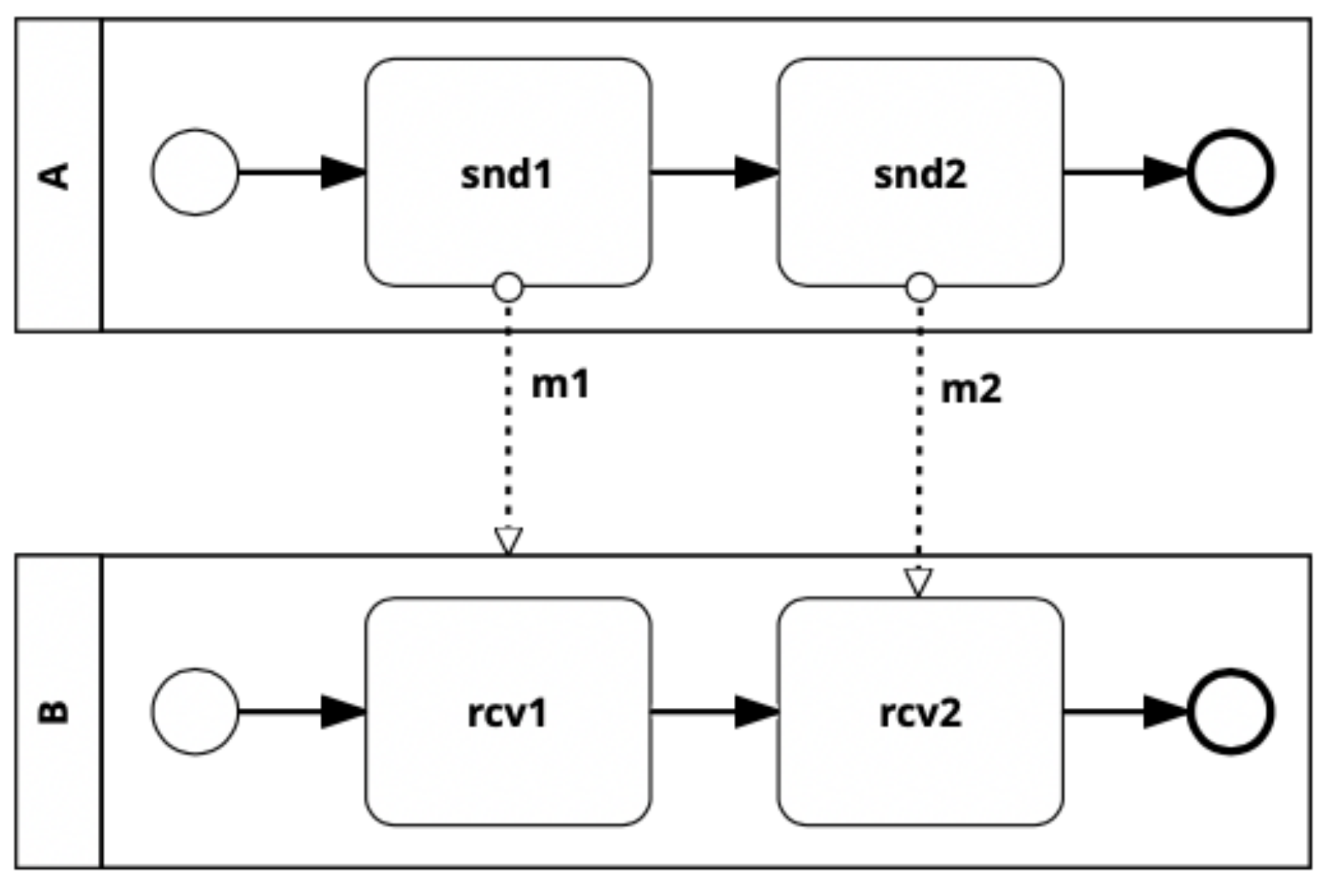}
 	 &
 	 \includegraphics[scale=0.3]{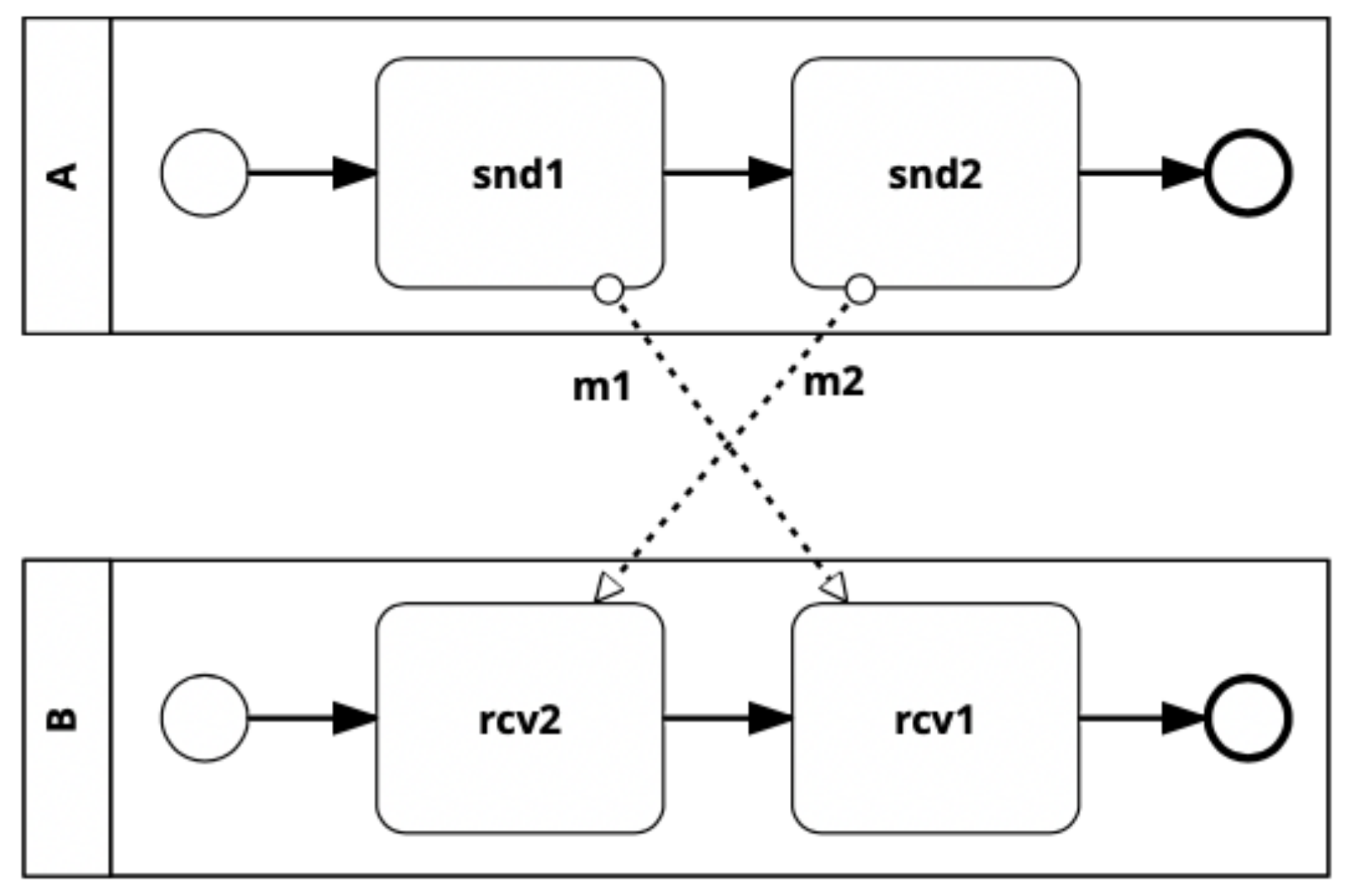}
 	 \\
 	 (a)
 	 &
 	 (b)
 	 \\[.5cm]
 	 \includegraphics[scale=0.3]{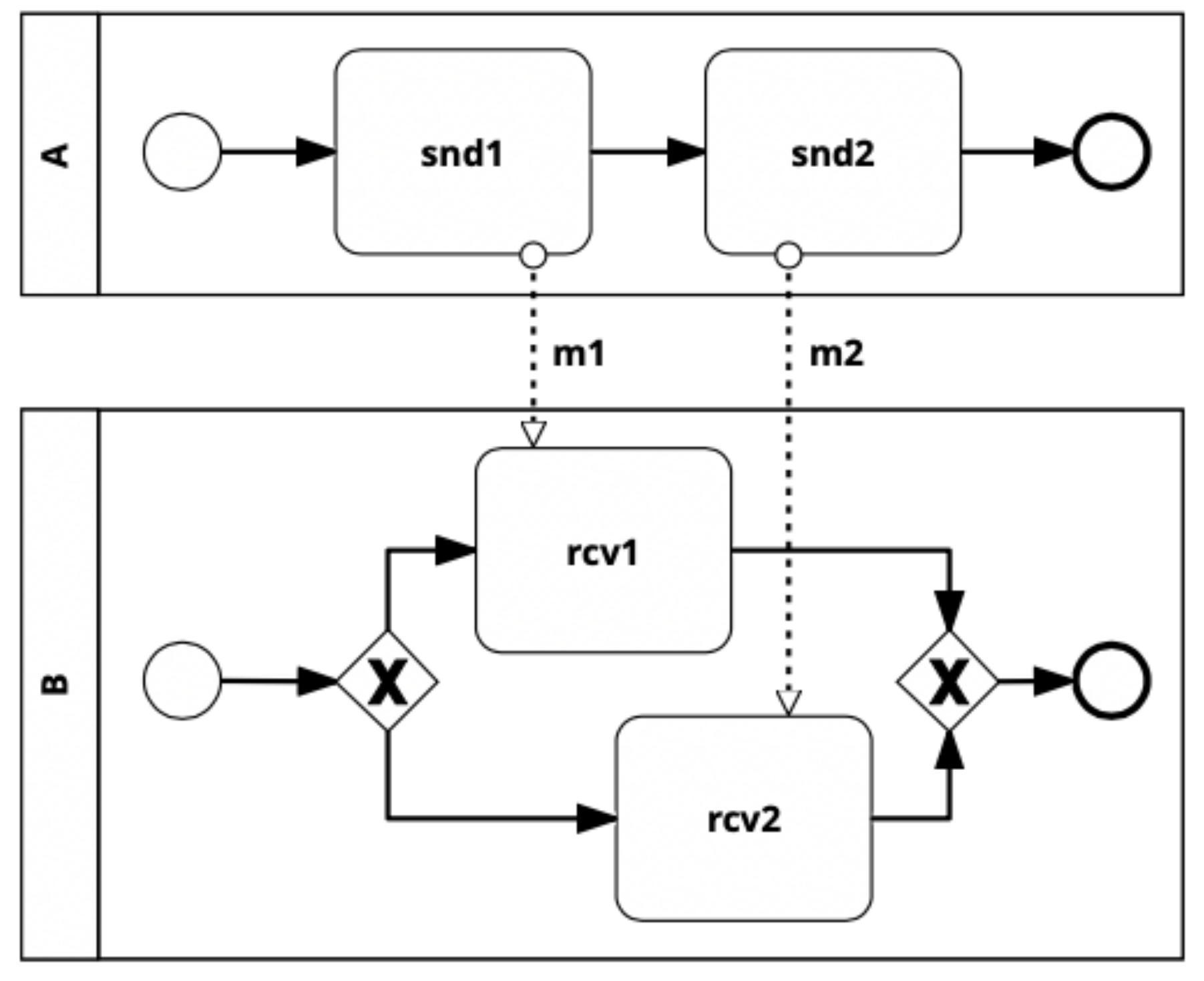}
 	 &
 	 \includegraphics[scale=0.3]{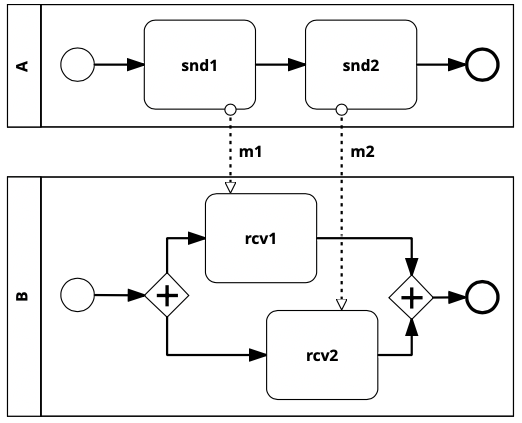}
 	 \\
 	 (c)
 	 &
 	 (d)
 	\end{tabular}
 	\vspace{-0.2cm}
 	\caption{Four examples of collaborations.}%
 	\label{fig:BadSendObservableCollaborations}
 	\vspace{-0.2cm}
 \end{figure}

\ft{In the previous sections we have shown how the different communication models of choreography and collaboration models affect the definition of our conformance checking, and the effects of its application to practical examples.

According to the BPMN standard, collaborations rely on an asynchronous communication model, as it typically happens in distributed systems. In such an asynchronous model of communication, to compare the behaviour of two systems, as observable are usually considered only the output capabilities of the systems, i.e.\ the sending actions. The intuition is that an `asynchronous observer' cannot directly observe the receipt of data that has been sent. This notion of observable action has led to the definition of asynchronous variants of behavioural equivalences (as, e.g., in the labeled bisimulation introduced for the asynchronous $\pi$-calculus~\cite{asyncPic}).

In this work, however, we are interested in comparing the behaviour of a collaboration model with that of a BPMN choreography model (not another BPMN collaboration). As explained at the beginning of this section, we consider the semantics of choreography diagrams as given in the BPMN standard. Therefore, our comparison between choreography and collaboration models is driven by the
specificities of BPMN and in particular, for what concern the communication model, by the fact that choreography tasks require synchronous communication.
Specifically, the standard states that a choreography task completes when the receiver participant reads the message~\cite[p. 315]{omg_business_2011}. Therefore, in our approach the effective completion of the message exchange in the choreography
is compared with the reception of the corresponding message in the collaboration.
This permits ensuring that the order of receives in the collaboration execution respects the order of message receptions prescribed by the choreography model, as requested by the BPMN standard. This would not be possible by only observing sending actions. For example, let us consider the choreography model in Fig.~\ref{fig:BadSendObservableChoreography},
where the participant $\mathsf{A}$ sends in sequence two messages, $\mathsf{m1}$ and $\mathsf{m2}$, to the participant $\mathsf{B}$.
If we compare this model with the collaborations in Fig.~\ref{fig:BadSendObservableCollaborations} considering only
send actions, the conformance checking is always successful, because the participant $\mathsf{A}$ sends
message $\mathsf{m2}$ after $\mathsf{m1}$ in all four cases. However, in case (b) the messages are always received in the opposite
order (hence, the collaboration execution corresponds to first completing the message exchange of $\mathsf{task2}$ before completing that of $\mathsf{task1}$), in case (c) one of the two messages is never received, and in case (d) the reception order may be correct or not depending on the interleaving of the $\mathsf{B}$'s tasks. Therefore, all these three latter cases lead to a conformance result undesirable according to the requirements of the BPMN standard. Instead, our conformance checking approach identifies these three collaborations as not conformant with the choreography.

The focus on the BPMN notation and its specificities, in particular those requiring to observe the reception of messages, is a distinctive aspect of our work. Indeed, related works either do not consider different communication models for choreography and collaboration specifications, or focus on choreography languages different from BPMN 2.0 choreography diagrams. For example,~\cite{basu_choreography_2011} abstracts from a specific choreography specification language, and considers as choreography specifications finite state machines, LTL formulae or CTL formulae, without any reference to the peculiarities of the BPMN choreography semantics.
}

\begin{figure}[t]
 \setlength{\belowcaptionskip}{-20pt}
 	\centering
 	\begin{tabular}{@{\!\!\!}c@{\quad}c@{}}
 	 \includegraphics[scale=0.32]{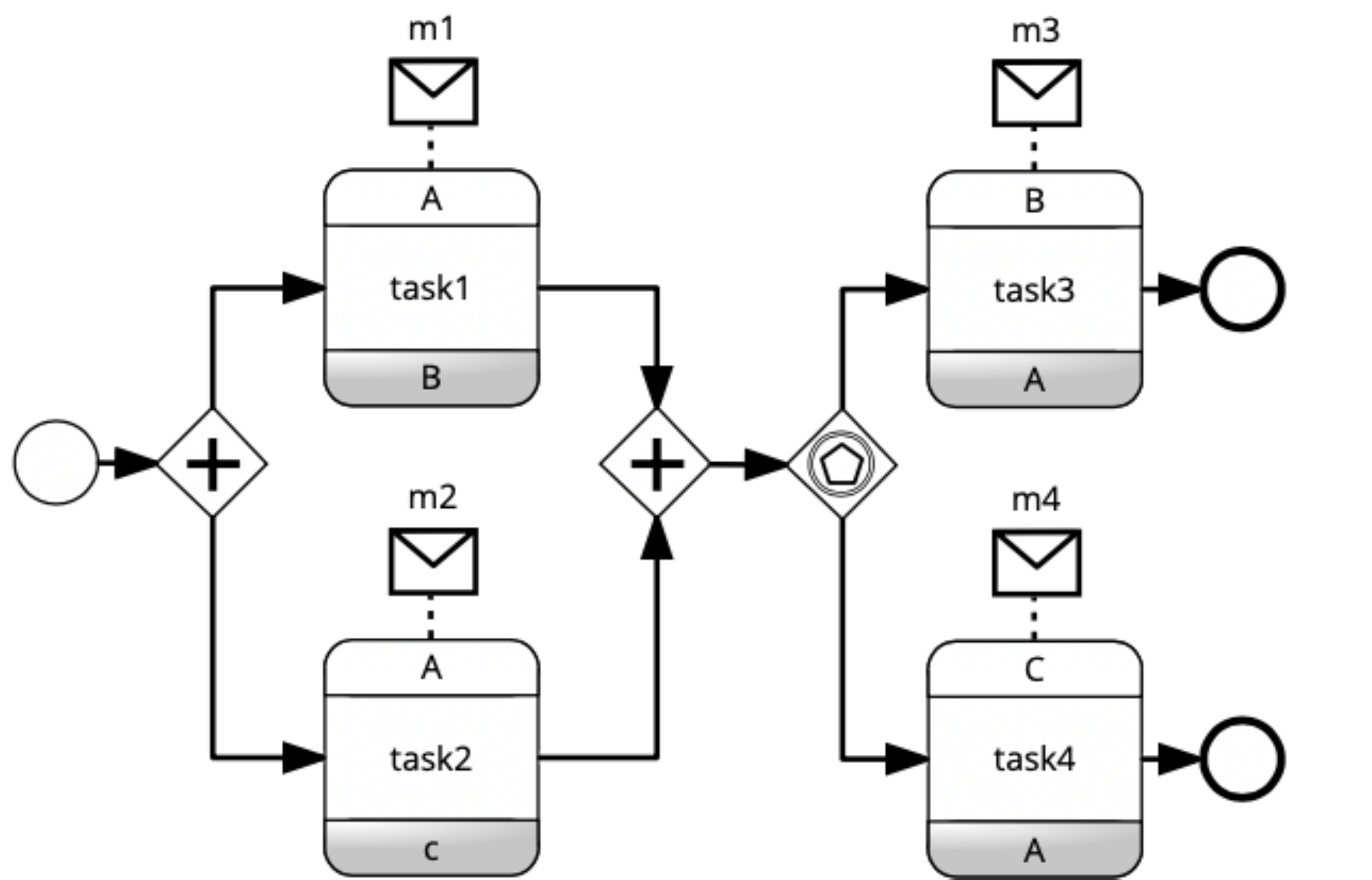}
 	 &
 	 \includegraphics[scale=0.37]{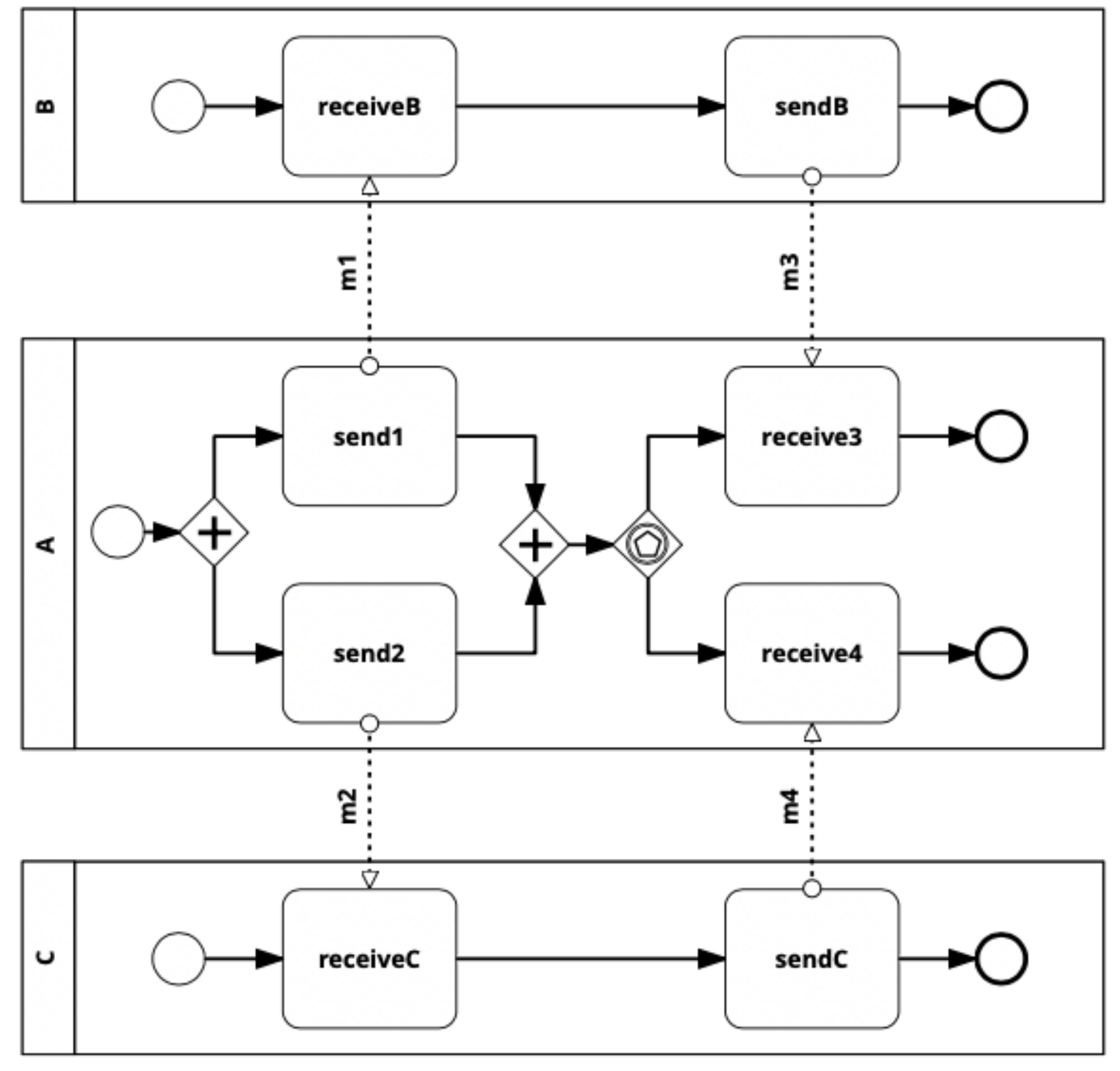}
 	 \\
 	 (a)
 	 &
 	 (b)
 	\end{tabular}
 	\vspace{-0.2cm}
 	\caption{Choreography and collaboration models involving the event-based gateway.}%
 	\label{fig:event-based}
 	\vspace{-0.2cm}
 \end{figure}

\ft{It is worth noticing that we only observe receiving actions in collaboration diagrams, rather than
both sends and receives. The motivation of this design choice is twofold.
On the one hand, we are driven by a pragmatic approach: observing both kinds of actions will lead
to a definition of conformance checking too discriminating for practical use. To explain this point,
let us consider again the choreography model in Fig.~\ref{fig:BadSendObservableChoreography},
and compare it with the collaboration (a) in Fig.~\ref{fig:BadSendObservableCollaborations}.
Observing both sending and receiving actions, the conformance checking would fail, because
we can have a collaboration execution where both $\mathsf{A}$'s sends are executed before
the receptions of the two messages by $\mathsf{B}$, while in this setting the sending of $\mathsf{m2}$
should take place after the reception of $\mathsf{m1}$. However, since there is no correlation
between the sending of $\mathsf{m2}$ and the reception of $\mathsf{m1}$, we believe that
this interpretation of conformance checking turns out to be too restrictive. Indeed, to fix the
issue we would add an extra-choreography message from $\mathsf{B}$ to $\mathsf{A}$ in order to
acknowledge that $\mathsf{m1}$ has been received. Our notion of conformance checking permits
avoiding to introduce such, unnecessary, ack message.
The other motivation concerning the observation of only receiving actions is more technical and, again,
strictly connected to the peculiarities of the BPMN standard. Indeed, differently from most
choreography languages, and in particular the formal ones, BPMN supports a wide variety of workflow operators, some of which
can complicate the formal treatment. This is, for instance, the case of the event-based gateway
(see, e.g.,~\cite{Corradini00T19}). If one would like to observe the sending actions,
many collaborations that are faithful to a given choreography involving an event-based gateway
would be discarded by the conformance checking. Let us consider for example the choreography in
Fig.~\ref{fig:event-based}(a), which represents a typical use of the event-based gateway: % chktex 36
the participant $\mathsf{A}$ sends in parallel a message to $\mathsf{B}$ and another
to $\mathsf{C}$, and then waits for one answer, the first message reception disables the
reception of the other one. The collaboration reported in Fig.~\ref{fig:event-based}(b) represents % chktex 36
a typical, and natural, implementation of this behaviour, where the particpant $\mathsf{A}$
exploits the event-based gateway. However, by observing the send actions, these two models
are not conformant, because at collaboration level there will be always one message between
$\mathsf{m3}$ and $\mathsf{m4}$ that will not be read by $\mathsf{A}$, and this message cannot
be matched by any task execution at choreography level. Instead, our conformance checking
approach properly recognizes these models as conformant.}

\begin{figure}[t]
 \setlength{\belowcaptionskip}{-20pt}
 	\centering
 	\begin{tabular}{@{\!\!}cc}
 	 \includegraphics[scale=0.4]{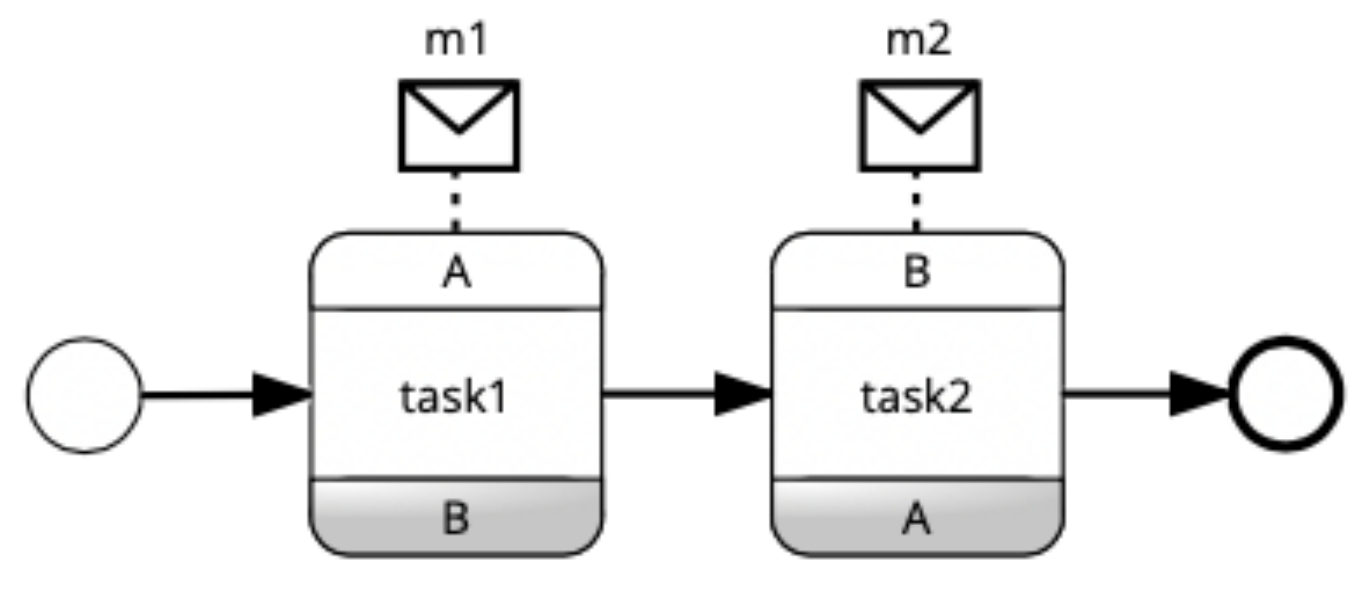}
 	 &
 	 \includegraphics[scale=0.32]{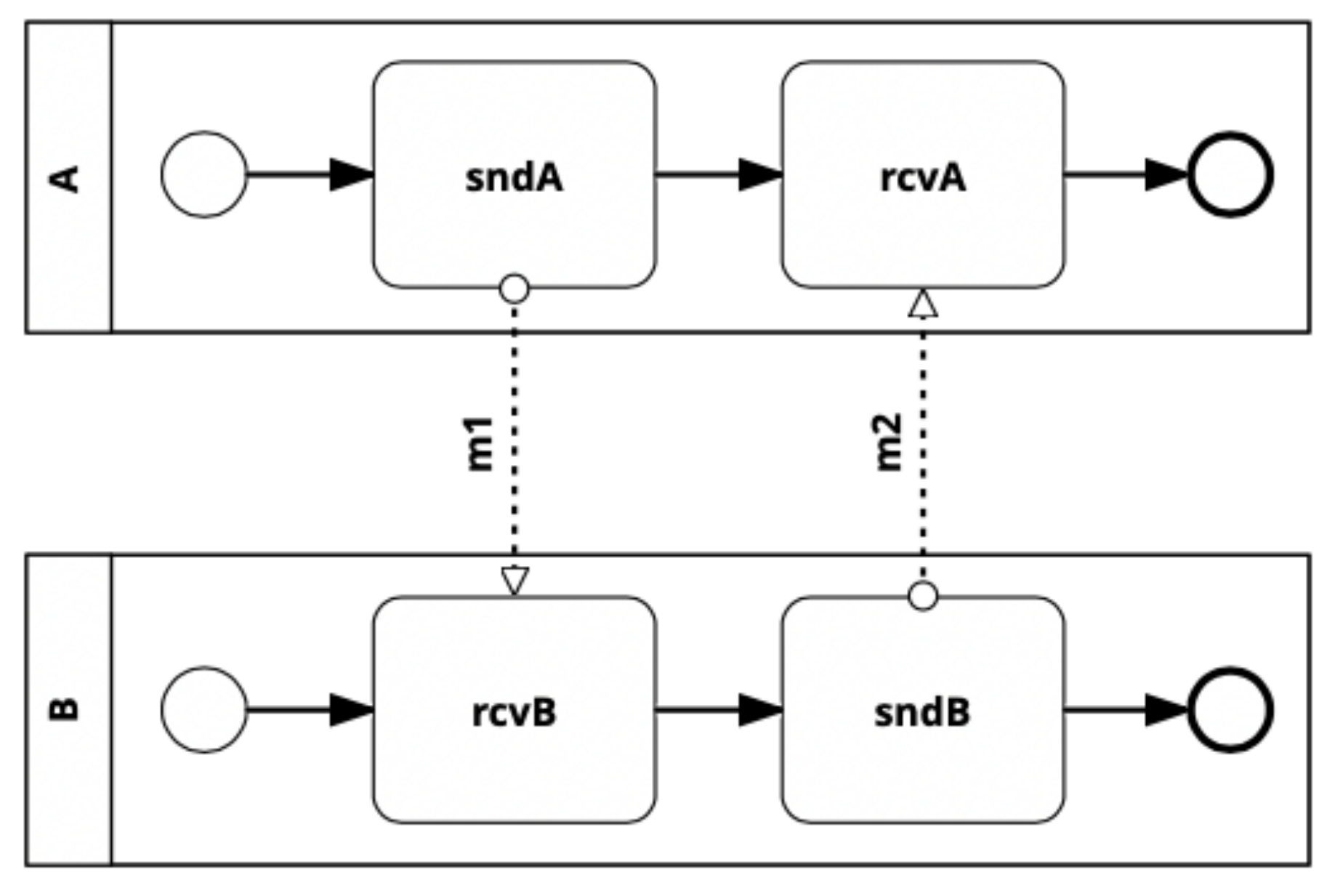}
 	 \\
 	 (a)
 	 &
 	 (b)
 	 \\[.5cm]
 	 \includegraphics[scale=0.31]{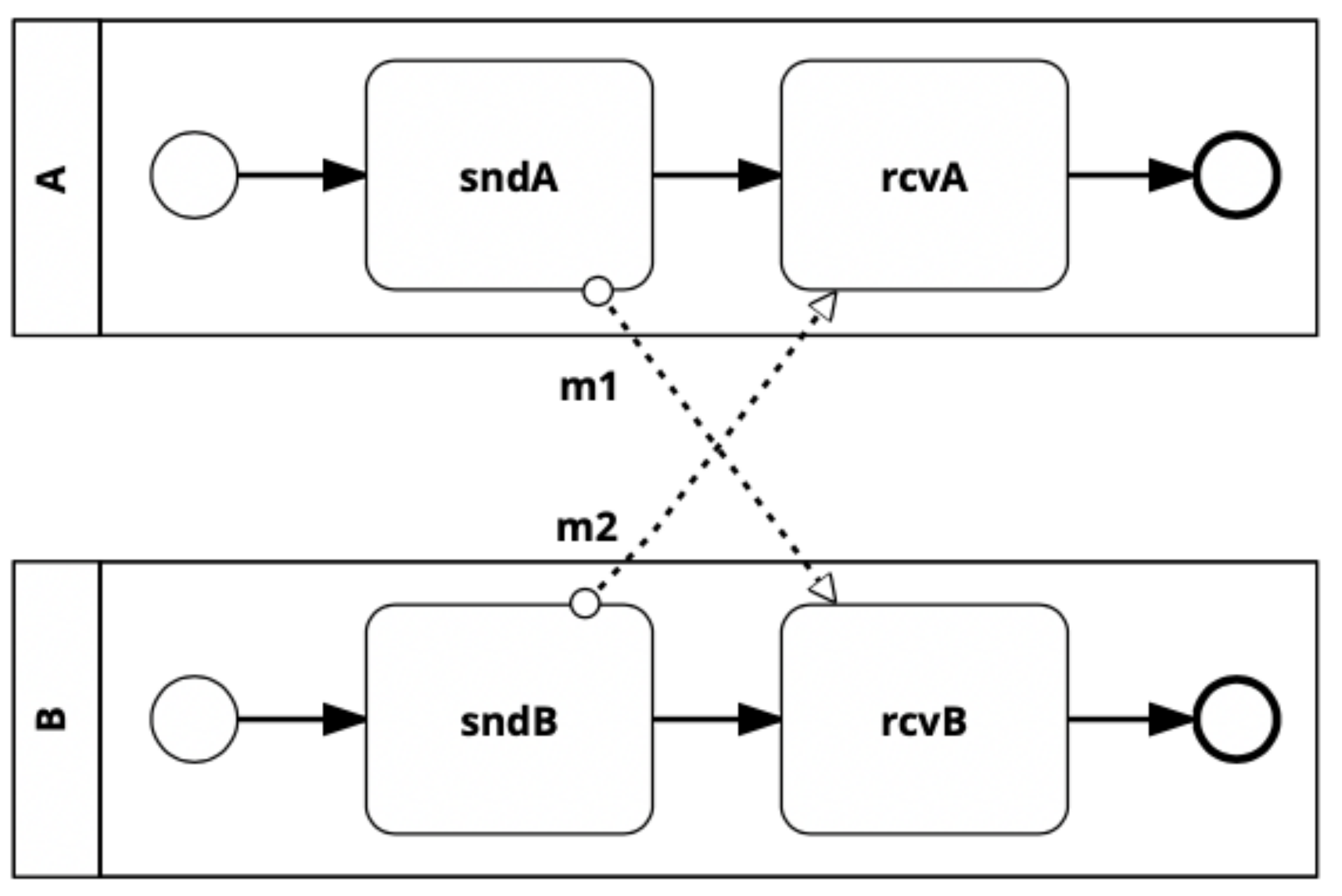}
 	 &
 	 \includegraphics[scale=0.31]{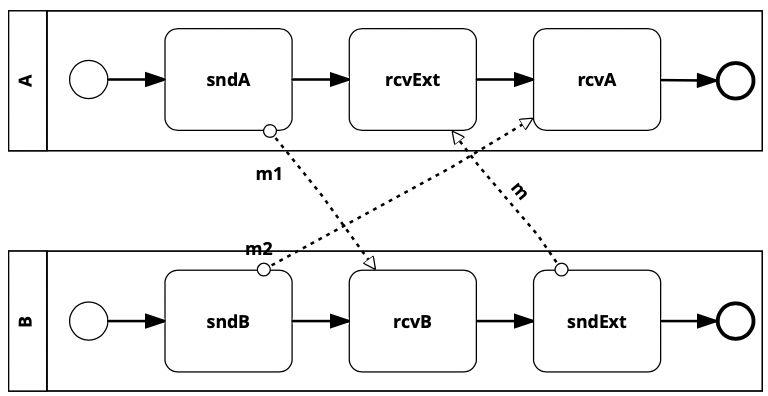}
 	 \\
 	 (c)
 	 &
 	 (d)
 	\end{tabular}
 	\vspace{-0.2cm}
 	\caption{Interplay between receive-only checking and interactions extra choreography.}%
 	\label{fig:req_resp}
 	\vspace{-0.2cm}
 \end{figure}

\ft{On the other hand, the interplay between the observation of only receives and the presence of
additional interactions in the collaboration that are not explicitly expressed
in the choreography may lead to subtle behaviours. We clarify this point with the example
in Fig.~\ref{fig:req_resp}, modelling a common request-response scenario.
The choreography (a) is clearly conformant with the collaboration (b), which is the expected
implementation of this simple behaviour, and not conformant with (c), because the response message
$\mathsf{m2}$ can be read before the request message $\mathsf{m1}$. However, the conformance is also
satisfied for the collaboration (d): in fact, despite $\mathsf{m2}$ can be sent before $\mathsf{m1}$,
the additional exchange of message $\mathsf{m}$ imposes the correct order of execution of the receives.
This behaviour can be considered acceptable when the message sent by $\mathsf{B}$ is not correlated
to that of $\mathsf{A}$, but it may be undesirable when $\mathsf{m2}$ has to be considered as a
response to the request $\mathsf{m1}$. Thus, this is an application-dependent aspect that, to be
possibly investigated, requires to consider the correlation between the content of the exchanged
messages (e.g., in the example we should check that part of the payload of $\mathsf{m2}$ results
from the evaluation of the information included in $\mathsf{m1}$).}

 \section{Related Works}%
\label{sec:related}
This section discusses the advantages and differences in our approach with respect to alternative proposals. The discussion is organized over different paragraphs; each one devoted to a specific aspect considered relevant for \name{}.

\smallskip
\noindent
\emph{\textbf{On the Choice of the Modelling Notation}}.
Researchers have worked in the definition and the study of modelling notations for the representation of collaborative systems for many years now.
In particular, the topic has received much attention in the field of service-oriented applications, where many modelling languages have been proposed.
Among the first proposals, we can certainly mention the OASIS standard WS-BPEL~\cite{WSBPEL}, for
the specification of collaborations referred as ``abstract orchestrations'',
and the W3C standard WS-CDL~\cite{WSCDL}, for the representation of choreographies.
These specifications have inspired the OMG standard \bpmn~\cite{omg_business_2011} that has inherited many concepts from theme.
In the literature, many proposals are available for conformance relations among models defined in different notations~\cite{papazoglou2007service,baldoni2005verifying,liu2007static,el2008business,rouached2012web,martens2003compatibility}.
However, the lack of a solid framework for the conformance verification related to the BPMN standard has motivated our effort.

\smallskip
\noindent
\emph{\textbf{From Choreographies to Code}}.
In the development of collaborative systems, much effort has been devoted to the study of model transformation approaches in particular in the line of model-driven engineering strategies. In particular, choreographies specifications have been fruitfully used in the generation of component stubs/skeletons embedding constraints of the message flow.
In~\cite{hofreiter2008model}, the authors propose a top-down approach, where, starting from a choreography they derive a UML profile respecting the global specification with the final aims to derive WS-BPEL processes.
In~\cite{nikaj2017semi} the authors provide a semi-automatic RESTful implementation of \bpmn{} choreographies interactions.
Nevertheless the above approaches do not provide any formal guarantee to ensure that the resulting system conforms to the interactions prescribed by the specification.
\ft{Another line of research works, focussing on session types and behavioural contracts~\cite{SessionTypes}, provides formal guarantees that permit
to achieve a correct-by-construction top-down approach. However, the formalisms used in these works
for the global specification of choreographies are much less rich than the BPMN notation (e.g., they do not support
arbitrary topology, such gateways as the event-based and parallel ones, different communication
models for collaborations and choreography, etc.), hence their practical application is far from
that enabled by the BPMN standard.}
\am{Furthermore, a top-down approach in a real context is not always the best choice, since organisations have already their processes regulating their activities.
For this reason, we propose a hybrid methodology that incentives the reuse and integration of already existing processes, providing a  complete framework able to deal directly with BPMN models and with the capability of providing formal evidence of their conformance.}

\smallskip
\noindent
\emph{\textbf{Conformance}}.
In the literature, conformance is referred with different terminologies depending on the context in which it is used. Possible synonymous are compliance or compatibility, but generally, they are never explicitly defined on the \bpmn{} standard.
Different works~\cite{governatori2006compliance,mandell2003bottom,schumm2010integrating,awad2008efficient,knuplesch2013enabling,ly2012enabling,knuplesch2012ensuring,knuplesch2012towards}  define notion of compliance between processes or collaborations exploiting domain-specific regulations and rules.
In general, these works express behavioural constraints using some form of temporal logics, rather than
using equivalences among models.
The usage of additional languages requires the system designer to study further technicalities in order to exploit the proposed checking technique.
Alternative approaches base the analysis only considering local views~\cite{weidlich2012behaviour,hidders2005two,dijkman2009graph}. Somehow this is a simplification of the problem that does not guarantee the conformance for the composition.

Our work differs from the ones mentioned above since it relies on
\bpmn\ models, at both global and local level, and propose a conformance check between different layers of specification abstraction without requiring any formal knowledge to the system designer.
\ft{Our conformance checking is based on LTSs, since our methodology aims at reusing standard bisimulation and trace equivalences supported by existing equivalence checkers. Thus, the novelty of our proposal is not in the equivalence checking between LTSs, per se, but it is in how these LTSs are produced starting from standard BPMN models. The passage from BPMN models and LTSs is indeed defined by our operational semantics, which takes into account the peculiarities of BPMN that instead are overlooked by other works (see Section~6). Furthermore, all the specific characteristics of our conformance relations (e.g., the use of hiding) have been already taken into account during the LTS generation. This relieves us of the duty of developing an ad-hoc conformance checker.}

\smallskip
\noindent
\emph{\textbf{Direct Semantics.}}
Many attempts trying to give a semantics to the \bpmn{} notation are present in the literature. The majority of these approaches provide semantics in terms of translation to other formalization languages such as Petri-nets or similar formalisms.
Providing semantics by translation can generate divergences with the expected behaviour of the elements.
Between the target specification languages, we found process algebras~\cite{busi2006choreography,salaun2009realizability,salaun2006describing},
formal languages~\cite{nguyen2012symbolic,poizat_checking_2012,molina2013establishing,gudemann2016verchor},
transition-based models (e.g., Petri-Nets)~\cite{dijkman_semantics_2008,basu_choreography_2011,bultan_tool_2009,corradini_inter-organisational_2015},
or session types~\cite{castagna2012global}.
\am{Although all these formalisms have been used to formalise the BPMN semantics, there is a general lack of support related to the communication model. The main issues are related to the difficulty to distinguish messages from control flow in the formalisation, and to deal with both synchronous and asynchronous communication at the same time. Therefore, studying these aspects directly on BPMN does not leave any room for ambiguity, and increases the potential for formal reasoning on BPMN\@. }
In \name\ we rely on a direct semantics for both collaboration and
choreography models.  The \name\ semantics is given in terms of features and constructs of \bpmn, rather than in terms of their low-level encoding
into another formalism that is equipped with its own syntax and
semantics. This permits to avoid any possible bias that would result from the encoding in another formalism.
The direct semantics proposed in this paper is inspired by~\cite{corradini2015operational}, and by its extended version in~\cite{CorradiniSCP}. Differences mainly refer to configuration states that are here
defined according to a global perspective. Moreover, the formalisation now
includes choreography diagrams, which were overlooked in the previous
semantics definition.

 \smallskip
 \noindent
 \emph{\textbf{Conformance vs Communication models.}}
The definition of semantics using other languages can bring to underspecified situations that can alter the conformance checking results. For this reason, we use a direct approach that permits to focus on specific features of \bpmn\ that would be ignored by using available Petri Nets-based semantics.
Evidence of this divergence can be noticed in the definition of gateways. Following the \bpmn\ mapping to
Petri Nets proposed in~\cite{dijkman_semantics_2008}, it
is not possible to distinguish different types of non-determinism
resulting from event-based or exclusive gateways~\cite{Corradini00T19}.
Indeed these two \bpmn\ elements have different effects: the
event-based gateway produces non-dominated non-determinism (roughly,
no one in the model has complete knowledge on the decision that will
be taken), while the exclusive gateway produces dominated
non-determinism (roughly, the decision is taken by one party and
followed by the
others).
 Differently from the translation-based approaches, our approach permits to distinguish the dominated and non-dominated
non-determinism produced by the gateways, as prescribed by the \bpmn\
standard. This is somehow similar to~\cite{busi2006choreography,qiu2007towards}, which rely on the concept
of internal and external choices defined in the CSP process algebra.
Notably, the different kinds of non-determinism have an impact on the conformance relations, as detailed in Table~\ref{tab:combination}.

Another fundamental aspect for the definition of the conformance is the asynchronous communication of collaborations versus the synchronous one of the choreographies.
In the literature, to deal with asynchronous communication, either additional constructs are used, such as buffers~\cite{poizat_checking_2012} or dedicate language structures~\cite{salaun2009realizability,gudemann2016verchor}, or simply reducing the
 asynchronous communication model to the synchronous one.
 The notions of conformances that we propose in \name{} allows the user to compare models assuming different communication strategies without making any assumption on the different management of the messages. Our conformance notion directly manages this aspect,
 \ft{as widely discussed in Section~\ref{sec:discussion}.}

\smallskip
\noindent
\emph{\textbf{Tool Support.}}
Verchor~\cite{gudemann2016verchor} is a tool similar to \name{}. In this case, the main objective is to use the conformance notion to check the realizability of a set of peers obtained from a projection of a given choreography.
Another tool, more focused on business properties, is VBPMN~\cite{krishna2017vbpmn}. Here the verification makes use of the well-known model checker CADP\@.
While VBPMN can deal only with the analysis of single processes,
\name{} is able to manage the conformance checking of collaborations w.r.t.\ choreographies.
\am{In \name{} the presence of an integrated system for the storing, design and verification of BPMN models
allows non-domain experts to check the conformance of their models without any previous notion of
the formal definitions.}

\section{Conclusions and Future Work}%
\label{conclusions}
This paper considers the theoretical and practical relevance of
checking the conformance of models related to the global specification
of application-level protocols (choreographies) and their possible
implementation through the composition of processes
(collaborations).
The specific context is that of the
\bpmn{} standard, which nowadays is the most used notation for
specifying inter-organisational processes. To
perform such conformance checking, the paper proposes a direct
semantics in the structural operational style, and defines two different equivalence
relations between choreographies and process collaborations. The
resulting formal framework has also been practically implemented in
the \name{} toolchain, which permits to support all
phases needed to derive inter-organisational process-based
systems. The tool is available as a web-based service, as well as a
standalone application.

In the next future, we intend to extend the framework further,
so to possibly check temporal properties on choreographies
and collaborations.
\new{Another extension of the framework we intend to investigate
concerns the treatment of additional BPMN elements, in order to
cover a more comprehensive set of elements.}
\new{Similarly, we also plan to investigate the
impact of using in our approach other notions of equivalences that are weaker than the weak bisimulation and stronger than the weak trace equivalence (e.g.\ weak failure equivalence).}
Moreover, concerning the repository storing the published choreographies and
the available processes, we foresee for this repository further developments
and a better integration with \name{}.
\new{Such integration would make it also possible to run, in a structured manner, an empirical evaluation of the proposed methodology so to complement the validity of the approach, which up to now is mainly based on comparison with the literature.
}
A further interesting line of
research that we intend to follow refers to the usage of
inter-organisational models to facilitate the integration of existing
services, so to make easier the development of the considered
collaborative systems.
\new{Moreover, we intend to investigate on the relation that may exist between the used notions of conformance and the ability to
solve issues detected in a collaboration. Indeed, as shown in the paragraph \textit{Conformance at work} in Section~\ref{atWork}, when BBC is not satisfied,
the satisfaction of TBC can leave some possibilities to solve the issue in the  collaboration in order to achieve the BBC conformance. However, the TBC could not provide enough guarantees to generate an adapter or define a refinement to solve the issue, and on the other hand there could be situations in which, even if the TBC is not satisfied, it is easy to solve the issue. Therefore, we believe that a deeper formal investigation is necessary to precisely define the requirements ensuring the existence of a solution for a given collaboration issue.}
\new{Finally, we aim at continuously improving the \name{ } implementation, in particular in relation to the introduction of (semi-)automatic mechanisms for the generation of adapters starting from the counterexamples returned by the conformance checking.} % chktex 36

\newcommand{\etalchar}[1]{$^{#1}$}

%%%%%%%%%%%%%%%%%%%%%%%%%%%%%%%%%%%%%%%%%%%%%%%%%

\newpage
\section*{Appendix}%
 \label{sec:appendix}

We report here the proof of the proposition concerning the well-composedness
of the collaboration produced by the function $\mathcal{C}$.

\restateprop*
\begin{proof}
The proof proceeds by contradiction. Suppose that there exist
$\tuple{\proc}$ and $\tuple{\orgname}$ such that
$\compFunction(\tuple{\proc},\tuple{\orgname})=\collabs$
and $\collabs$ is not well-composed.
By Def.~\ref{def1:composition}, we have six cases
(below notation $\in^n$ means that the number of times an element
occurs in the multiset is $n$):
\begin{enumerate}
\item
$
\exists \msgEdge{\orgname}{\orgname}{\messagename} \in \outF{C}.
$
This means that $C$ contains a term
$\tasksendG{\edgeG{\edgename_i}}{\edgeG{\edgename_o}}{\msgEdge{\orgname}{\orgname}{\messagename}}$
or a term
$\evInterSndG{\edgeG{\edgename_i}}{\edgeG{\edgename_o}}{\msgEdge{\orgname}{\orgname}{\messagename}}$. Le us consider the former case, the other is similar.
According to the definitions of $\compFunction$ and $\mathcal{N}$, the sending task results from
the evaluation of
$\tasksendG{\edgeG{\edgename_i}}{\edgeG{\edgename_o}}{
(\sndFunction(\messagename),\rcvFunction(\messagename),\messagename)}
$ for some
$\sndFunction$ and $\rcvFunction$ such that
$\sndFunction(\messagename)=\rcvFunction(\messagename)=\orgname$.
However, by definition of $\mathcal{N}$,
$\sndFunction(\messagename)\neq\rcvFunction(\messagename)$, which
is a contradiction.

\item
$
\exists \msgEdge{\orgname_1}{\orgname_2}{\messagename} \in \outF{C} \ . \ \not\!\!\exists \msgEdge{\orgname_1}{\orgname_2}{\messagename} \in \inF{C}
$
This means that $C$ contains a term
$\tasksendG{\edgeG{\edgename_i}}{\edgeG{\edgename_o}}{\msgEdge{\orgname_1}{\orgname_2}{\messagename}}$
or a term
$\evInterSndG{\edgeG{\edgename_i}}{\edgeG{\edgename_o}}{\msgEdge{\orgname_1}{\orgname_2}{\messagename}}$.
Le us consider the former case, the other is similar.
According to the definitions of $\compFunction$ and $\mathcal{N}$, the sending task results from
the evaluation of
$\tasksendG{\edgeG{\edgename_i}}{\edgeG{\edgename_o}}{
(\sndFunction(\messagename),\rcvFunction(\messagename),\messagename)}
$ for some
$\sndFunction$ and $\rcvFunction$ such that
$\sndFunction(\messagename)=\orgname_1$ and $\rcvFunction(\messagename)=\orgname_2$.
The latter means that $\rcvFunction$ has the form $\{\mpair{\messagename}{\orgname_2}\}\sqcup\rcvFunction'$ for some $\rcvFunction'$; according to the definition of $\mathcal{R}$,
the pair $\{\mpair{\messagename}{\orgname_2}\}$ is produced by either
$\rcvFunctionC{\taskreceiveG{\edgeG{\edgename_i'}}{\edgeG{\edgename_o'}}{\messagename}}{\orgname_2}$, $\rcvFunctionC{\evInterRcvG{\edgeG{\edgename_i'}}{\edgeG{\edgename_o'}}{{\messagename}}}{\orgname_2}$ or
$\rcvFunctionC{\eventbasedG{\edgeG{\edgename_i''}}{(\msgEdgeEventP{\orgname_1}{\orgname_2}{\messagename}{\edgename_o''},\msgList)} }{\orgname_2}$,
with ${\sf taskRcv}$, ${\sf interRcv}$ or ${\sf eventBased}$ term in $C$.
Let us consider the case where $C$ contains the term $\taskreceiveG{\edgeG{\edgename_i'}}{\edgeG{\edgename_o'}}{\messagename}$; the other two
cases are similar.
By definition of $\mathcal{N}$, and using the functions $\sndFunction$ and $\rcvFunction$ above,
we have
$
\namingFunction{\taskreceiveG{\edgeG{\edgename_i'}}{\edgeG{\edgename_o'}}{\messagename},\sndFunction,\rcvFunction}=
\taskreceiveG{\edgeG{\edgename_i'}}{\edgeG{\edgename_o'}}{
(\sndFunction(\messagename),\rcvFunction(\messagename),\messagename)}=
\taskreceiveG{\edgeG{\edgename_i'}}{\edgeG{\edgename_o'}}{
(\orgname_1,\orgname_2,\messagename)}
$.
By applying function $in$ to this term, we obtain
$\inF{\taskreceiveG{\edgeG{\edgename_i'}}{\edgeG{\edgename_o'}}{
(\orgname_1,\orgname_2,\messagename)}}=\{\msgEdge{\orgname_1}{\orgname_2}{\messagename}\}$,
hence $\msgEdge{\orgname_1}{\orgname_2}{\messagename} \in \inF{C}$,
which is a contradiction.

\item
$
\exists \msgEdge{\orgname_1}{\orgname_2}{\messagename} \in \outF{C} \ . \
\exists \msgEdge{\orgname_1}{\orgname_2}{\messagename} \in^2 \inF{C}
$.
According to the definition of function $in$, this means that two terms of the form
$\taskreceiveG{\edgeG{\edgename_i}}{\edgeG{\edgename_o}}{\msgEdge{\orgname_1}{\orgname_2}{\messagename}}$,
$\evInterRcvG{\edgeG{\edgename_i}}{\edgeG{\edgename_o}}{\msgEdge{\orgname_1}{\orgname_2}{\messagename}}$ or
$\eventbasedG{\edgeG{\edgename_i}}{(\msgEdgeEvent{\orgname_1}{\orgname_2}{\messagename}{\edgename_o},\msgList)}$
are in $C$. Thus, according to the definition of $\mathcal{N}$,
two terms of the form
$\taskreceiveG{\edgeG{\edgename_i}}{\edgeG{\edgename_o}}{\messagename}$,
$\evInterRcvG{\edgeG{\edgename_i}}{\edgeG{\edgename_o}}{\messagename}$ or
$\eventbasedG{\edgeG{\edgename_i}}{(\msgEdgeEventP{\orgname_1}{\orgname_2}{\messagename}{\edgename_o},\msgList)}$
are in $\tuple{\proc}$,
and there exists $\sndFunction$ and $\rcvFunction$ such that
$\sndFunction(\messagename)=\orgname_1$ and $\rcvFunction(\messagename)=\orgname_2$.
Function $\rcvFunction$ must result from the application of function $\mathcal{R}$
to $\tuple{\proc}$. By applying $\mathcal{R}$ to $\tuple{\proc}$, since $\tuple{\proc}$
contains two terms of the form
$\taskreceiveG{\edgeG{\edgename_i}}{\edgeG{\edgename_o}}{\messagename}$,
$\evInterRcvG{\edgeG{\edgename_i}}{\edgeG{\edgename_o}}{\messagename}$ or
$\eventbasedG{\edgeG{\edgename_i}}{(\msgEdgeEventP{\orgname_1}{\orgname_2}{\messagename}{\edgename_o},\msgList)}$, we obtain
$\{\mpair{\messagename}{\orgname_2}\}\sqcup\{\mpair{\messagename}{\orgname_2}\}\sqcup\rcvFunction'$ for some $\rcvFunction'$. However, operator $\sqcup$
is defined only when they arguments have disjoint domain. Hence,
$\{\mpair{\messagename}{\orgname_2}\}\sqcup\{\mpair{\messagename}{\orgname_2}\}$
is undefined. As consequence, $\compFunction(\tuple{\proc},\tuple{\orgname})$ is undefined,
which is a contradiction.

\item
$
\exists \msgEdge{\orgname}{\orgname}{\messagename} \in \inF{C}
$.
Similar to case (1).

\item
$
\exists \msgEdge{\orgname_1}{\orgname_2}{\messagename} \in \inF{C} \ . \
\not\exists \msgEdge{\orgname_1}{\orgname_2}{\messagename} \in \outF{C}
$.
Similar to case (2).

\item
$
\exists \msgEdge{\orgname_1}{\orgname_2}{\messagename} \in \inF{C} \ . \
\exists \msgEdge{\orgname_1}{\orgname_2}{\messagename} \in^2 \outF{C}
$.
Similar to case (3).
\qedhere
\end{enumerate}
\end{proof}

\end{document}

%% file: macroChoreography.tex
\newcommand{\bpmn}{BPMN}

\newcommand{\name}{$C^{4}$}

\usepackage{numprint}

\newcommand{\collabs}{C}

\newcommand{\new}[1]{{\color{black} #1}}
\newcommand{\ap}[1]{{#1}}
\newcommand{\ft}[1]{{#1}}
\newcommand{\am}[1]{{#1}}
\newcommand{\qsep}{.}

\newcommand{\tuple}[1]{\overline{#1}}
\newcommand{\tuplel}[1]{\mid{#1}\mid}
\newcommand{\proj}[1]{\downarrow_{#1}}
\newcommand{\sndFunction}{S}
\newcommand{\rcvFunction}{R}
\newcommand{\compFunction}{\mathcal{C}}
\newcommand{\namingFunction}[1]{\ensuremath{\mathcal{N}(#1)}}
\newcommand{\sndFunctionC}[2]{\ensuremath{\mathcal{S}(#1,#2)}}
\newcommand{\rcvFunctionC}[2]{\ensuremath{\mathcal{R}(#1,#2)}}
\newcommand{\mpair}[2]{\ensuremath{#1 \mapsto #2}}

\newcommand{\evStartG}[1]{\ensuremath{{\sf start}(#1)}}
\newcommand{\evEndG}[1]{\ensuremath{{\sf end}(#1)}}

\newcommand{\evInterRcvG}[3]{\ensuremath{{\sf interRcv}(#1,#2, #3)}}
\newcommand{\evInterSndG}[3]{\ensuremath{{\sf interSnd}(#1,#2, #3)}}
\newcommand{\andSplitG}[2]{\ensuremath{{\sf andSplit}(#1,#2)}}
\newcommand{\andJoinG}[2]{\ensuremath{{\sf andJoin}(#1,#2)}}
\newcommand{\xorSplitG}[2]{\ensuremath{{\sf xorSplit}(#1,#2)}}
\newcommand{\xorJoinG}[2]{\ensuremath{{\sf xorJoin}(#1,#2)}}

\newcommand{\taskG}[2]{\ensuremath{{\sf task}(#1,#2)}}
\newcommand{\tasksendG}[3]{\ensuremath{{\sf taskSnd}(#1,#2,#3)}}
\newcommand{\taskreceiveG}[3]{\ensuremath{{\sf taskRcv}(#1,#2,#3)}}
\newcommand{\eventbasedG}[2]{\ensuremath{{\sf eventBased}(#1,#2)}}

\newcommand{\gtransitionCho}{\transitionCho{\tau}{}}
\newcommand{\transitionCho}[2]{\xrightarrow[#2]{#1}}
\newcommand{\transitionColl}[2]{\xrightarrow[#2]{#1}}

\newcommand{\transition}[2]{\xrightarrow[#2]{#1}}
\newcommand{\Transition}[2]{\xRightarrow[#2]{#1}}

\newcommand{\edgeG}[1]{#1}
\newcommand{\incToken}[2]{\ensuremath{inc(#1,#2)}}
\newcommand{\decToken}[2]{\ensuremath{dec(#1,#2)}}
\newcommand{\incTokenName}{inc}
\newcommand{\decTokenName}{dec}
\newcommand{\stateedge}{\sigma}
\newcommand{\statemsg}{\delta}

\newcommand{\msgEdge}[3]{(#1,#2,#3)}
\newcommand{\msgEdgeEvent}[4]{(#1,#2,#3,#4)}
\newcommand{\msgEdgeEventP}[4]{(#3,#4)}
\newcommand{\initstateedge}{\sigma_0}

\newcommand{\stateupd}[3]{\ensuremath{#1\cdot\{#2\mapsto #3\}}}

\newcommand{\statesSet}{\mathbb{S}}
\newcommand{\statefunc}[1]{\sigma(#1)}

\newcommand{\edgeSetE}{\mathbb{E}}

\newcommand{\msgSetMSG}{\mathbb{M}}

\newcommand{\chos}{Ch}
\newcommand{\choevStartG}[1]{\ensuremath{{\sf start}(#1)}}
\newcommand{\choevEndG}[1]{\ensuremath{{\sf end}(#1)}}
\newcommand{\choandSplitG}[2]{\ensuremath{{\sf andSplit}(#1,#2)}}
\newcommand{\choandJoinG}[2]{\ensuremath{{\sf andJoin}(#1,#2)}}
\newcommand{\choxorSplitG}[2]{\ensuremath{{\sf xorSplit}(#1,#2)}}
\newcommand{\choxorJoinG}[2]{\ensuremath{{\sf xorJoin}(#1,#2)}}

\newcommand{\chotaskOWG}[6]{\ensuremath{{\sf task}(#1,#2,#3,#4,#5)}}

\newcommand{\choeventbasedG}[3]{\ensuremath{{\sf eventBased}(#1,#2,#3)}}

\newcommand{\chotasklistOWG}[5]{\ensuremath{{\sf }(#1,#2,#3,#4)}}

\newcommand{\coSet}{\mathbb{C}}
\newcommand{\choSet}{\mathbb{C}h}
\newcommand{\proc}{P}
\newcommand{\edgename}{{\sf e}}
\newcommand{\edgeList}{E}
\newcommand{\messagename}{{\sf m}}
\newcommand{\msgList}{M}
\newcommand{\msgListE}{\mathit{ME}}
\newcommand{\taskList}{T}

\newcommand{\taskname}{{\sf t}}
\newcommand{\token}{{\sf n}}
\newcommand{\collabspar}{\mid}
\newcommand{\chopar}{\mid}
\newcommand{\orgname}{{\sf p}}

\newcommand{\outF}[1]{\ensuremath{{out}(#1)}}
\newcommand{\inF}[1]{\ensuremath{{in}(#1)}}

%% file: macroProoftree.tex
\newdimen\proofrulebreadth \proofrulebreadth=.04em
\newdimen\proofdotseparation \proofdotseparation=1.25ex
\newdimen\proofrulebaseline \proofrulebaseline=2ex
\newcount\proofdotnumber \proofdotnumber=3
\let\then\relax
\def\hfi{\hskip0pt plus.0001fil}
\mathchardef\squigto="3A3B
\newif\ifinsideprooftree\insideprooftreefalse
\newif\ifonleftofproofrule\onleftofproofrulefalse
\newif\ifproofdots\proofdotsfalse
\newif\ifdoubleproof\doubleprooffalse
\let\wereinproofbit\relax
\newdimen\shortenproofleft
\newdimen\shortenproofright
\newdimen\proofbelowshift
\newbox\proofabove
\newbox\proofbelow
\newbox\proofrulename
\def\shiftproofbelow{\let\next\relax\afterassignment\setshiftproofbelow\dimen0 }
\def\shiftproofbelowneg{\def\next{\multiply\dimen0 by-1 }%
\afterassignment\setshiftproofbelow\dimen0 }
\def\setshiftproofbelow{\next\proofbelowshift=\dimen0 }
\def\setproofrulebreadth{\proofrulebreadth}

\def\prooftree{% NESTED ZERO (\ifonleftofproofrule)
\ifnum  \lastpenalty=1 \then   \unpenalty \else
\onleftofproofrulefalse \fi
\ifonleftofproofrule \else   \ifinsideprooftree
        \then   \hskip.5em plus1fil
        \fi
\fi
\bgroup% NESTED ONE (\proofbelow, \proofrulename, \proofabove,
\setbox\proofbelow=\hbox{}\setbox\proofrulename=\hbox{}%
\let\justifies\proofover\let\leadsto\proofoverdots\let\Justifies\proofoverdbl
\let\using\proofusing\let\[\prooftree
\ifinsideprooftree\let\]\endprooftree\fi
\proofdotsfalse\doubleprooffalse
\let\thickness\setproofrulebreadth
\let\shiftright\shiftproofbelow \let\shift\shiftproofbelow
\let\shiftleft\shiftproofbelowneg
\let\ifwasinsideprooftree\ifinsideprooftree
\insideprooftreetrue
\setbox\proofabove=\hbox\bgroup$\displaystyle % NESTED TWO
\let\wereinproofbit\prooftree
\shortenproofleft=0pt \shortenproofright=0pt \proofbelowshift=0pt
\onleftofproofruletrue\penalty1 }

\def\eproofbit{% NESTED TWO
\ifx    \wereinproofbit\prooftree \then   \ifcase \lastpenalty
        \then   \shortenproofright=0pt  % 0: some other object, no indentation
        \or     \unpenalty\hfil         % 1: empty hypotheses, just glue
        \or     \unpenalty\unskip       % 2: just had a tree, remove glue
        \else   \shortenproofright=0pt  % eh?
        \fi
\fi
\global\dimen0=\shortenproofleft \global\dimen1=\shortenproofright
\global\dimen2=\proofrulebreadth \global\dimen3=\proofbelowshift
\global\dimen4=\proofdotseparation
$\egroup  % NESTED ONE
\shortenproofleft=\dimen0 \shortenproofright=\dimen1
\proofrulebreadth=\dimen2 \proofbelowshift=\dimen3
\proofdotseparation=\dimen4
}

\def\proofover{% NESTED TWO
\eproofbit % NESTED ONE
\setbox\proofbelow=\hbox\bgroup % NESTED TWO
\let\wereinproofbit\proofover
$\displaystyle
}%
\def\proofoverdbl{% NESTED TWO
\eproofbit % NESTED ONE
\doubleprooftrue
\setbox\proofbelow=\hbox\bgroup % NESTED TWO
\let\wereinproofbit\proofoverdbl
$\displaystyle
}%
\def\proofoverdots{% NESTED TWO
\eproofbit % NESTED ONE
\proofdotstrue
\setbox\proofbelow=\hbox\bgroup % NESTED TWO
\let\wereinproofbit\proofoverdots
$\displaystyle
}%
\def\proofusing{% NESTED TWO
\eproofbit % NESTED ONE
\setbox\proofrulename=\hbox\bgroup % NESTED TWO
\let\wereinproofbit\proofusing
\kern0.3em$ }

\def\endprooftree{% NESTED TWO
\eproofbit % NESTED ONE
  \dimen5 =0pt% spread of hypotheses
\dimen0=\wd\proofabove \advance\dimen0-\shortenproofleft
\advance\dimen0-\shortenproofright
\dimen1=.5\dimen0 \advance\dimen1-.5\wd\proofbelow \dimen4=\dimen1
\advance\dimen1\proofbelowshift \advance\dimen4-\proofbelowshift
\ifdim  \dimen1<0pt \then   \advance\shortenproofleft\dimen1
        \advance\dimen0-\dimen1
        \dimen1=0pt
        \ifdim  \shortenproofleft<0pt
        \then   \setbox\proofabove=\hbox{%
                        \kern-\shortenproofleft\unhbox\proofabove}%
                \shortenproofleft=0pt
        \fi
\fi
\ifdim  \dimen4<0pt \then   \advance\shortenproofright\dimen4
        \advance\dimen0-\dimen4
        \dimen4=0pt
\fi
\ifdim  \shortenproofright<\wd\proofrulename \then
\shortenproofright=\wd\proofrulename \fi
\dimen2=\shortenproofleft \advance\dimen2 by\dimen1
\dimen3=\shortenproofright\advance\dimen3 by\dimen4
\ifproofdots \then
        \dimen6=\shortenproofleft \advance\dimen6 .5\dimen0
        \setbox1=\vbox to\proofdotseparation{\vss\hbox{$\cdot$}\vss}
        \setbox0=\hbox{%
                \kern\dimen6
                $\vcenter to\proofdotnumber\proofdotseparation
                        {\leaders\box1\vfill}$%
                \unhbox\proofrulename}%
\else   \dimen6=\fontdimen22\the\textfont2 % height of maths axis
        \dimen7=\dimen6
        \advance\dimen6by.5\proofrulebreadth
        \advance\dimen7by-.5\proofrulebreadth
        \setbox0=\hbox{%
                \kern\shortenproofleft
                \ifdoubleproof
                \then   \hbox to\dimen0{%
                        $\mathsurround0pt\mathord=\mkern-6mu%
                        \cleaders\hbox{$\mkern-2mu=\mkern-2mu$}\hfill
                        \mkern-6mu\mathord=$}%
                \else   \vrule height\dimen6 depth-\dimen7 width\dimen0
                \fi
                \unhbox\proofrulename}%
        \ht0=\dimen6 \dp0=-\dimen7
\fi
\let\doll\relax
\ifwasinsideprooftree \then   \let\VBOX\vbox \else
\ifmmode\else$\let\doll=$\fi
        \let\VBOX\vcenter
\fi
\VBOX   {\baselineskip\proofrulebaseline \lineskip.2ex
        \expandafter\lineskiplimit\ifproofdots0ex\else-0.6ex\fi
        \hbox   spread\dimen5   {\hfi\unhbox\proofabove\hfi}%
        \hbox{\box0}%
        \hbox   {\kern\dimen2 \box\proofbelow}}\doll%
\global\dimen2=\dimen2 \global\dimen3=\dimen3
\egroup % NESTED ZERO
\ifonleftofproofrule \then   \shortenproofleft=\dimen2 \fi
\shortenproofright=\dimen3
\onleftofproofrulefalse \ifinsideprooftree \then   \hskip.5em plus
1fil \penalty2 \fi }